\documentclass[notitlepage,nofootinbib,aps,prd,a4paper,11pt,amsfonts,amssymb,amsmath]{revtex4-1}
\usepackage[colorlinks,hyperindex,plainpages=false]{hyperref}
\usepackage[utf8]{inputenc}
\usepackage{graphicx}
\usepackage{bigints}

\begin{document}
\title{Post-Newtonian parameter $\gamma$ and the deflection of light\\in ghost-free massive bimetric gravity}

\author{Manuel Hohmann}
\email{manuel.hohmann@ut.ee}
\affiliation{Laboratory of Theoretical Physics, Institute of Physics, University of Tartu, W. Ostwaldi 1, 50411 Tartu, Estonia}

\begin{abstract}
We consider the parametrized post-Newtonian (PPN) limit of ghost-free massive bimetric gravity with two mutually non-interacting matter sectors coupled to the two metrics. Making use of a gauge-invariant differential decomposition of the metric perturbations, we solve the field equations up to the linear PPN order for a static, point-like mass source. From the result we derive the PPN parameter \(\gamma\) for spherically symmetric systems, which describes the gravitational deflection of light by visible matter. By a comparison to its value measured in the solar system we obtain bounds on the parameters of the theory. We further discuss the deflection of light by dark matter and find an agreement with the observed light deflection by galaxies. We finally speculate about a possible explanation for the observed distribution of dark matter in galactic mergers such as Abell 520 and Abell 3827.
\end{abstract}
\maketitle

\section{Motivation}\label{sec:motivation}
Current observations, together with their interpretation according to the standard $\Lambda$CDM model of cosmology, indicate that visible matter constitutes only 5\% of the total matter content of the universe, while the remaining constituents are given by 26\% dark matter and 69\% dark energy~\cite{Ade:2015xua}. These dark components of the universe arise purely from phenomenology: dark energy offers a potential explanation for the observed accelerating expansion of the universe~\cite{Peebles:2002gy,Riess:1998cb,Perlmutter:1998np,Kowalski:2008ez,Amanullah:2010vv,Suzuki:2011hu}, while dark matter potentially explains the rotation curves of galaxies~\cite{Rubin:1970zza,Rubin:1980zd,Persic:1995ru}, the formation of large scale structures in the early universe~\cite{Peebles:1980,Padmanabhan:1993,Springel:2005nw} or the lensing and peculiar motion in galaxy clusters~\cite{Wambsganss:1998gg,Zwicky:1933gu,Zwicky:1937zza}. Note that all of these observations are based on the gravitational influence of the dark universe on visible matter. Despite significant effort, no direct, non-gravitational interaction of visible matter and dark components has been observed. This raises the question whether any such non-gravitational interaction exists, or whether the coupling between the dark and visible sectors is purely gravitational, or even gravity itself constitutes at least part of the dark sector.

In this article we discuss a model which naturally features a purely gravitational coupling prescription for dark matter, while at the same time being able to accommodate dark energy. This model is based on the idea that the geometry of spacetime, which mediates the gravitational interaction, is described not by a single metric as in general relativity, but by two separate metrics. Further, each matter field couples to only one of these metrics, and there is no direct, non-gravitational coupling between matter fields associated to different metrics. These assumptions imply the existence of two different matter sectors, whose mutual interaction is mediated only by an interaction between the two metrics, so that they appear mutually dark. However attractive in its phenomenology, this idea also leads to potential theoretical issues, since a coupling between them requires at least one of the corresponding gravitons to be massive~\cite{Boulanger:2000rq}, and such massive gravity theories generally suffer from the existence of a ghost instability~\cite{Boulware:1973my}.

While it has been believed for several decades that the aforementioned ghost instability completely excludes any theories with gravitationally interacting massive spin 2 particles, it has been discovered recently that this is not the case, and that a particular, narrow class of theories avoids the ghost~\cite{deRham:2010kj,Hassan:2011vm,Hassan:2011hr,deRham:2011rn,Hassan:2011tf,Hassan:2011zd,deRham:2013tfa}; see~\cite{Hinterbichler:2011tt,deRham:2014zqa,Schmidt-May:2015vnx} for a number of reviews. The most simple class of such theories indeed features two metric tensors, and allows for two separate, mutually non-gravitationally non-interacting classes of matter fields, each of which couples exclusively to one metric, and which interact with each other only through an interaction between the two metrics~\cite{Baccetti:2012bk,Hassan:2014vja,Yamashita:2014fga,deRham:2014naa}. The interpretation of one of them as corresponding to dark matter, as we discussed above, has also been studied~\cite{Aoki:2013joa,Aoki:2014cla}, possibly involving an additional ``graviphoton'' vector field and reproducing modified Newtonian dynamics (MOND) on galactic scales~\cite{Bernard:2014psa,Blanchet:2015sra,Blanchet:2015bia,Blanchet:2017duj}. Note, however, that in contrast to the latter we do not introduce a graviphoton, or aim to model dark matter as a gravitational effect as in MOND. We further remark that in massive gravity theories also the massive graviton is a potential dark matter candidate, an idea which has only recently been considered~\cite{Schmidt-May:2016hsx,Babichev:2016hir,Aoki:2016zgp,Babichev:2016bxi}.

Besides providing possible explanations for the observed dark sector of the universe, any viable theory of gravity must of course also pass tests in the solar system. An important tool for testing metric gravity theories with high-precision data from solar system experiments is the parametrized post-Newtonian (PPN) formalism~\cite{Nordtvedt:1968qs,Thorne:1970wv,Will:1971zzb,Will:1971wt,Will:1993ns,Will:2014kxa}. The main idea of the PPN formalism is to express the metric tensor as a perturbation around a flat background, and then expand the perturbation in terms of certain integrals over the gravitating matter distribution. The coefficients of these potentials in the metric perturbation are characteristic for a given gravity theory, and can directly be linked to observable quantities. In this article we focus on a particular PPN parameter, conventionally denoted \(\gamma\), which has been measured to high precision in numerous solar system experiments~\cite{Will:2014kxa}, in particular through very long baseline interferometry~\cite{Shapiro:2004zz,Fomalont:2009zg,Lambert:2009xy,Lambert:2011}, via the Shapiro delay of radio signals~\cite{Bertotti:2003rm} and using combined observations of the motion of bodies in the solar system~\cite{Fienga:2011qh,Pitjeva:2013chs,Verma:2013ata,Fienga:2014}. To present all observations are in full agreement with the general relativity value \(\gamma = 1\)~\cite{Will:2014kxa}.

While all of the aforementioned experiments observed the gravitational interaction within the visible sector, it should be noted that the PPN parameter \(\gamma\) has also been determined through the deflection of visible light by galaxies, whose total gravitating matter content contains a significant contribution from dark matter~\cite{Bolton:2006yz,Schwab:2009nz,Cao:2017nnq}. Also these observations, although less precise, are in agreement with the general relativity value \(\gamma = 1\). It is needless to say that understanding the light deflection by dark matter is essential for a correct interpretation of observations where the dark matter distribution is reconstructed from lensing under the assumption that dark matter deflects light in the same way as visible matter. This is particularly important in the case of galactic mergers, such as the so-called ``Bullet Cluster'' 1E0657-558~\cite{Tucker:1998tp,Markevitch:2003at,Clowe:2003tk,Clowe:2006eq,Randall:2007ph}, the ``Train Wreck Cluster'' Abell 520~\cite{Mahdavi:2007yp,Jee:2014hja}, MACS J0025.4-1222~\cite{Bradac:2008eu} or Abell 3827~\cite{Massey:2015dkw}, where visible and dark matter appear clearly separated from each other. However, it is not a priori clear that this assumption is valid in a theory in which dark and visible matter couple differently to gravity.

As mentioned above, the PPN formalism in its standard form requires a single dynamical metric for the description of gravity. In order to discuss theories with multiple metric tensors, we need an extension of this standard PPN formalism. A possible extension, which features massless and massive gravity modes, but includes only one type of gravitating source matter, has been introduced and applied in~\cite{Clifton:2010hz}. A complementary extension to multiple metrics and a corresponding number of matter sectors, but including only massless gravity modes, has been developed and applied in~\cite{Hohmann:2010ni,Hohmann:2013oca}. In this article we choose to make use of the latter, and to extend it to also allow calculating the PPN parameter \(\gamma\) for both dark and visible matter in massive gravity. This is the simplest possible extension, and the first step towards a fully general extension of the formalism to massive gravity theories; the latter would allow for a calculation of all PPN parameters.

We remark that the perturbative expansion of the metric in a weak field limit, which is an important ingredient to the PPN formalism, is not always valid in the context of bimetric gravity due to the Vainshtein mechanism~\cite{Vainshtein:1972sx,Babichev:2013pfa,Babichev:2013usa}, and that a full, non-linear treatment is required in order to determine the gravitational dynamics close to the source mass. This non-linear mechanism typically suppresses all deviations from general relativity within a given radius around the source mass, called the Vainshtein radius. A perturbative treatment is valid only outside this radius. We will not discuss the Vainshtein mechanism in this article, and restrict ourselves to the case of theories in which the Vainshtein radius is sufficiently small so that the perturbative treatment is valid on solar system scales and above.

The outline of this article is as follows. In section~\ref{sec:action} we briefly review the action and field equations of ghost-free massive bimetric gravity. We then perform a perturbative expansion of these field equations in section~\ref{sec:ppn}, using an adapted version of the PPN formalism. We further simplify the obtained equations using gauge-invariant perturbation theory in section~\ref{sec:gidecomp}. This will yield us a set of equations, which we will solve for a static, point-like mass source in section~\ref{sec:solution}, and thus determine the effective Newtonian gravitational constant and PPN parameter \(\gamma\). We will connect our result to observations, in particular of the deflection of light, in section~\ref{sec:observation}. We end with a conclusion in section~\ref{sec:conclusion}. A few lengthy calculations are displayed in the appendix. In appendix~\ref{app:linpot} we derive the linearized interaction potential connecting the two metrics. In appendix~\ref{app:yukawa} we list derivatives of the Yukawa potential. We show how to check our solution of the field equations in appendix~\ref{app:feqcheck}.

\section{Action and field equations}\label{sec:action}
In this section we start our discussion of the post-Newtonian limit of bimetric gravity with a brief review of its action and gravitational field equations, which are derived by variation with respect to the two metrics. We then trace-reverse the field equations, as this will be more convenient when we construct their solution. These trace-reversed field equations will be the main ingredient for our calculation.

The starting point for our derivation is the action functional
\begin{multline}\label{eqn:action}
S = \bigintss_Md^4x\Bigg[\frac{m_g^2}{2}\sqrt{-\det g}R^g + \frac{m_f^2}{2}\sqrt{-\det f}R^f - m^4\sqrt{-\det g}\sum_{n = 0}^4\beta_ne_n\left(\sqrt{g^{-1}f}\right)\\
+ \sqrt{-\det g}\mathcal{L}_m^g(g,\Phi^g) + \sqrt{-\det f}\mathcal{L}_m^f(f,\Phi^f)\Bigg]
\end{multline}
for two metric tensors \(g_{\mu\nu}, f_{\mu\nu}\) and two sets of matter fields \(\Phi^{g,f}\), each of which couples to only one metric tensor, and between which there is no direct, non-gravitational interaction. One may thus interpret, e.g., \(\Phi^g\) as visible matter, constituted by the standard model fields and governed by the standard matter Lagrangian \(\mathcal{L}^g\), and \(\Phi^f\) as dark matter, constituted by a distinct set of fields with possibly different structure of the Lagrangian \(\mathcal{L}^f\). However, we will not make any assumptions on the constituting fields of the two matter types here, or on their Lagrangians, as these will not be relevant for our calculation.

Note that since there are two metrics, there is no canonical prescription for raising or lowering tensor indices. Therefore, we will not raise or lower indices automatically, but provide definitions for all tensor fields with fixed index positions. In the action~\eqref{eqn:action} this applies to the Ricci scalars, each of which is defined solely through its corresponding metric, such that they are related to the Ricci tensors by
\begin{equation}
R^g = g^{\mu\nu}R^g_{\mu\nu}\,, \quad R^f = f^{\mu\nu}R^f_{\mu\nu}\,.
\end{equation}
Note further the appearance of the $(1,1)$ tensor field \(g^{-1}f\), which we assume to have a square root \(A\) such that
\begin{equation}\label{eqn:sqrt}
A^{\mu}{}_{\sigma}A^{\sigma}{}_{\nu} = g^{\mu\sigma}f_{\sigma\nu}\,.
\end{equation}
This is certainly the case in a sufficiently small neighborhood of the flat proportional background metrics \(g_{\mu\nu} = \eta_{\mu\nu}, f_{\mu\nu} = c^2\eta_{\mu\nu}\), which we will henceforth consider. The functions \(e_0, \ldots, e_4\) in the action are the matrix invariants
\begin{equation}\label{eqn:matrixinv}
e_k(A) = A^{\mu_1}{}_{[\mu_1} \cdots A^{\mu_k}{}_{\mu_k]} = \frac{1}{k!(4 - k)!}\epsilon^{\mu_1 \cdots \mu_k \lambda_1 \cdots \lambda_{4 - k}}\epsilon_{\nu_1 \cdots \nu_k \lambda_1 \cdots \lambda_{4 - k}}A^{\nu_1}{}_{\mu_1} \cdots A^{\nu_k}{}_{\mu_k}
\end{equation}
of this square root, while their coefficients \(\beta_0, \ldots, \beta_4\) are constant, dimensionless parameters to the action. The remaining parameters are the Planck masses \(m_g, m_f\) and the interaction mass \(m\), all of which are of mass dimension. Any choice of the constant parameters determines a particular action, and hence a particular theory.

By variation with respect to the metric tensors we obtain the gravitational field equations
\begin{equation}\label{eqn:fieldeq}
m_g^2\left(R^g_{\mu\nu} - \frac{1}{2}g_{\mu\nu}R^g\right) + m^4V^g_{\mu\nu} = T^g_{\mu\nu}\,, \quad
m_f^2\left(R^f_{\mu\nu} - \frac{1}{2}f_{\mu\nu}R^f\right) + m^4V^f_{\mu\nu} = T^f_{\mu\nu}\,.
\end{equation}
Here we have defined the energy-momentum tensors as usual through
\begin{equation}
T^g_{\mu\nu} = -\frac{2}{\sqrt{-\det g}}\frac{\delta\left(\sqrt{-\det g}\mathcal{L}_m^g(g,\Phi_g)\right)}{\delta g^{\mu\nu}}\,, \quad
T^f_{\mu\nu} = -\frac{2}{\sqrt{-\det f}}\frac{\delta\left(\sqrt{-\det f}\mathcal{L}_m^f(f,\Phi_f)\right)}{\delta f^{\mu\nu}}\,.
\end{equation}
The potential terms \(V^{g,f}_{\mu\nu}\) are given by
\begin{equation}\label{eqn:potential}
V^g_{\mu\nu} = g_{\mu\rho}\sum_{n = 0}^3(-1)^n\beta_nY_n{}^{\rho}{}_{\nu}(A)\,, \quad
V^f_{\mu\nu} = f_{\mu\rho}\sum_{n = 0}^3(-1)^n\beta_{4 - n}Y_n{}^{\rho}{}_{\nu}(A^{-1})\,,
\end{equation}
where the functions \(Y_0, \ldots, Y_3\) are defined as
\begin{equation}\label{eqn:potsum}
Y_n(A) = \sum_{k = 0}^n(-1)^ke_k(A)A^{n - k}
\end{equation}
and analogously for \(Y_n(A^{-1})\). We remark that the action~\eqref{eqn:action}, and hence also the field equations~\eqref{eqn:fieldeq}, are fully symmetric under a simultaneous exchange \(g_{\mu\nu} \leftrightarrow f_{\mu\nu}\) and \(\beta_n \leftrightarrow \beta_{4 - n}\).

While we could work directly with the field equations~\eqref{eqn:fieldeq}, it turns out to be more convenient to use the trace-reversed equations instead, which read
\begin{equation}\label{eqn:trrevfeq}
m_g^2R^g_{\mu\nu} + m^4\bar{V}^g_{\mu\nu} = \bar{T}^g_{\mu\nu}\,, \quad
m_f^2R^f_{\mu\nu} + m^4\bar{V}^f_{\mu\nu} = \bar{T}^f_{\mu\nu}\,.
\end{equation}
Here we have trace-reversed each term with its corresponding metric, i.e., we have applied the definitions
\begin{subequations}\label{eqn:trrevpot}
\begin{align}
\bar{V}^g_{\mu\nu} &= V^g_{\mu\nu} - \frac{1}{2}g_{\mu\nu}g^{\rho\sigma}V^g_{\rho\sigma}\,, &
\bar{T}^g_{\mu\nu} &= T^g_{\mu\nu} - \frac{1}{2}g_{\mu\nu}g^{\rho\sigma}T^g_{\rho\sigma}\,,\\
\bar{V}^f_{\mu\nu} &= V^f_{\mu\nu} - \frac{1}{2}f_{\mu\nu}f^{\rho\sigma}V^f_{\rho\sigma}\,, &
\bar{T}^f_{\mu\nu} &= T^f_{\mu\nu} - \frac{1}{2}f_{\mu\nu}f^{\rho\sigma}T^f_{\rho\sigma}\,.
\end{align}
\end{subequations}
The field equations~\eqref{eqn:trrevfeq} are the equations we will be working with during the remainder of this article. In order to calculate the post-Newtonian limit, we will need a perturbative expansion of these equations. This will be done in the next section.

\section{Post-Newtonian approximation}\label{sec:ppn}
We now come to a perturbative expansion of the field equations~\eqref{eqn:trrevfeq} displayed in the previous section. For this purpose, we first briefly review the notion of velocity orders in section~\ref{ssec:ppnorder}, and label the relevant components of the metric and energy-momentum tensors. We then discuss the metric ansatz, essentially following the construction developed in~\cite{Hohmann:2010ni,Hohmann:2013oca}, in section~\ref{ssec:ppnmetric}. Finally, we apply these constructions to the field equations under consideration. We derive and solve the field equations at the zeroth velocity order in section~\ref{ssec:background}, and derive the second order equations in section~\ref{ssec:2ndorder}.

\subsection{Expansion in velocity orders}\label{ssec:ppnorder}
A central ingredient of the PPN formalism is the assumption that the gravitating source matter is constituted by a perfect fluid. Since there are two different types of matter \(\Phi^{g,f}\) in the theory we consider, which interact only gravitationally, we apply this assumption to each of them. Their energy-momentum tensors therefore take the form
\begin{subequations}\label{eqn:enmomten}
\begin{align}
T^{g\,\mu\nu} &= (\rho^g + \rho^g\Pi^g + p^g)u^{g\,\mu}u^{g\,\nu} + p^gg^{\mu\nu}\,,\\
T^{f\,\mu\nu} &= (\rho^f + \rho^f\Pi^f + p^f)u^{f\,\mu}u^{f\,\nu} + p^ff^{\mu\nu}\,,
\end{align}
\end{subequations}
with rest energy densities \(\rho^{g,f}\), specific internal energies \(\Pi^{g,f}\), pressures \(p^{g,f}\) and four-velocities \(u^{g,f\,\mu}\). Note that the four-velocities are normalized with their corresponding metrics,
\begin{equation}
u^{g\,\mu}u^{g\,\nu}g_{\mu\nu} = u^{f\,\mu}u^{f\,\nu}f_{\mu\nu} = -1\,.
\end{equation}
We further assume that the source matter is slow-moving within our chosen frame of reference, so that the velocity components satisfy
\begin{equation}
v^{g,f\,i} = \frac{u^{g,f\,i}}{u^{g,f\,0}} \ll 1\,.
\end{equation}
We then assign orders of magnitude \(\mathcal{O}(n) \propto |\vec{v}|^n\) to all dynamical quantities. For the matter variables we assign \(\rho^{g,f} \sim \Pi^{g,f} \sim \mathcal{O}(2)\) and \(p^{g,f} \sim \mathcal{O}(4)\), based on their values for the matter constituting the solar system. For the metrics we assume a small perturbation around a flat, proportional background solution, where we expand the perturbation in velocity orders in the form
\begin{subequations}\label{eqn:metperturb}
\begin{align}
g_{\mu\nu} = \eta_{\mu\nu} + h_{\mu\nu} &= \eta_{\mu\nu} + h^{(1)}_{\mu\nu} + h^{(2)}_{\mu\nu} + h^{(3)}_{\mu\nu} + h^{(4)}_{\mu\nu} + \mathcal{O}(5)\,,\\
c^{-2}f_{\mu\nu} = \eta_{\mu\nu} + e_{\mu\nu} &= \eta_{\mu\nu} + e^{(1)}_{\mu\nu} + e^{(2)}_{\mu\nu} + e^{(3)}_{\mu\nu} + e^{(4)}_{\mu\nu} + \mathcal{O}(5)
\end{align}
\end{subequations}
with constant \(c > 0\). Not all of these components are relevant for the post-Newtonian limit, while others vanish due to symmetries and conservation laws. The only non-vanishing components that are relevant for our calculation of the PPN parameter \(\gamma\) in this article are
\begin{equation}\label{eqn:relcomps}
h^{(2)}_{00}\,, \quad h^{(2)}_{ij}\,, \quad e^{(2)}_{00}\,, \quad e^{(2)}_{ij}\,.
\end{equation}
Further, we only consider quasi-static solutions, so that changes of the metric are induced only by the motion of the source matter. We therefore assign another velocity order \(\partial_0\) to any time derivative. We finally assume that the source matter is located in a bounded region and that the metrics are asymptotically flat, so that the metric perturbations and their derivatives vanish at infinity.

Since we are interested only in the second order metric perturbations~\eqref{eqn:relcomps}, and hence need to solve the field equations only up to the second velocity order, it is also sufficient to expand the energy-momentum tensors~\eqref{eqn:enmomten} to the second velocity order. The only relevant components are
\begin{equation}
T^{g(2)\,00} = \rho^g\,, \quad T^{f(2)\,00} = \frac{\rho^f}{c^2}\,, \quad T^{g(2)\,ij} = T^{f(2)\,ij} = 0\,.
\end{equation}
In order to use them in the field equations~\eqref{eqn:fieldeq}, we need to lower their indices with their corresponding metrics, which yields
\begin{equation}
T^{g(2)}_{00} = \rho^g\,, \quad T^{f(2)}_{00} = c^2\rho^f\,, \quad T^{g(2)}_{ij} = T^{f(2)}_{ij} = 0\,.
\end{equation}
Finally, we also need to trace-reverse these terms with their corresponding metrics, from which we obtain
\begin{equation}\label{eqn:enmom2nd}
\bar{T}^{g(2)}_{00} = \frac{1}{2}\rho^g\,, \quad
\bar{T}^{f(2)}_{00} = \frac{c^2}{2}\rho^f\,, \quad
\bar{T}^{g(2)}_{ij} = \frac{1}{2}\rho^g\delta_{ij}\,, \quad
\bar{T}^{f(2)}_{ij} = \frac{c^2}{2}\rho^f\delta_{ij}\,.
\end{equation}
These terms will enter the trace-reversed field equations~\eqref{eqn:trrevfeq}.

\subsection{Metric ansatz and PPN parameters}\label{ssec:ppnmetric}
Another important ingredient of the PPN formalism is an ansatz for the metric in terms of potentials, which are integrals over the source matter distribution. Their coefficients in the metric are observable quantities which allow a characterization of the gravity theory under examination. For single metric theories there is a standard form for this PPN metric ansatz~\cite{Will:1993ns,Will:2014kxa}. Here we use a generalization to multimetric theories developed in~\cite{Hohmann:2010ni,Hohmann:2013oca}. Our ansatz for the second order metric perturbation reads
\begin{subequations}\label{eqn:ppnmetric}
\begin{align}
h^{(2)}_{00} &= -\alpha^{gg}\triangle\chi^g - \alpha^{gf}\triangle\chi^f\,,\\
h^{(2)}_{ij} &= 2\theta^{gg}\chi^g_{,ij} + 2\theta^{gf}\chi^f_{,ij} - \left[(\gamma^{gg} + \theta^{gg})\triangle\chi^g + (\gamma^{gf} + \theta^{gf})\triangle\chi^f\right]\delta_{ij}\,,\\
e^{(2)}_{00} &= -\alpha^{fg}\triangle\chi^g - \alpha^{ff}\triangle\chi^f\,,\\
e^{(2)}_{ij} &= 2\theta^{fg}\chi^g_{,ij} + 2\theta^{ff}\chi^f_{,ij} - \left[(\gamma^{fg} + \theta^{fg})\triangle\chi^g + (\gamma^{ff} + \theta^{ff})\triangle\chi^f\right]\delta_{ij}\,,
\end{align}
\end{subequations}
where \(\triangle = \partial^i\partial_i\) and indices are raised and lowered with the flat metric \(\eta_{\mu\nu}\). The PPN potentials we have introduced here are second order derivatives of the superpotentials
\begin{equation}\label{eqn:superpot}
\chi^g(t, \vec{x}) = -\int\rho^g(t, \vec{x}')\,|\vec{x} - \vec{x}'|\,d^3x'\,, \quad
\chi^f(t, \vec{x}) = -c^3\int\rho^f(t, \vec{x}')\,|\vec{x} - \vec{x}'|\,d^3x'\,.
\end{equation}
The definition of \(\chi^f\) contains a factor \(c^3\), which originates from the volume element of the spatial part of the unperturbed contribution \(c^2\eta_{\mu\nu}\) of the metric \(f_{\mu\nu}\). In the PPN metric we further have twelve PPN parameters \(\alpha^{g,f\,g,f}, \gamma^{g,f\,g,f}, \theta^{g,f\,g,f}\). Note that these are not independent, as we have not yet fixed a gauge for the metric. The gauge freedom allows us to apply a diffeomorphism generated by a vector field \(\xi^{\mu}\), provided that it preserves the perturbation ansatz~\eqref{eqn:metperturb}. This means that the vector field \(\xi^{\mu}\) must be of the same order as the metric perturbations. Recall that under a diffeomorphism the metrics change according to
\begin{equation}
\delta_{\xi}g_{\mu\nu} = (\mathcal{L}_{\xi}g)_{\mu\nu} = 2g_{\sigma(\mu}\nabla^g_{\nu)}\xi^{\sigma}\,, \quad \delta_{\xi}f_{\mu\nu} = (\mathcal{L}_{\xi}f)_{\mu\nu} = 2f_{\sigma(\mu}\nabla^f_{\nu)}\xi^{\sigma}\,,
\end{equation}
where \(\mathcal{L}\) denotes the Lie derivative. At the linear perturbation level, which is sufficient for our calculation here, this yields the transformation of the metric perturbations
\begin{equation}
\delta_{\xi}h_{\mu\nu} = \delta_{\xi}e_{\mu\nu} = 2\eta_{\sigma(\mu}\partial_{\nu)}\xi^{\sigma}\,.
\end{equation}
Further demanding consistency with the PPN metric ansatz~\eqref{eqn:ppnmetric} we find that the only allowed and relevant vector field is of second velocity order and can be written as~\cite{Will:1993ns,Hohmann:2013oca}
\begin{equation}
\xi_0 = 0\,, \quad \xi_i = \lambda^g\chi^g_{,i} + \lambda^f\chi^f_{,i}\,,
\end{equation}
with two constants \(\lambda^{g,f}\), and where we have defined \(\xi_{\mu} = \eta_{\mu\nu}\xi^{\nu}\). Under a diffeomorphism generated by this vector field the metric perturbations change by
\begin{equation}
\delta_{\xi}h^{(2)}_{00} = \delta_{\xi}e^{(2)}_{00} = 0\,, \quad \delta_{\xi}h^{(2)}_{ij} = \delta_{\xi}e^{(2)}_{ij} = 2\lambda^g\chi^g_{,ij} + 2\lambda^f\chi^f_{,ij}\,.
\end{equation}
By choosing \(\lambda^g = -\theta^{gg}\) and \(\lambda^f = -\theta^{ff}\) we can always eliminate these two PPN parameters from the metric ansatz. In the remainder of our calculation we will adopt this gauge, in which \(\theta^{gg} = \theta^{ff} = 0\), as this turns out to be compatible with the standard PPN gauge for single metric theories~\cite{Hohmann:2013oca}.

\subsection{Background solution}\label{ssec:background}
Recall from section~\ref{ssec:ppnorder} that we have expanded the metrics around a flat Minkowski background, as usual in the PPN formalism. For the PPN formalism to be applicable in this form, it is necessary that this background is a solution of the field equations at the zeroth velocity order. Since both the Ricci tensors \(R^{g,f(0)}_{\mu\nu}\) and the energy-momentum tensors \(\bar{T}^{g,f(0)}_{\mu\nu}\) at the zeroth velocity order vanish, these simply reduce to
\begin{equation}
m^4\bar{V}^{g(0)}_{\mu\nu} = 0\,, \quad m^4\bar{V}^{f(0)}_{\mu\nu} = 0\,,
\end{equation}
where we assume \(m > 0\). In order to determine the potential terms \(\bar{V}^{g,f(0)}_{\mu\nu}\), and also the second velocity order in the next section, it is useful to first linearize the potential in the metric perturbations. Since this is a rather lengthy procedure, we have deferred it to appendix~\ref{app:linpot}. Here we make use of the result~\eqref{eqn:lintrpot}, from which we read off the zeroth order contribution
\begin{subequations}
\begin{align}
\bar{V}^{g(0)}_{\mu\nu} &= -(\tilde{\beta}_0 + 3\tilde{\beta}_1 + 3\tilde{\beta}_2 + \tilde{\beta}_3)\eta_{\mu\nu} = -(\beta_0 + 3c\beta_1 + 3c^2\beta_2 + c^3\beta_3)\eta_{\mu\nu}\,,\\
\bar{V}^{f(0)}_{\mu\nu} &= -(\tilde{\beta}_1 + 3\tilde{\beta}_2 + 3\tilde{\beta}_3 + \tilde{\beta}_4)c^{-2}\eta_{\mu\nu} = -(\beta_1 + 3c\beta_2 + 3c^2\beta_3 + c^3\beta_4)c^{-1}\eta_{\mu\nu}\,,
\end{align}
\end{subequations}
where we used the abbreviations \(\tilde{\beta}_k = c^k\beta_k\). We require that these equations, which are polynomial in \(c\), possess at least one common positive solution \(c > 0\). Note that a particular fixed \(c\) solves both equations if and only if the parameters in the action~\eqref{eqn:action} satisfy
\begin{equation}
\beta_0 = -3c\beta_1 - 3c^2\beta_2 - c^3\beta_3\,, \quad
\beta_4 = -c^{-3}\beta_1 - 3c^{-2}\beta_2 - 3c^{-1}\beta_3\,.
\end{equation}
The condition that the background equations are solved by proportional flat metrics therefore completely determines the two parameters \(\beta_0\) and \(\beta_4\) in the action in terms of a new free parameter \(c > 0\). In the following we will therefore replace \(\beta_0\) and \(\beta_4\), and hence \(\tilde{\beta}_0\) and \(\tilde{\beta}_4\), using
\begin{equation}
\tilde{\beta}_0 = -3\tilde{\beta}_1 - 3\tilde{\beta}_2 - \tilde{\beta}_3\,, \quad
\tilde{\beta}_4 = -\tilde{\beta}_1 - 3\tilde{\beta}_2 - 3\tilde{\beta}_3\,,
\end{equation}
and keep \(\beta_1, \beta_2, \beta_3\) and \(c\) as free parameters of the class of theories we discuss.

\subsection{Second order field equations}\label{ssec:2ndorder}
For the remainder of our calculation and in order to determine the second order metric perturbations we will need to expand the field equations~\eqref{eqn:trrevfeq} to the second velocity order. The only relevant components are given by
\begin{subequations}\label{eqn:2ndorder}
\begin{align}
m_g^2R^{g(2)}_{00} + m^4\bar{V}^{g(2)}_{00} &= \bar{T}^{g(2)}_{00}\,, &
m_g^2R^{g(2)}_{ij} + m^4\bar{V}^{g(2)}_{ij} &= \bar{T}^{g(2)}_{ij}\,,\\
m_f^2R^{f(2)}_{00} + m^4\bar{V}^{f(2)}_{00} &= \bar{T}^{f(2)}_{00}\,, &
m_f^2R^{f(2)}_{ij} + m^4\bar{V}^{f(2)}_{ij} &= \bar{T}^{f(2)}_{ij}\,.
\end{align}
\end{subequations}
We have already calculated the necessary components~\eqref{eqn:enmom2nd} of the energy-momentum tensor at the second velocity order. The components of the Ricci tensor are easily obtained and yield the standard textbook result~\cite{Will:1993ns}
\begin{subequations}\label{eqn:ricci2nd}
\begin{align}
R^{g(2)}_{00} &= -\frac{1}{2}\triangle h^{(2)}_{00}\,, &
R^{g(2)}_{ij} &= -\frac{1}{2}\left(\triangle h^{(2)}_{ij} - h^{(2)}_{00,ij} + h^{(2)}_{kk,ij} - h^{(2)}_{ik,jk} - h^{(2)}_{jk,ik}\right)\,,\\
R^{f(2)}_{00} &= -\frac{1}{2}\triangle e^{(2)}_{00}\,, &
R^{f(2)}_{ij} &= -\frac{1}{2}\left(\triangle e^{(2)}_{ij} - e^{(2)}_{00,ij} + e^{(2)}_{kk,ij} - e^{(2)}_{ik,jk} - e^{(2)}_{jk,ik}\right)\,.
\end{align}
\end{subequations}
Finally, we also need the second velocity order contribution from the potential terms. Using the result~\eqref{eqn:lintrpot} derived in appendix~\eqref{app:linpot} one finds the components
\begin{subequations}\label{eqn:pot2nd}
\begin{align}
\bar{V}^{g(2)}_{00} &= \frac{1}{4}\tilde{\beta}\left(3h^{(2)}_{00} - 3e^{(2)}_{00} - h^{(2)}_{ii} + e^{(2)}_{ii}\right)\,,\\
\bar{V}^{g(2)}_{ij} &= \frac{1}{4}\tilde{\beta}\left[2h^{(2)}_{ij} - 2e^{(2)}_{ij} + \left(h^{(2)}_{kk} - e^{(2)}_{kk} - h^{(2)}_{00} + e^{(2)}_{00}\right)\delta_{ij}\right]\,,\\
\bar{V}^{f(2)}_{00} &= \frac{1}{4c^2}\tilde{\beta}\left(3e^{(2)}_{00} - 3h^{(2)}_{00} - e^{(2)}_{ii} + h^{(2)}_{ii}\right)\,,\\
\bar{V}^{f(2)}_{ij} &= \frac{1}{4c^2}\tilde{\beta}\left[2e^{(2)}_{ij} - 2h^{(2)}_{ij} + \left(e^{(2)}_{kk} - h^{(2)}_{kk} - e^{(2)}_{00} + h^{(2)}_{00}\right)\delta_{ij}\right]\,,
\end{align}
\end{subequations}
where we introduced the abbreviation
\begin{equation}
\tilde{\beta} = \tilde{\beta}_1 + 2\tilde{\beta}_2 + \tilde{\beta}_3\,.
\end{equation}
These are the field equations we will be using during the remainder of this article. However, directly working with these equations poses two difficulties. First, the field equations possess a gauge freedom, as discussed in section~\ref{ssec:ppnmetric}, and so the solution will be unique only after gauge fixing. Second, the equations turn out to be involved and cumbersome to solve due to the mixing of tensor components. Both of these difficulties can be solved straightforwardly by performing a gauge-invariant differential decomposition of the metric perturbations. We will detail this formalism in the next section.

\section{Gauge-invariant differential decomposition}\label{sec:gidecomp}
In the previous section we have performed an expansion of the gravitational field equations~\eqref{eqn:trrevfeq} into velocity orders and obtained the second order equations~\eqref{eqn:2ndorder}. Instead of solving them directly for the metric perturbations~\eqref{eqn:metperturb}, we will first bring them into a significantly simpler form in this section. For this purpose, we employ the formalism of gauge-invariant perturbations, which is well-known from cosmology~\cite{Bardeen:1980kt,Malik:2008im,Stewart:1990fm}. We apply this procedure in several steps. First, we decompose the metric perturbations into gauge-invariant potentials in section~\ref{ssec:metricdec}. In section~\ref{ssec:invpotvel} we further decompose these potentials into velocity orders as required by the PPN formalism. Using the expressions obtained, we then decompose the Ricci tensors~\eqref{eqn:ricci2nd} in section~\ref{ssec:riccidec}, the potentials~\eqref{eqn:pot2nd} in section~\ref{ssec:potdec} and the energy-momentum tensors~\eqref{eqn:enmom2nd} in section~\ref{ssec:enmomdec}. This will finally yield us a full decomposition of the field equations~\eqref{eqn:2ndorder} in section~\ref{ssec:feqdec}.

\subsection{Decomposition of the metrics}\label{ssec:metricdec}
We start with a differential decomposition of the metric perturbations. Using the split into time and space components, we introduce the decomposition
\begin{subequations}\label{eqn:metricdec}
\begin{align}
h_{00} &= -2\phi^g, &
h_{0i} &= \partial_iB^g + B^g_i, &
h_{ij} &= -2\psi^g\delta_{ij} + 2\triangle_{ij}E^g + 4\partial_{(i}E^g_{j)} + 2E^g_{ij}\,,\\
e_{00} &= -2\phi^f, &
e_{0i} &= \partial_iB^f + B^f_i, &
e_{ij} &= -2\psi^f\delta_{ij} + 2\triangle_{ij}E^f + 4\partial_{(i}E^f_{j)} + 2E^f_{ij}
\end{align}
\end{subequations}
into four scalars \(\phi^{g,f}, \psi^{g,f}, B^{g,f}, E^{g,f}\), two divergence-free vectors \(B^{g,f}_i, E^{g,f}_i\) and one trace-free, divergence-free tensor \(E^{g,f}_{ij}\). Here \(\triangle_{ij}\) denotes the trace-free second derivative \(\triangle_{ij} = \partial_i\partial_j - \frac{1}{3}\delta_{ij}\triangle\). From these quantities we further derive the potentials
\begin{gather}
I_1^{g,f} = \phi^{g,f} + \partial_0B^{g,f} - \partial_0^2E^{g,f}\,, \quad
I_3^{g,f} = B^{g,f}\,, \quad
I_4^{g,f} = E^{g,f}\,,\nonumber\\
I_2^{g,f} = \psi^{g,f} + \frac{1}{3}\triangle E^{g,f}\,, \quad
I^{g,f}_i = B^{g,f}_i - 2\partial_0E^{g,f}_i\,, \quad
I'^{g,f}_i = E^{g,f}_i\,, \quad
I^{g,f}_{ij} = E^{g,f}_{ij}\,.\label{eqn:metricinv}
\end{gather}
The advantage of using these potentials becomes apparent when we consider gauge transformations of the metric, i.e., diffeomorphisms generated by a vector field \(\xi^{\mu}\) which preserve the perturbation ansatz~\eqref{eqn:metperturb} as discussed in section~\ref{ssec:ppnmetric}. Here we introduce a differential decomposition for \(\xi_{\mu} = \eta_{\mu\nu}\xi^{\nu}\) of the form
\begin{equation}
\xi_0 = X\,, \quad \xi_i = \partial_iX' + X_i
\end{equation}
into two scalars \(X, X'\) and one divergence-free vector \(X_i\). One now easily computes from the decomposition~\eqref{eqn:metricdec} the transformations
\begin{gather}
\delta_{\xi}\phi^{g,f} = -\partial_0X\,, \quad
\delta_{\xi}\psi^{g,f} = -\frac{1}{3}\triangle X'\,, \quad
\delta_{\xi}B^{g,f} = \partial_0X' + X\,, \quad
\delta_{\xi}E^{g,f} = X'\,,\nonumber\\
\delta_{\xi}B^{g,f}_i = \partial_0X_i\,, \quad
\delta_{\xi}E^{g,f}_i = \frac{1}{2}X_i\,, \quad
\delta_{\xi}E^{g,f}_{ij} = 0\,.\label{eqn:comptrans}
\end{gather}
The potentials~\eqref{eqn:metricinv} hence transform as
\begin{gather}
\delta_{\xi}I_1^{g,f} = \delta_{\xi}I_2^{g,f} = 0\,, \quad
\delta_{\xi}I_3^{g,f} = \partial_0X' + X\,, \quad
\delta_{\xi}I_4^{g,f} = X'\,,\nonumber\\
\delta_{\xi}I^{g,f}_i = 0\,, \quad
\delta_{\xi}I'^{g,f}_i = \frac{1}{2}X_i\,, \quad
\delta_{\xi}I^{g,f}_{ij} = 0\,.\label{eqn:invtrans}
\end{gather}
Finally, defining the linearly related potentials
\begin{gather}
I_1^{\pm} = I_1^g \pm I_1^f\,, \quad
I_2^{\pm} = I_2^g \pm I_2^f\,, \quad
I_3^{\pm} = I_3^g \pm I_3^f\,, \quad
I_4^{\pm} = I_4^g \pm I_4^f\,,\nonumber\\
I^{\pm}_i = I^g_i \pm I^f_i\,, \quad
I'^{\pm}_i = I'^g_i \pm I^f_i\,, \quad
I^{\pm}_{ij} = I^g_{ij} \pm I^f_{ij}\,,\label{eqn:invpotpm}
\end{gather}
we see that the six scalar potentials \(I_1^{\pm}, I_2^{\pm}, I_3^-, I_4^-\), the three vectors \(I^{\pm}_i, I'^-_i\) and the two tensors \(I^{\pm}_{ij}\) are invariant under gauge transformations, while the remaining two scalars \(I_3^+, I_4^+\) and the vector \(I'^+_i\) are pure gauge degrees of freedom corresponding to the two scalars and the vector constituting the diffeomorphisms. The only physical degrees of freedom are the gauge invariant potentials. Since the gravitational field equations are derived from a diffeomorphism invariant action, we can fully express them in terms of these gauge invariants. In the following we will do so by introducing a suitable differential decomposition of the Ricci tensors, potentials and energy-momentum tensors.

\subsection{Gauge invariant potentials and velocity orders}\label{ssec:invpotvel}
Recall from section~\ref{sec:ppn} that we have decomposed the metric perturbations into velocity orders and that only the components~\eqref{eqn:relcomps} are relevant for the calculation we present in this article. We now apply the decomposition into velocity orders to the gauge invariant potentials above in order to determine which of them will be relevant for our calculation. A comparison of the relevant components~\eqref{eqn:relcomps} with the differential decomposition~\eqref{eqn:metricdec} shows that only the quantities
\begin{equation}
\phi^{g,f(2)}\,, \quad \psi^{g,f(2)}\,, \quad E^{g,f(2)}\,, \quad E^{g,f(2)}_i\,, \quad E^{g,f(2)}_{ij}
\end{equation}
at the second velocity order will be relevant. Using the relations~\eqref{eqn:metricinv}, while taking into account that time derivatives are weighted with an additional velocity order \(\mathcal{O}(1)\), then yields the relevant potentials
\begin{gather}
I_1^{g,f(2)} = \phi^{g,f(2)}\,, \quad
I_2^{g,f(2)} = \psi^{g,f(2)} + \frac{1}{3}\triangle E^{g,f(2)}\,, \quad
I_4^{g,f(2)} = E^{g,f(2)}\,,\nonumber\\
I'^{g,f(2)}_i = E^{g,f(2)}_i\,, \quad
I^{g,f(2)}_{ij} = E^{g,f(2)}_{ij}\,.
\end{gather}
Finally, transitioning to the linearly related potentials~\eqref{eqn:invpotpm} shows that the only relevant gauge-invariant potentials are the five scalars \(I_1^{\pm(2)}, I_2^{\pm(2)}, I_4^{-(2)}\), the vector \(I'^{-(2)}_i\) and the two tensors \(I^{\pm(2)}_{ij}\), while the scalar \(I_4^{+(2)}\) and the vector \(I'^{+(2)}_i\) are pure gauge quantities. From the former we can now calculate the relevant components of the Ricci tensor and the potential.

\subsection{Decomposition of the Ricci tensors}\label{ssec:riccidec}
We now perform a differential decomposition of the Ricci tensors, similar to the differential decomposition~\eqref{eqn:metricdec} of the metric introduced above. Here we use the defining relations
\begin{equation}\label{eqn:riccidec}
R^{g,f}_{00} = K_1^{g,f}\,, \quad
R^{g,f}_{0i} = \partial_iK_3^{g,f} + K^{g,f}_i\,, \quad
R^{g,f}_{ij} = \frac{1}{3}K_2^{g,f}\delta_{ij} + \triangle_{ij}K_4^{g,f} + 2\partial_{(i}K'^{g,f}_{j)} + K^{g,f}_{ij}\,.
\end{equation}
From this definition and the second order field equations~\eqref{eqn:2ndorder} follows that the only relevant components for our calculation are given by
\begin{gather}
K_1^{g,f(2)} = \triangle I_1^{g,f(2)}\,, \quad
K_2^{g,f(2)} = 4\triangle I_2^{g,f(2)} - \triangle I_1^{g,f(2)}\,,\nonumber\\
K_4^{g,f(2)} = I_2^{g,f(2)} - I_1^{g,f(2)}\,, \quad
K'^{g,f(2)}_i = 0\,, \quad
K^{g,f(2)}_{ij} = -\triangle I^{g,f(2)}_{ij}\,.
\end{gather}
Comparing these expressions with the gauge transformations~\eqref{eqn:invtrans} we see that they contain only gauge-invariant potentials, as expected from the fact that they originate from a diffeomorphism invariant action.

\subsection{Decomposition of the potentials}\label{ssec:potdec}
For the trace-reversed potentials we proceed in full analogy to the decomposition~\eqref{eqn:riccidec} of the Ricci tensors. Here we use the decomposition
\begin{equation}
\bar{V}^{g,f}_{00} = U_1^{g,f}\,, \quad
\bar{V}^{g,f}_{0i} = \partial_iU_3^{g,f} + U^{g,f}_i\,, \quad
\bar{V}^{g,f}_{ij} = \frac{1}{3}U_2^{g,f}\delta_{ij} + \triangle_{ij}U_4^{g,f} + 2\partial_{(i}U'^{g,f}_{j)} + U^{g,f}_{ij}\,.
\end{equation}
From the second order field equations~\eqref{eqn:2ndorder} we read off that the relevant components are given by
\begin{subequations}
\begin{align}
U_1^{g(2)} = -c^2U_1^{f(2)} &= -\frac{1}{2}\tilde{\beta}(3I_1^{-(2)} - 3I_2^{-(2)} + \triangle I_4^{-(2)})\,,\\
U_2^{g(2)} = -c^2U_2^{f(2)} &= \frac{1}{2}\tilde{\beta}(3I_1^{-(2)} - 15I_2^{-(2)} + 5\triangle I_4^{-(2)})\,,\\
U_4^{g(2)} = -c^2U_4^{f(2)} &= \tilde{\beta}I_4^{-(2)}\,,\\
U'^{g(2)}_i = -c^2U'^{f(2)}_i &= \tilde{\beta}I'^{-(2)}_i\,,\\
U^{g(2)}_{ij} = -c^2U^{f(2)}_{ij} &= \tilde{\beta}I^{-(2)}_{ij}\,.
\end{align}
\end{subequations}
Again we see that these depend only on gauge invariant potentials, as expected.

\subsection{Decomposition of the energy-momentum tensors}\label{ssec:enmomdec}
We finally also need to perform a differential decomposition of the energy-momentum tensors. Following the same prescription as for the Ricci tensors and the potentials we define
\begin{equation}
\bar{T}^{g,f}_{00} = Q_1^{g,f}\,, \quad
\bar{T}^{g,f}_{0i} = \partial_iQ_3^{g,f} + Q^{g,f}_i\,, \quad
\bar{T}^{g,f}_{ij} = \frac{1}{3}Q_2^{g,f}\delta_{ij} + \triangle_{ij}Q_4^{g,f} + 2\partial_{(i}Q'^{g,f}_{j)} + Q^{g,f}_{ij}\,.
\end{equation}
For the expressions~\eqref{eqn:enmom2nd} for the second order trace-reversed energy-momentum tensors of the perfect fluid then follow the relevant components
\begin{gather}
Q_1^{g(2)} = \frac{1}{2}\rho^g\,, \quad
Q_1^{f(2)} = \frac{1}{2}c^2\rho^f\,, \quad
Q_2^{g(2)} = \frac{3}{2}\rho^g\,, \quad
Q_2^{f(2)} = \frac{3}{2}c^2\rho^f\,,\nonumber\\
Q_4^{g,f(2)} = 0\,, \quad
Q'^{g,f(2)}_i = 0\,, \quad
Q^{g,f(2)}_{ij} = 0\,.
\end{gather}
These are all expressions we need for the field equations~\eqref{eqn:2ndorder} at the second velocity order.

\subsection{Decomposition of the field equations}\label{ssec:feqdec}
We now have all expressions at hand which are necessary to perform a differential decomposition of the second order field equations~\eqref{eqn:2ndorder} and to fully express them in terms of gauge-invariant quantities. It is an important feature of the differential decomposition that it is unique and bijective under the boundary conditions mentioned in section~\ref{ssec:ppnorder}, which imply that all metric perturbations and their derivatives vanish at infinity. It thus follows that the field equations~\eqref{eqn:2ndorder} are equivalent to the decomposed field equations
\begin{subequations}
\begin{align}
m_g^2K_1^{g(2)} + m^2U_1^{g(2)} &= Q_1^{g(2)}\,, &
m_f^2K_1^{f(2)} + m^2U_1^{f(2)} &= Q_1^{f(2)}\,,\label{eqn:decscal1}\\
m_g^2K_2^{g(2)} + m^2U_2^{g(2)} &= Q_2^{g(2)}\,, &
m_f^2K_2^{f(2)} + m^2U_2^{f(2)} &= Q_2^{f(2)}\,,\label{eqn:decscal2}\\
m_g^2K_4^{g(2)} + m^2U_4^{g(2)} &= Q_4^{g(2)}\,, &
m_f^2K_4^{f(2)} + m^2U_4^{f(2)} &= Q_4^{f(2)}\,,\label{eqn:decscal4}\\
m_g^2K'^{g(2)}_i + m^2U'^{g(2)}_i &= Q'^{g(2)}_i\,, &
m_f^2K'^{f(2)}_i + m^2U'^{f(2)}_i &= Q'^{f(2)}_i\,,\label{eqn:decvect}\\
m_g^2K^{g(2)}_{ij} + m^2U^{g(2)}_{ij} &= Q^{g(2)}_{ij}\,, &
m_f^2K^{f(2)}_{ij} + m^2U^{f(2)}_{ij} &= Q^{f(2)}_{ij}\,.\label{eqn:dectens}
\end{align}
\end{subequations}
We can now insert the expressions for the differential components of the Ricci tensors, the potentials and the energy-momentum tensors which we derived above. We start with the trace-free, divergence-free tensor equations~\eqref{eqn:dectens}. Inserting the components \(K^{g,f(2)}_{ij}, U^{g,f(2)}_{ij}, Q^{g,f(2)}_{ij}\) yields the equations
\begin{subequations}
\begin{align}
-\frac{m_g^2}{2}\left(\triangle I^{+(2)}_{ij} + \triangle I^{-(2)}_{ij}\right) + m^4\tilde{\beta} I^{-(2)}_{ij} &= 0\,,\\
-\frac{m_f^2}{2}\left(\triangle I^{+(2)}_{ij} - \triangle I^{-(2)}_{ij}\right) - \frac{m^4\tilde{\beta}}{c^2}I^{-(2)}_{ij} &= 0\,.
\end{align}
\end{subequations}
Note that together with the boundary conditions they yield the trivial solution \(I^{\pm(2)}_{ij} = 0\). We then continue with the divergence-free vector equations~\eqref{eqn:decvect}. Using the expressions for \(K'^{g,f(2)}_i, U'^{g,f(2)}_i, Q'^{g,f(2)}_i\) we obtain
\begin{equation}
m^4\tilde{\beta}I'^{-(2)}_i = 0\,, \quad -\frac{m^4\tilde{\beta}}{c^2}I'^{-(2)}_i = 0\,.
\end{equation}
These equations are equivalent as a consequence of the Bianchi identities, which follow from the diffeomorphism invariance of the action~\eqref{eqn:action}. Also these equations yield a trivial solution \(I'^{-(2)}_i = 0\). We are thus left with the scalar equations~\eqref{eqn:decscal1}, \eqref{eqn:decscal2} and~\eqref{eqn:decscal4}, which take the form
\begin{subequations}\label{eqn:scalar}
\begin{align}
\frac{1}{2}\rho^g &= \frac{m_g^2}{2}\left(\triangle I_1^{+(2)} + \triangle I_1^{-(2)}\right) - \frac{m^4\beta}{2}\left(3I_1^{-(2)} - 3I_2^{-(2)} + \triangle I_4^{-(2)}\right)\,,\label{eqn:scalg1}\\
\frac{c^2}{2}\rho^f &= \frac{m_f^2}{2}\left(\triangle I_1^{+(2)} - \triangle I_1^{-(2)}\right) + \frac{m^4\beta}{2c^2}\left(3I_1^{-(2)} - 3I_2^{-(2)} + \triangle I_4^{-(2)}\right)\,,\label{eqn:scalf1}\\
\frac{3}{2}\rho^g &= \frac{m_g^2}{2}\left(4\triangle I_2^{+(2)} + 4\triangle I_2^{-(2)} - \triangle I_1^{+(2)} - \triangle I_1^{-(2)}\right) + \frac{m^4\beta}{2}\left(3I_1^{-(2)} - 15I_2^{-(2)} + 5\triangle I_4^{-(2)}\right)\,,\label{eqn:scalg2}\\
\frac{3c^2}{2}\rho^f &= \frac{m_f^2}{2}\left(4\triangle I_2^{+(2)} - 4\triangle I_2^{-(2)} - \triangle I_1^{+(2)} + \triangle I_1^{-(2)}\right) - \frac{m^4\beta}{2c^2}\left(3I_1^{-(2)} - 15I_2^{-(2)} + 5\triangle I_4^{-(2)}\right)\,,\label{eqn:scalf2}\\
0 &= \frac{m_g^2}{2}\left(I_2^{+(2)} + I_2^{-(2)} - I_1^{+(2)} - I_1^{-(2)}\right) + m^4\beta I_4^{-(2)}\,,\label{eqn:scalg4}\\
0 &= \frac{m_f^2}{2}\left(I_2^{+(2)} - I_2^{-(2)} - I_1^{+(2)} + I_1^{-(2)}\right) - \frac{m^4\beta}{c^2}I_4^{-(2)}\,.\label{eqn:scalf4}
\end{align}
\end{subequations}
Note that also these equations are not independent, but are related to each other as a consequence of the Bianchi identities. Indeed, one easily checks that
\begin{gather}
K_2^{g,f(2)} - 3K_1^{g,f(2)} - 4\triangle K_4^{g,f(2)} = 0\,, \quad
Q_2^{g,f(2)} - 3Q_1^{g,f(2)} - 4\triangle Q_4^{g,f(2)} = 0\,,\nonumber\\
U_2^{g(2)} - 3U_1^{g(2)} - 4\triangle U_4^{g(2)} = 6m^4\tilde{\beta}\left(\triangle I_1^{-(2)} - 2\triangle I_2^{-(2)}\right) = -c^2\left(U_2^{f(2)} - 3U_1^{f(2)} - 4\triangle U_4^{f(2)}\right)\,,
\end{gather}
which shows that the corresponding linear combinations of the scalar equations become identical. Symbolically, this can be written as
\begin{equation}\label{eqn:lincom}
\text{\eqref{eqn:scalg2}} - 3\text{\eqref{eqn:scalg1}} - 4\triangle\text{\eqref{eqn:scalg4}} = -c^2\left[\text{\eqref{eqn:scalf2}} - 3\text{\eqref{eqn:scalf1}} - 4\triangle\text{\eqref{eqn:scalf4}}\right]\,.
\end{equation}
Hence, one of the equations~\eqref{eqn:scalar} is redundant and can be omitted. The remaining five equations then determine the five gauge-invariant scalar potentials \(I_1^{\pm(2)}, I_2^{\pm(2)}, I_4^{-(2)}\). We will solve these equations in the following section for the special case of a static point mass source.

\section{Static spherically symmetric solution}\label{sec:solution}
Using the gauge invariant field equations~\eqref{eqn:scalar} derived in the preceding section we are now in the position to construct an explicit solution. The starting point will be a point mass, which we discuss in section~\ref{ssec:pointmass}. We will then determine a solution for the gauge invariant potentials in section~\ref{ssec:invpotsol}. From these we will derive the metric components in section~\ref{ssec:metricsol}, reversing the procedure detailed in section~\ref{ssec:metricdec}. By comparison with the metric ansatz~\eqref{eqn:ppnmetric} we read off the PPN parameters in section~\ref{ssec:ppnpar}. We finally discuss a few limiting cases in section~\ref{ssec:limits}.

\subsection{Point-mass source}\label{ssec:pointmass}
The matter source we consider for our solution is a static point mass located at the origin of our coordinate system, which is constituted by masses \(M^g\) and \(M^f\) with respect to the two matter sectors. Invoking the interpretation of the matter sectors as visible and dark matter, this would correspond to a source containing both visible and dark matter, unless one of the masses vanishes. This choice is the most general one, and includes the physically relevant case of a galaxy with a dark matter component, as we discuss later in section~\ref{ssec:darkmatter}. A source of this type is characterized by the matter variables
\begin{equation}\label{eqn:pointmass}
\rho^g = M^g\delta(\vec{x})\,, \quad \rho^f = M^f\frac{\delta(\vec{x})}{c^3}\,, \quad \Pi^{g,f} = 0\,, \quad p^{g,f} = 0\,, \quad v^{g,f}_i = 0\,,
\end{equation}
where we have normalized the delta function in \(\rho^f\) with the spatial volume element \(c^3\) of the unperturbed metric \(f^{(0)}_{\mu\nu} = c^2\eta_{\mu\nu}\). Note that this factor cancels the volume element in the corresponding superpotential~\eqref{eqn:superpot}. Using isotropic spherical coordinates, the superpotentials thus read
\begin{equation}\label{eqn:pmsuperpot}
\chi^{g,f} = -M^{g,f}r\,.
\end{equation}
For later convenience we also list the second order derivatives of the superpotentials, which take the form
\begin{equation}
\chi^{g,f}_{,ij} = M^{g,f}\left(\frac{x_ix_j}{r^3} - \frac{\delta_{ij}}{r}\right)\,, \quad \triangle\chi^{g,f} = -2\frac{M^{g,f}}{r}\,.
\end{equation}
These will be used when we read off the PPN parameters in section~\ref{ssec:ppnpar}.

\subsection{Gauge-invariant potentials}\label{ssec:invpotsol}
We will now determine the gauge-invariant potentials \(I_1^{\pm(2)}, I_2^{\pm(2)}, I_4^{-(2)}\) by solving the scalar part~\eqref{eqn:scalar} of the field equations at the second velocity order, where we assume the matter source given by the point mass introduced above. It will turn out to be convenient to use rescaled mass units
\begin{equation}\label{eqn:massrescale}
\tilde{m}_g = m_g\,, \quad \tilde{m}_f = cm_f\, \quad \tilde{M}^g = M^g\,, \quad \tilde{M}^f = cM^f\,.
\end{equation}
We then start with the purely algebraic equations~\eqref{eqn:scalg4} and~\eqref{eqn:scalf4}. Using the definitions above these take the form
\begin{subequations}
\begin{align}
0 &= \frac{\tilde{m}_g^2}{2}\left(I_2^{+(2)} + I_2^{-(2)} - I_1^{+(2)} - I_1^{-(2)}\right) + m^4\tilde{\beta}I_4^{-(2)}\,,\\
0 &= \frac{\tilde{m}_f^2}{2}\left(I_2^{+(2)} - I_2^{-(2)} - I_1^{+(2)} + I_1^{-(2)}\right) - m^4\tilde{\beta}I_4^{-(2)}\,.
\end{align}
\end{subequations}
Here we choose to solve these equations for the potentials \(I_2^{\pm(2)}\). The solutions are given by
\begin{equation}\label{eqn:i2elim}
I_2^{\pm(2)} = I_1^{\pm(1)} - \left(\frac{1}{\tilde{m}_g^2} \mp \frac{1}{\tilde{m}_f^2}\right)m^4\tilde{\beta}I_4^{-(2)}\,.
\end{equation}
We can use this relation to eliminate \(I_2^{\pm(2)}\) from the remaining equations. Using the linear combination~\eqref{eqn:lincom}, which together with the boundary conditions yields
\begin{equation}\label{eqn:i1i2rel}
I_1^{-(2)} = 2I_2^{-(2)}\,,
\end{equation}
we can then solve for \(I_4^{-(2)}\) and obtain the solution
\begin{equation}\label{eqn:i4elim}
I_4^{-(2)} = \frac{1}{2\mu^2}I_1^{-(2)}\,,
\end{equation}
where we have defined the mass parameter
\begin{equation}
\mu = m^2\sqrt{\tilde{\beta}\left(\frac{1}{\tilde{m}_f^2} + \frac{1}{\tilde{m}_g^2}\right)}\,.
\end{equation}
We now take a suitable linear combination of the scalar equations~\eqref{eqn:scalg1} and~\eqref{eqn:scalf1}, so that the terms involving \(I_1^{+(2)}\) cancel. Eliminating \(I_2^{-(2)}\) and \(I_4^{-(2)}\) with the relations~\eqref{eqn:i2elim} and~\eqref{eqn:i4elim} we obtain
\begin{equation}
\triangle I_1^{-(2)} - \mu^2I_1^{-(2)} = \frac{2}{3}\left(\frac{\tilde{M}^g}{\tilde{m}_g^2} - \frac{\tilde{M}^f}{\tilde{m}_f^2}\right)\delta(\vec{x})\,,
\end{equation}
which is a screened Poisson equation for \(I_1^{-(2)}\). The solution is given by
\begin{equation}
I_1^{-(2)} = -\left(\frac{\tilde{M}^g}{\tilde{m}_g^2} - \frac{\tilde{M}^f}{\tilde{m}_f^2}\right)\frac{e^{-\mu r}}{6\pi r}\,.
\end{equation}
From the relations~\eqref{eqn:i1i2rel} and~\eqref{eqn:i4elim} then immediately follows
\begin{equation}
I_2^{-(2)} = -\left(\frac{\tilde{M}^g}{\tilde{m}_g^2} - \frac{\tilde{M}^f}{\tilde{m}_f^2}\right)\frac{e^{-\mu r}}{12\pi r}\,, \quad
I_4^{-(2)} = -\left(\frac{\tilde{M}^g}{\tilde{m}_g^2} - \frac{\tilde{M}^f}{\tilde{m}_f^2}\right)\frac{e^{-\mu r}}{12\pi\mu^2r}\,.
\end{equation}
Inserting these results into the scalar equation~\eqref{eqn:scalg1}, we obtain an equation for \(I_1^{+(2)}\), which is most conveniently expressed as
\begin{equation}
\triangle\left(I_1^{+(2)} + \frac{\tilde{m}_g^2 - \tilde{m}_f^2}{\tilde{m}_g^2 + \tilde{m}_f^2}I_1^{-(2)}\right) = \frac{\tilde{M}^g + \tilde{M}^f}{\tilde{m}_g^2 + \tilde{m}_f^2}\delta(\vec{x})\,.
\end{equation}
This is an ordinary Poisson equation and one immediately reads off the solution
\begin{equation}
I_1^{+(2)} = -\frac{\tilde{M}^g + \tilde{M}^f}{\tilde{m}_g^2 + \tilde{m}_f^2}\frac{1}{4\pi r} - \frac{\tilde{m}_g^2 - \tilde{m}_f^2}{\tilde{m}_g^2 + \tilde{m}_f^2}I_1^{-(2)} = -\frac{\tilde{M}^g + \tilde{M}^f}{\tilde{m}_g^2 + \tilde{m}_f^2}\frac{1}{4\pi r} + \frac{\tilde{m}_g^2 - \tilde{m}_f^2}{\tilde{m}_g^2 + \tilde{m}_f^2}\left(\frac{\tilde{M}^g}{\tilde{m}_g^2} - \frac{\tilde{M}^f}{\tilde{m}_f^2}\right)\frac{e^{-\mu r}}{6\pi r}\,.
\end{equation}
Finally, making use of the relation~\eqref{eqn:i2elim} yields
\begin{equation}
I_2^{+(2)} = -\frac{\tilde{M}^g + \tilde{M}^f}{\tilde{m}_g^2 + \tilde{m}_f^2}\frac{1}{4\pi r} + \frac{\tilde{m}_g^2 - \tilde{m}_f^2}{\tilde{m}_g^2 + \tilde{m}_f^2}\left(\frac{\tilde{M}^g}{\tilde{m}_g^2} - \frac{\tilde{M}^f}{\tilde{m}_f^2}\right)\frac{e^{-\mu r}}{12\pi r}\,.
\end{equation}
This completes the solution of the field equations in terms of gauge-invariant potentials.

\subsection{Metric components}\label{ssec:metricsol}
Before we calculate the metric components from the solution for the gauge-invariant potentials, it is convenient to introduce the abbreviations
\begin{equation}\label{eqn:potconst}
\mathcal{I}_M = \frac{\tilde{M}^g + \tilde{M}^f}{8\pi(\tilde{m}_g^2 + \tilde{m}_f^2)}\,, \quad \mathcal{I}_{\pm} = -\frac{1}{24\pi\mu^2}\left(\frac{\tilde{M}^g}{\tilde{m}_g^2} \pm \frac{\tilde{M}^f}{\tilde{m}_f^2}\right)\,, \quad \mathcal{D} = -\frac{\tilde{m}_g^2 - \tilde{m}_f^2}{\tilde{m}_g^2 + \tilde{m}_f^2}
\end{equation}
for a few frequently occurring constants. Further, we use the shorthand notation
\begin{equation}\label{eqn:yukcoul}
\mathcal{Y}_{\mu}(r) = \frac{e^{-\mu r}}{r}\,, \quad \mathcal{Y}_0(r) = \frac{1}{r}
\end{equation}
for the Yukawa and Coulomb potentials. Using these abbreviations, the solution derived above takes the simple form
\begin{gather}
I_1^{-(2)} = 4\mu^2\mathcal{I}_-\mathcal{Y}_{\mu}\,, \quad
I_2^{-(2)} = 2\mu^2\mathcal{I}_-\mathcal{Y}_{\mu}\,, \quad
I_4^{-(2)} = 2\mathcal{I}_-\mathcal{Y}_{\mu}\,,\nonumber\\
I_1^{+(2)} = -2\mathcal{I}_M\mathcal{Y}_0 + 4\mu^2\mathcal{D}\mathcal{I}_-\mathcal{Y}_{\mu}\,, \quad
I_2^{+(2)} = -2\mathcal{I}_M\mathcal{Y}_0 + 2\mu^2\mathcal{D}\mathcal{I}_-\mathcal{Y}_{\mu}\,.
\end{gather}
In order to separate these potentials into the potentials for the individual metrics, we further need to fix the pure gauge potential \(I_4^{+(2)}\). A convenient choice, which turns out to be compatible with the standard PPN gauge, is given by
\begin{equation}\label{eqn:i4gauge}
I_4^{+(2)} = 2\mathcal{I}_+\mathcal{Y}_{\mu}\,.
\end{equation}
Together with the relations~\eqref{eqn:invpotpm} we then obtain the potentials
\begin{subequations}
\begin{align}
I_1^{g,f(2)} &= -\mathcal{I}_M\mathcal{Y}_0 + 2\mu^2(\mathcal{D} \pm 1)\mathcal{I}_-\mathcal{Y}_{\mu}\,,\\
I_2^{g,f(2)} &= -\mathcal{I}_M\mathcal{Y}_0 + \mu^2(\mathcal{D} \pm 1)\mathcal{I}_-\mathcal{Y}_{\mu}\,,\\
I_4^{g,f(2)} &= (\mathcal{I}_+ \pm \mathcal{I}_-)\mathcal{Y}_{\mu}\,.
\end{align}
\end{subequations}
Now using the relations~\eqref{eqn:metricinv} we obtain the quantities
\begin{subequations}
\begin{align}
\phi^{g,f(2)} &= -\mathcal{I}_M\mathcal{Y}_0 + 2\mu^2(\mathcal{D} \pm 1)\mathcal{I}_-\mathcal{Y}_{\mu}\,,\\
\psi^{g,f(2)} &= -\mathcal{I}_M\mathcal{Y}_0 + \mu^2(\mathcal{D} \pm 1)\mathcal{I}_-\mathcal{Y}_{\mu} - \frac{1}{3}(\mathcal{I}_+ \pm \mathcal{I}_-)\triangle\mathcal{Y}_{\mu}\,,\\
\tilde{E}^{g,f(2)} &= (\mathcal{I}_+ \pm \mathcal{I}_-)\mathcal{Y}_{\mu}\,,
\end{align}
\end{subequations}
and finally using their definition~\eqref{eqn:metricdec} yields the components of the metric perturbations
\begin{subequations}\label{eqn:metsoly}
\begin{align}
h^{(2)}_{00} &= 2\mathcal{I}_M\mathcal{Y}_0 - 4\mu^2(\mathcal{D} + 1)\mathcal{I}_-\mathcal{Y}_{\mu}\,,\\
e^{(2)}_{00} &= 2\mathcal{I}_M\mathcal{Y}_0 - 4\mu^2(\mathcal{D} - 1)\mathcal{I}_-\mathcal{Y}_{\mu}\,,\\
h^{(2)}_{ij} &= 2[\mathcal{I}_M\mathcal{Y}_0 - \mu^2(\mathcal{D} + 1)\mathcal{I}_-\mathcal{Y}_{\mu}]\delta_{ij} + 2(\mathcal{I}_+ + \mathcal{I}_-)\partial_i\partial_j\mathcal{Y}_{\mu}\,,\\
e^{(2)}_{ij} &= 2[\mathcal{I}_M\mathcal{Y}_0 - \mu^2(\mathcal{D} - 1)\mathcal{I}_-\mathcal{Y}_{\mu}]\delta_{ij} + 2(\mathcal{I}_+ - \mathcal{I}_-)\partial_i\partial_j\mathcal{Y}_{\mu}\,.
\end{align}
\end{subequations}
For later use we now insert the constants~\eqref{eqn:potconst} and the Yukawa and Coulomb potentials~\eqref{eqn:yukcoul}. Note that second order derivatives of these potentials contain also delta functions, which must be taken into account for deriving further quantities from the metric perturbations. We have listed the relevant formulas in appendix~\ref{app:yukawa}. Using these formulas we obtain
\begin{subequations}\label{eqn:metsolr}
\begin{align}
h^{(2)}_{00} &= \frac{\tilde{M}^g + \tilde{M}^f}{4\pi(\tilde{m}_g^2 + \tilde{m}_f^2)r} + \frac{\tilde{m}_f^2\tilde{M}^g - \tilde{m}_g^2\tilde{M}^f}{3\pi\tilde{m}_g^2(\tilde{m}_g^2 + \tilde{m}_f^2)r}e^{-\mu r}\,,\\
e^{(2)}_{00} &= \frac{\tilde{M}^g + \tilde{M}^f}{4\pi(\tilde{m}_g^2 + \tilde{m}_f^2)r} - \frac{\tilde{m}_f^2\tilde{M}^g - \tilde{m}_g^2\tilde{M}^f}{3\pi \tilde{m}_f^2(\tilde{m}_g^2 + \tilde{m}_f^2)r}e^{-\mu r}\,,\\
h^{(2)}_{ij} &= \left[\frac{\tilde{M}^g + \tilde{M}^f}{4\pi(\tilde{m}_g^2 + \tilde{m}_f^2)r} - \frac{2\tilde{m}_g^2\tilde{m}_f^2\tilde{M}^f - \tilde{m}_g^4\tilde{M}^f - 3\tilde{m}_f^4\tilde{M}^g}{18\pi \tilde{m}_g^2\tilde{m}_f^2(\tilde{m}_g^2 + \tilde{m}_f^2)r}e^{-\mu r} - \frac{2\tilde{M}^f}{9\tilde{m}_f^2\mu^2}\delta(\vec{x})\right]\delta_{ij}\nonumber\\
&\phantom{=}+ \frac{[\mu r(\mu r + 3) + 3]\tilde{M}^f}{6\pi \tilde{m}_f^2\mu^2r^5}e^{-\mu r}\left(x_ix_j - \frac{1}{3}r^2\delta_{ij}\right)\,,\\
e^{(2)}_{ij} &= \left[\frac{\tilde{M}^g + \tilde{M}^f}{4\pi(\tilde{m}_g^2 + \tilde{m}_f^2)r} - \frac{2\tilde{m}_g^2\tilde{m}_f^2\tilde{M}^g - \tilde{m}_f^4\tilde{M}^g - 3\tilde{m}_g^4\tilde{M}^f}{18\pi \tilde{m}_g^2\tilde{m}_f^2(\tilde{m}_g^2 + \tilde{m}_f^2)r}e^{-\mu r} - \frac{2\tilde{M}^g}{9\tilde{m}_g^2\mu^2}\delta(\vec{x})\right]\delta_{ij}\nonumber\\
&\phantom{=}+ \frac{[\mu r(\mu r + 3) + 3]\tilde{M}^g}{6\pi\tilde{m}_g^2\mu^2r^5}e^{-\mu r}\left(x_ix_j - \frac{1}{3}r^2\delta_{ij}\right)\,.
\end{align}
\end{subequations}
Note that the off-diagonal contribution of \(h^{(2)}_{ij}\) depends only on the mass \(\tilde{M}^f\), while the off-diagonal contribution of \(e^{(2)}_{ij}\) contains only the mass \(\tilde{M}^g\). This is a consequence of our gauge choice~\eqref{eqn:i4gauge}, and the reason for making this choice.

\subsection{PPN parameters}\label{ssec:ppnpar}
We can now read off the PPN parameters by comparing the solution~\eqref{eqn:metsolr} to the PPN metric ansatz~\eqref{eqn:ppnmetric}. Since we have used rescaled mass units~\eqref{eqn:massrescale}, it is convenient to replace the superpotentials~\eqref{eqn:superpot}, which for a point mass source take the form~\eqref{eqn:pmsuperpot}, by the correspondingly rescaled superpotentials
\begin{equation}
\tilde{\chi}^g = \chi^g = -M^gr = -\tilde{M}^gr\,, \quad \tilde{\chi}^f = c\chi^f = -cM^fr = -\tilde{M}^fr\,.
\end{equation}
We thus use the modified PPN metric ansatz
\begin{subequations}
\begin{align}
h^{(2)}_{00} &= 2\frac{\tilde{\alpha}^{gg}\tilde{M}^g + \tilde{\alpha}^{gf}\tilde{M}^f}{r}\,, &
h^{(2)}_{ij} &= 2\frac{\tilde{\gamma}^{gg}\tilde{M}^g + \tilde{\gamma}^{gf}\tilde{M}^f}{r}\delta_{ij} + 2\frac{\tilde{\theta}^{gg}\tilde{M}^g + \tilde{\theta}^{gf}\tilde{M}^f}{r^3}x_ix_j\,,\\
e^{(2)}_{00} &= 2\frac{\tilde{\alpha}^{fg}\tilde{M}^g + \tilde{\alpha}^{ff}\tilde{M}^f}{r}\,, &
e^{(2)}_{ij} &= 2\frac{\tilde{\gamma}^{fg}\tilde{M}^g + \tilde{\gamma}^{ff}\tilde{M}^f}{r}\delta_{ij} + 2\frac{\tilde{\theta}^{fg}\tilde{M}^g + \tilde{\theta}^{ff}\tilde{M}^f}{r^3}x_ix_j\,.
\end{align}
\end{subequations}
Note that the observable parameters \(\alpha^{gg} = \tilde{\alpha}^{gg}\), \(\gamma^{gg} = \tilde{\gamma}^{gg}\) and \(\theta^{gg} = \tilde{\theta}^{gg}\), which govern the gravitational interaction within the visible matter sector, are unaffected by this rescaling, and that only the PPN parameters involving the dark sector receive constant factors. We then read off the PPN parameters
\begin{subequations}\label{eqn:ppnparsol}
\begin{align}
\tilde{\alpha}^{gg} &= \frac{3\tilde{m}_g^2 + 4\tilde{m}_f^2e^{-\mu r}}{24\pi\tilde{m}_g^2(\tilde{m}_f^2 + \tilde{m}_g^2)}\,, &
\tilde{\alpha}^{gf} &= \frac{3 - 4e^{-\mu r}}{24\pi(\tilde{m}_f^2 + \tilde{m}_g^2)}\,,\\
\tilde{\alpha}^{ff} &= \frac{3\tilde{m}_f^2 + 4\tilde{m}_g^2e^{-\mu r}}{24\pi \tilde{m}_f^2(\tilde{m}_f^2 + \tilde{m}_g^2)}\,, &
\tilde{\alpha}^{fg} &= \frac{3 - 4e^{-\mu r}}{24\pi(\tilde{m}_f^2 + \tilde{m}_g^2)}\,,\\
\tilde{\gamma}^{gg} &= \frac{3\tilde{m}_g^2 + 2\tilde{m}_f^2e^{-\mu r}}{24\pi\tilde{m}_g^2(\tilde{m}_f^2 + \tilde{m}_g^2)}\,, &
\tilde{\gamma}^{gf} &= \frac{9\tilde{m}_f^2 + 2(\tilde{m}_g^2 - 2\tilde{m}_f^2)e^{-\mu r}}{72\pi \tilde{m}_f^2(\tilde{m}_f^2 + \tilde{m}_g^2)} - \frac{\mu r(\mu r + 3) + 3}{36\pi \tilde{m}_f^2\mu^2r^2}e^{-\mu r}\,,\\
\tilde{\gamma}^{ff} &= \frac{3\tilde{m}_f^2 + 4\tilde{m}_g^2e^{-\mu r}}{24\pi \tilde{m}_f^2(\tilde{m}_f^2 + \tilde{m}_g^2)}\,, &
\tilde{\gamma}^{fg} &= \frac{9\tilde{m}_g^2 + 2(\tilde{m}_f^2 - 2\tilde{m}_g^2)e^{-\mu r}}{72\pi\tilde{m}_g^2(\tilde{m}_f^2 + \tilde{m}_g^2)} - \frac{\mu r(\mu r + 3) + 3}{36\pi\tilde{m}_g^2\mu^2r^2}e^{-\mu r}\,,\\
\tilde{\theta}^{gg} &= 0\,, &
\tilde{\theta}^{gf} &= \frac{\mu r(\mu r + 3) + 3}{12\pi \tilde{m}_f^2\mu^2r^2}e^{-\mu r}\,,\\
\tilde{\theta}^{ff} &= 0\,, &
\tilde{\theta}^{fg} &= \frac{\mu r(\mu r + 3) + 3}{12\pi\tilde{m}_g^2\mu^2r^2}e^{-\mu r}\,.
\end{align}
\end{subequations}
We find that the gauge condition \(\tilde{\theta}^{gg} = \tilde{\theta}^{ff} = 0\), which we have introduced in section~\ref{ssec:ppnorder}, is satisfied, due to our choice~\eqref{eqn:i4gauge}. From these parameters we can in particular derive the observable quantities
\begin{equation}\label{eqn:geffgamma}
G_{\text{eff}} = \tilde{\alpha}^{gg} = \frac{3\tilde{m}_g^2 + 4\tilde{m}_f^2e^{-\mu r}}{24\pi\tilde{m}_g^2(\tilde{m}_f^2 + \tilde{m}_g^2)}\,, \quad
\gamma = \frac{\tilde{\gamma}^{gg}}{\tilde{\alpha}^{gg}} = \frac{3\tilde{m}_g^2 + 2\tilde{m}_f^2e^{-\mu r}}{3\tilde{m}_g^2 + 4\tilde{m}_f^2e^{-\mu r}}\,,
\end{equation}
which are the effective Newtonian constant and the usual PPN parameter \(\gamma\). Both quantities depend on the distance \(r\) between the mass source and the location where the gravitational field is probed, in contrast to general relativity, where both quantities are constant. It is further remarkable that \(\gamma\) depends only on the ratio \(\tilde{m}_f/\tilde{m}_g\) of the two Planck masses and the graviton mass \(\mu\), and that this result essentially resembles the observable parameters of scalar-tensor theory with a general potential~\cite{Hohmann:2013rba,Scharer:2014kya}, or the more general Horndeski class of theories~\cite{Hohmann:2015kra}, which depend on the Brans-Dicke parameter \(\omega\) and the scalar field mass.

\subsection{Limiting cases}\label{ssec:limits}
We finally discuss a few interesting limiting cases for the mass parameters \(\tilde{m}_{g,f}\) and \(\mu\) and their consequences for the PPN parameters. These are in particular:

\begin{itemize}
\item
It is well known that in the limit \(\tilde{m}_f \to 0\), while keeping the parameters \(m\) and \(\beta_k\) in the interaction potential fixed, one obtains the general relativity limit for the visible sector~\cite{Schmidt-May:2015vnx}. Note that in this limit we also have \(\mu \to \infty\). The PPN parameters~\eqref{eqn:ppnparsol} then take the form
\begin{subequations}
\begin{align}
\tilde{\alpha}^{gg} = \tilde{\alpha}^{gf} = \tilde{\alpha}^{fg} = \tilde{\alpha}^{ff} &= \frac{1}{8\pi\tilde{m}_g^2}\,,\\
\tilde{\gamma}^{gg} = \tilde{\gamma}^{gf} = \tilde{\gamma}^{fg} = \tilde{\gamma}^{ff} &= \frac{1}{8\pi\tilde{m}_g^2}\,,\\
\tilde{\theta}^{gg} = \tilde{\theta}^{gf} = \tilde{\theta}^{fg} = \tilde{\theta}^{ff} &= 0\,,
\end{align}
\end{subequations}
while the observable parameters~\eqref{eqn:geffgamma} are given by
\begin{equation}
G_{\text{eff}} = \frac{1}{8\pi\tilde{m}_g^2}\,, \quad \gamma = 1\,,
\end{equation}
as usual in general relativity.

\item
For equal Planck mass parameters \(\tilde{m}_g = \tilde{m}_f\) one obtains the PPN parameters
\begin{subequations}
\begin{align}
\tilde{\alpha}^{gg} = \tilde{\alpha}^{ff} &= \frac{3 + 4e^{-\mu r}}{48\pi\tilde{m}_g^2}\,, &
\tilde{\alpha}^{gf} = \tilde{\alpha}^{fg} &= \frac{3 - 4e^{-\mu r}}{48\pi\tilde{m}_g^2}\,,\\
\tilde{\gamma}^{gg} = \tilde{\gamma}^{ff} &= \frac{3 + 2e^{-\mu r}}{48\pi\tilde{m}_g^2}\,, &
\tilde{\gamma}^{gf} = \tilde{\gamma}^{fg} &= \frac{3 - 2e^{-\mu r}}{48\pi\tilde{m}_g^2} - \frac{(\mu r + 1)e^{-\mu r}}{12\pi\tilde{m}_g^2\mu^2r^2}\,,\\
\tilde{\theta}^{gg} = \tilde{\theta}^{ff} &= 0\,, &
\tilde{\theta}^{gf} = \tilde{\theta}^{fg} &= \frac{[\mu r(\mu r + 3) + 3]e^{-\mu r}}{12\pi\tilde{m}_g^2\mu^2r^2}
\end{align}
\end{subequations}
and the observable parameters
\begin{equation}\label{eqn:eqpmgamma}
G_{\text{eff}} = \frac{3 + 4e^{-\mu r}}{48\pi\tilde{m}_g^2}\,, \quad
\gamma = \frac{3 + 2e^{-\mu r}}{3 + 4e^{-\mu r}}\,.
\end{equation}
We remark that this result is similar to the PPN parameter \(\gamma\) in higher-order gravity, except for an additional scalar contribution and a different sign due to the massive graviton being a ghost in the latter class of theories~\cite{Bueno:2016ypa}. Note that the effective Planck mass for the visible sector,
\begin{equation}\label{eqn:planckmass}
m_{\text{Pl}}^2 = \lim_{r \to \infty}\frac{1}{8\pi G_{\text{eff}}}\,,
\end{equation}
is given by \(m_{\text{Pl}}^2 = 2\tilde{m}_g^2\).

\item
In the limit \(\mu \to \infty\) of a highly massive graviton we find the PPN parameters
\begin{subequations}
\begin{align}
\tilde{\alpha}^{gg} = \tilde{\alpha}^{gf} = \tilde{\alpha}^{fg} = \tilde{\alpha}^{ff} &= \frac{1}{8\pi(\tilde{m}_g^2 + \tilde{m}_f^2)}\,,\\
\tilde{\gamma}^{gg} = \tilde{\gamma}^{gf} = \tilde{\gamma}^{fg} = \tilde{\gamma}^{ff} &= \frac{1}{8\pi(\tilde{m}_g^2 + \tilde{m}_f^2)}\,,\\
\tilde{\theta}^{gg} = \tilde{\theta}^{gf} = \tilde{\theta}^{fg} = \tilde{\theta}^{ff} &= 0\,,
\end{align}
\end{subequations}
from which follow the observable parameters
\begin{equation}
G_{\text{eff}} = \frac{1}{8\pi(\tilde{m}_g^2 + \tilde{m}_f^2)}\,, \quad \gamma = 1\,.
\end{equation}
In this case the effective Planck mass~\eqref{eqn:planckmass} turns out to be \(m_{\text{Pl}}^2 = \tilde{m}_g^2 + \tilde{m}_f^2\).
\end{itemize}

This concludes our discussion of the post-Newtonian limit of ghost-free bimetric gravity for a static point mass. The PPN parameters we have obtained now allow us to discuss observable effects, and in particular the deflection of light by both dark and visible matter. This will be done in the following section.

\section{Confrontation with observations}\label{sec:observation}
In the previous section we obtained both a general result and a number of limiting cases for the effective gravitational constant and the PPN parameter \(\gamma\), as well as additional PPN parameters which govern effects involving a second, dark type of matter. We can now compare our results with observations, in particular of the deflection of light. We will restrict ourselves to visible matter in section~\ref{ssec:solsys} and derive bounds on the parameters of ghost-free massive bimetric gravity from solar system experiments. In section~\ref{ssec:darkmatter} we will discuss the deflection of visible light by dark matter and its consistency with observations of lensing effects by galaxies. We will further speculate on a possible explanation for the lensing effects observed in the vicinity of galactic mergers, in particular Abell 520 and Abell 3827.

\subsection{Solar system consistency}\label{ssec:solsys}
\begin{figure}[htbp]
\includegraphics[width=10cm]{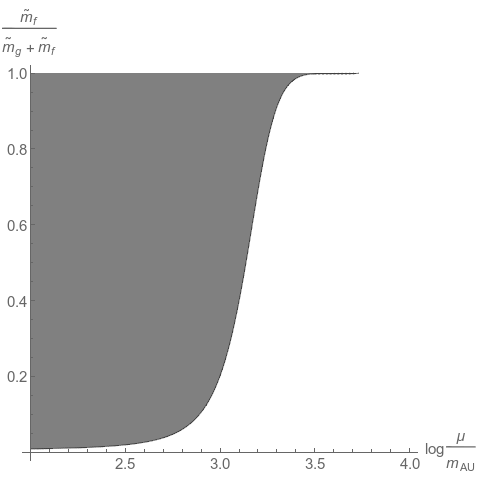}
\caption{Two-dimensional section of the parameter space of ghost-free massive bimetric gravity showing only the Planck mass ratio \(\tilde{m}_f/\tilde{m}_g\) (rescaled to map the interval \((0, \infty)\) into \((0, 1)\)) and graviton mass \(\mu\) in inverse astronomical units \(m_{\mathrm{AU}} = 1\mathrm{AU}^{-1} \approx 1.32 \cdot 10^{-18}\mathrm{eV}/c^2\). The region excluded by the Cassini tracking experiment at \(2\sigma\) confidence level is shown in gray.}
\label{fig:cassini}
\end{figure}

We have remarked in section~\ref{ssec:ppnpar} that our result~\eqref{eqn:geffgamma} for the effective Newtonian constant \(G_{\text{eff}}\) and the PPN parameter \(\gamma\) has essentially the same form as the corresponding result for scalar-tensor gravity with a general potential~\cite{Hohmann:2013rba,Scharer:2014kya}, or the more general Horndeski class of theories~\cite{Hohmann:2015kra}. Hence the experimental constraints on the parameters of these theories derived from measurements of \(\gamma\) can directly be translated to constraints on the parameters of ghost-free massive bimetric gravity, and in particular to the ratio \(\tilde{m}_f/\tilde{m}_g\) of the Planck masses and the graviton mass \(\mu\). An important obstacle that must be taken into account is the fact that \(\gamma\) is not constant, but depends exponentially on the distance \(r\) between the gravitating mass source and the observer. This restricts the possible experimental tests of \(\gamma\) to those for which such an interaction distance can be defined. The most precise observation of \(\gamma\) which satisfies this condition is the measurement of the Shapiro time delay of radio signals between Earth and the Cassini spacecraft on its way to Saturn, from which a value \(\gamma - 1 = (2.1 \pm 2.3) \cdot 10^{-5}\) was obtained~\cite{Bertotti:2003rm}. These were passing by the sun at a distance of \(1.6\) solar radii, so that we define the interaction distance \(r_0 \approx 7.44 \cdot 10^{-3}\mathrm{AU}\). Following the same procedure as detailed in~\cite{Hohmann:2013rba}, we find that the area of the parameter space shown in figure~\ref{fig:cassini} is excluded at \(2\sigma\) confidence level. Note, however, that the assumption of a constant interaction distance for this experiment is only an approximation, and that more accurate results are obtained from a thorough treatment of light propagation in the solar gravitational field~\cite{Deng:2016moh}.

\subsection{Light deflection by dark matter}\label{ssec:darkmatter}
The full set~\eqref{eqn:ppnparsol} of PPN parameters, which we derived in section~\ref{ssec:ppnpar}, allows us to discuss also the gravitational interaction of dark matter. We first consider the parameter \(\tilde{\alpha}^{gf}\), which can be interpreted as an effective Newtonian constant for the gravitational influence of dark matter \(\Phi^f\) on visible matter \(\Phi^g\). For short distances, \(\mu r < \ln(4/3)\), we see that \(\tilde{\alpha}^{gf}\) becomes negative, so that the gravitational interaction between dark and visible matter becomes repulsive; however, taking into account the bounds shown in figure~\ref{fig:cassini}, we see that this is possible only on scales significantly smaller than the solar system, and hence does not play any role for the observed dark matter concentrations. On the scales of galaxies or even galactic clusters we can safely assume \(\mu r \gg 1\), and thus use the PPN parameters obtained in the limit \(\mu \to \infty\) in section~\ref{ssec:limits}. In this limit, the gravitational effects on both test masses and light become indistinguishable between visible and dark matter sources. In particular, it follows that the deflection of visible light by dark matter is likewise governed by a PPN parameter
\begin{equation}
\bar{\gamma} = \frac{\tilde{\gamma}^{gf}}{\tilde{\alpha}^{gf}} \to 1
\end{equation}
in the limit of large scales. This agrees with observations of the deflection of light by galaxies, which contain significant amounts of dark matter in addition to the visible mass~\cite{Bolton:2006yz,Schwab:2009nz,Cao:2017nnq}.

Our result plays an important role in particular for the observed light deflection by galactic mergers, such as most prominently the so-called ``Bullet Cluster'' 1E0657-558~\cite{Markevitch:2003at,Clowe:2003tk,Clowe:2006eq,Randall:2007ph} or more recently MACS J0025.4-1222~\cite{Bradac:2008eu}. Measurements of the mass distribution in these and other mergers using weak lensing together with x-ray imaging show that the gas component of the merger, which is heated by the collision and which constitutes the major amount of visible matter, is not at the same location as the dominant gravitating matter contribution, and that the motion of the latter is largely unaffected by the collision. This leads to the conclusion that their dark matter content is non-interacting, so that the dark matter components of the colliding objects pass through each other~\cite{Harvey:2015hha}. However, observations of the so-called ``Train Wreck Cluster'' Abell 520~\cite{Mahdavi:2007yp,Jee:2014hja} or Abell 3827~\cite{Massey:2015dkw} show a more differentiated picture. While also Abell 520 shows evidence for dark matter components which have passed through each other unaffectedly, one has further identified another dark mass concentration in the central region, which is difficult to explain if dark matter is non-interacting. Similar stress on the non-interacting dark matter model is put by an observed separation between stellar and dark matter in Abell 3827. A possible explanation for these observations is to assume that dark matter also possesses a component which interacts non-gravitationally~\cite{Heikinheimo:2015kra,Kahlhoefer:2015vua,Sepp:2016tfs}.

The bimetric class of theories we studied in this article allows for an interesting tentative model for the aforementioned observations, which hint towards the existence of both interacting and non-interacting dark matter components. Invoking the interpretation of the matter sector \(\Phi^f\) as dark matter, as suggested in~\cite{Aoki:2014cla,Bernard:2014psa,Blanchet:2015sra,Blanchet:2015bia}, and further assuming that \(\Phi^f\) contains an interacting component, would suggest that the central dark matter concentration in Abell 520 and the separated dark matter concentration in Abell 3827 result from a collision of these interacting components, while any dark matter constituted by massive gravitons, as suggested in~\cite{Schmidt-May:2016hsx,Babichev:2016hir,Aoki:2016zgp,Babichev:2016bxi}, would pass the merger unaffectedly, and could thus account for the dark matter concentrations away from the center of Abell 520 or the unaffected dark matter halos in Abell 3827. Future extensions of our work presented here will be necessary in order to quantitatively assert the viability of such models.

\section{Conclusion}\label{sec:conclusion}
We have considered the post-Newtonian limit of ghost-free massive bimetric gravity with two mutually non-interacting matter sectors. From the assumption that the vacuum field equations are solved by two flat metrics proportional to the Minkowski metric, we have derived restrictions on the parameters in the action. For this restricted class of theories we have derived the field equations up to the second velocity order by making use of a suitable extension of the PPN formalism to multiple metrics. We have solved these equations for a point-like mass source using a gauge-invariant differential decomposition of the metric perturbations. From this solution we have read off the effective gravitational constant \(G_{\text{eff}}\) and the PPN parameter \(\gamma\) for the visible matter sector. By comparing our result to the observed value determined by the Cassini tracking experiment we have derived combined bounds on two parameters of the theory, namely on the mass of the massive graviton and on the ratio of the Planck masses occurring in the bimetric action.

We have further discussed the interpretation of the additional matter sector as a possible constituent of dark matter. From our experimental bounds we then concluded that on scales significantly larger than the solar system, and hence in particular on the observationally relevant scales of galaxies and clusters, the gravitational effects caused by visible and dark matter become indistinguishable from each other. It thus follows that dark matter should deflect light in the same way as visible matter does, in agreement with measurements of the PPN parameter \(\gamma\) through the lensing effect of galaxies, which contain a significant dark matter component. Another possible experimental test of this result could be performed by searching for possible (non-)correlations between the ratio of dark to visible matter of a galaxy and its light deflection. Such an analysis would be most effective with data of higher precision than available to date~\cite{Bolton:2008xf,Cao:2015qja}.

On a more speculative note, we have considered that besides the second matter sector also massive gravitons could contribute to the observed dark matter content of the universe. The assumption that the former contains non-gravitational self-interactions, while the latter interacts only gravitationally, then provides a tentative explanation for the observed separation of apparently different dark matter components in galactic mergers such as Abell 520 and Abell 3827. The question arises whether such different dark matter constituents could be distinguished also in other processes besides galactic mergers, for example, by their light deflection properties. An extension of our work presented here to the light deflection caused by massive graviton concentrations might answer this question.

There are also other possibilities to further extend the theoretical analysis we presented in this article. While we have studied only linear perturbations of flat vacuum solutions, considering also the quadratic perturbation order would allow us to calculate the PPN parameter \(\beta\), and thus open the possibility for additional tests using solar system observations. This would ultimately lead to a full generalization of the formalism developed in~\cite{Hohmann:2010ni,Hohmann:2013oca} to massive gravity theories. Further, one may also include cosmological corrections to the PPN formalism along the lines of~\cite{Sanghai:2016tbi}, and thus relax the condition of a flat background. Finally, one may consider more general theories with \(N > 2\) metric tensors and a corresponding number of matter sectors~\cite{Hinterbichler:2012cn,Hassan:2012wt,Nomura:2012xr,Noller:2014sta,Scargill:2014wya,Hinterbichler:2015yaa,Noller:2015eda}, or involving an effective metric~\cite{Akrami:2013ffa,deRham:2014fha,Huang:2015yga,Melville:2015dba}, both of which allow for ghost-free matter coupling prescriptions~\cite{Schmidt-May:2015vnx}. We intend to study these generalizations in future research.

\begin{acknowledgments}
The author is happy to thank the members of the Laboratory of Theoretical Physics at the University of Tartu for fruitful discussions. He gratefully acknowledges the full financial support of the Estonian Research Council through the Startup Research Grant PUT790 and the European Regional Development Fund through the Center of Excellence TK133 ``The Dark Side of the Universe''.
\end{acknowledgments}

\appendix
\section{Linearization of the potential}\label{app:linpot}
In this appendix we show how to obtain the linearized potentials, which enter the gravitational field equations at the zeroth and second velocity order as shown in sections~\ref{ssec:background} and~\ref{ssec:2ndorder}. The starting point for our derivation is a linear perturbation ansatz for the metrics, which we write in the form
\begin{equation}
g_{\mu\nu} = \eta_{\mu\nu} + h_{\mu\nu}\,,\quad
f_{\mu\nu} = c^2\left(\eta_{\mu\nu} + e_{\mu\nu}\right)\,.
\end{equation}
Up to the linear perturbation order, we can then write their inverses as
\begin{equation}
g^{\mu\nu} = \eta^{\mu\nu} - \eta^{\mu\rho}\eta^{\nu\sigma}h_{\rho\sigma} + \mathcal{O}(h^2)\,, \quad
f^{\mu\nu} = \frac{1}{c^2}\left(\eta^{\mu\nu} - \eta^{\mu\rho}\eta^{\nu\sigma}e_{\rho\sigma}\right) + \mathcal{O}(e^2)\,.
\end{equation}
For their product we find
\begin{equation}
g^{\mu\rho}f_{\rho\nu} = c^2(\delta^{\mu}_{\nu} - D^{\mu}{}_{\nu}) + \mathcal{O}(\{h,e\}^2)\,,
\end{equation}
where we introduced the perturbation tensor
\begin{equation}
D^{\mu}{}_{\nu} = \eta^{\mu\rho}(h_{\rho\nu} - e_{\rho\nu})\,.
\end{equation}
Since the matrix \(g^{\mu\rho}f_{\rho\nu}\) is given as a perturbation of the Kronecker symbol \(\delta^{\mu}_{\nu}\), we can find its square root \(A^{\mu}{}_{\nu}\) as defined in~\eqref{eqn:sqrt} using a series expansion analogously to the well-known Taylor series
\begin{equation}
\sqrt{1 + x} = 1 + \frac{x}{2} + \mathcal{O}(x^2)\,.
\end{equation}
This series expansion yields
\begin{equation}
A^{\mu}{}_{\nu} = c\left(\delta^{\mu}_{\nu} - \frac{1}{2}D^{\mu}{}_{\nu}\right) + \mathcal{O}(\{h,e\}^2)\,.
\end{equation}
For later use we also need to expand powers of \(A\) into linear perturbations. These are given by
\begin{equation}
(A^k)^{\mu}{}_{\nu} = c^k\left(\delta^{\mu}_{\nu} - \frac{k}{2}D^{\mu}{}_{\nu}\right) + \mathcal{O}(\{h,e\}^2)\,.
\end{equation}
The matrix invariants \(e_k(A)\) defined by~\eqref{eqn:matrixinv} then take the form
\begin{subequations}
\begin{align}
e_0(A) &= 1\,,\\
e_1(A) &= A^{\mu}{}_{\mu} = c\left(4 - \frac{1}{2}D^{\mu}{}_{\mu}\right) + \mathcal{O}(\{h,e\}^2)\,,\\
e_2(A) &= \frac{1}{2}(A^{\mu}{}_{\mu}A^{\nu}{}_{\nu} - A^{\mu}{}_{\nu}A^{\nu}{}_{\mu}) = c^2\left(6 - \frac{3}{2}D^{\mu}{}_{\mu}\right) + \mathcal{O}(\{h,e\}^2)\,,\\
e_3(A) &= \frac{1}{6}(A^{\mu}{}_{\mu}A^{\nu}{}_{\nu}A^{\rho}{}_{\rho} - 3A^{\mu}{}_{\nu}A^{\nu}{}_{\mu}A^{\rho}{}_{\rho} + 2A^{\mu}{}_{\nu}A^{\nu}{}_{\rho}A^{\rho}{}_{\mu}) = c^3\left(4 - \frac{3}{2}D^{\mu}{}_{\mu}\right) + \mathcal{O}(\{h,e\}^2)\,,\\
e_4(A) &= \frac{1}{24}(A^{\mu}{}_{\mu}A^{\nu}{}_{\nu}A^{\rho}{}_{\rho}A^{\sigma}{}_{\sigma} - 6A^{\mu}{}_{\mu}A^{\nu}{}_{\nu}A^{\rho}{}_{\sigma}A^{\sigma}{}_{\rho} + 3A^{\mu}{}_{\nu}A^{\nu}{}_{\mu}A^{\rho}{}_{\sigma}A^{\sigma}{}_{\rho} + 8A^{\mu}{}_{\nu}A^{\nu}{}_{\rho}A^{\rho}{}_{\mu}A^{\sigma}{}_{\sigma}\\
&\phantom{=}- 6A^{\mu}{}_{\nu}A^{\nu}{}_{\rho}A^{\rho}{}_{\sigma}A^{\sigma}{}_{\mu}) = c^4\left(1 - \frac{1}{2}D^{\mu}{}_{\mu}\right) + \mathcal{O}(\{h,e\}^2)\,.
\end{align}
\end{subequations}
For the matrices \(Y_n\) defined via~\eqref{eqn:potsum} we then find the expressions
\begin{subequations}
\begin{align}
Y_0^{\mu}{}_{\nu}(A) &= \delta^{\mu}_{\nu} + \mathcal{O}(\{h,e\}^2)\,,\\
Y_1^{\mu}{}_{\nu}(A) &= c\left(-3\delta^{\mu}_{\nu} - \frac{1}{2}D^{\mu}{}_{\nu} + \frac{1}{2}D^{\rho}{}_{\rho}\delta^{\mu}_{\nu}\right) + \mathcal{O}(\{h,e\}^2)\,,\\
Y_2^{\mu}{}_{\nu}(A) &= c^2\left(3\delta^{\mu}_{\nu} + D^{\mu}{}_{\nu} - D^{\rho}{}_{\rho}\delta^{\mu}_{\nu}\right) + \mathcal{O}(\{h,e\}^2)\,,\\
Y_3^{\mu}{}_{\nu}(A) &= c^3\left(-\delta^{\mu}_{\nu} - \frac{1}{2}D^{\mu}{}_{\nu} + \frac{1}{2}D^{\rho}{}_{\rho}\delta^{\mu}_{\nu}\right) + \mathcal{O}(\{h,e\}^2)\,.
\end{align}
\end{subequations}
In order to obtain the corresponding expressions for \(A^{-1} = \sqrt{f^{-1}g}\) instead of \(A\), one simply replaces \(D\) by \(-D\) and \(c\) by \(c^{-1}\). We can now calculate the potentials~\eqref{eqn:potential}. Using renormalized parameters \(\tilde{\beta}_k = c^k\beta_k\) we obtain
\begin{subequations}
\begin{align}
V^g_{\mu\nu} &= \bigg[\left(\tilde{\beta}_0 + 3\tilde{\beta}_1 + 3\tilde{\beta}_2 + \tilde{\beta}_3\right)\eta_{\mu\nu} - \left(\frac{1}{2}\tilde{\beta}_1 + \tilde{\beta}_2 + \frac{1}{2}\tilde{\beta}_3\right)\eta_{\mu\nu}\eta^{\rho\sigma}(h_{\rho\sigma} - e_{\rho\sigma})\nonumber\\
&\phantom{=}+ \left(\tilde{\beta}_0 + \frac{7}{2}\tilde{\beta}_1 + 4\tilde{\beta}_2 + \frac{3}{2}\tilde{\beta}_3\right)h_{\mu\nu} - \left(\frac{1}{2}\tilde{\beta}_1 + \tilde{\beta}_2 + \frac{1}{2}\tilde{\beta}_3\right)e_{\mu\nu}\bigg] + \mathcal{O}(\{h,e\}^2)\,,\\
V^f_{\mu\nu} &= \frac{1}{c^2}\bigg[\left(\tilde{\beta}_1 + 3\tilde{\beta}_2 + 3\tilde{\beta}_3 + \tilde{\beta}_4\right)\eta_{\mu\nu} + \left(\frac{1}{2}\tilde{\beta}_1 + \tilde{\beta}_2 + \frac{1}{2}\tilde{\beta}_3\right)\eta_{\mu\nu}\eta^{\rho\sigma}(h_{\rho\sigma} - e_{\rho\sigma})\nonumber\\
&\phantom{=}+ \left(\frac{3}{2}\tilde{\beta}_1 + 4\tilde{\beta}_2 + \frac{7}{2}\tilde{\beta}_3 + \tilde{\beta}_4\right)e_{\mu\nu} - \left(\frac{1}{2}\tilde{\beta}_1 + \tilde{\beta}_2 + \frac{1}{2}\tilde{\beta}_3\right)h_{\mu\nu}\bigg] + \mathcal{O}(\{h,e\}^2)\,.
\end{align}
\end{subequations}
Finally, we calculate the trace-reversed potentials~\eqref{eqn:trrevpot}, which are given by
\begin{subequations}\label{eqn:lintrpot}
\begin{align}
\bar{V}^g_{\mu\nu} &= \bigg[-\left(\tilde{\beta}_0 + 3\tilde{\beta}_1 + 3\tilde{\beta}_2 + \tilde{\beta}_3\right)\eta_{\mu\nu} + \left(\frac{1}{4}\tilde{\beta}_1 + \frac{1}{2}\tilde{\beta}_2 + \frac{1}{4}\tilde{\beta}_3\right)\eta_{\mu\nu}\eta^{\rho\sigma}(h_{\rho\sigma} - e_{\rho\sigma})\nonumber\\
&\phantom{=}- \left(\tilde{\beta}_0 + \frac{5}{2}\tilde{\beta}_1 + 2\tilde{\beta}_2 + \frac{1}{2}\tilde{\beta}_3\right)h_{\mu\nu} - \left(\frac{1}{2}\tilde{\beta}_1 + \tilde{\beta}_2 + \frac{1}{2}\tilde{\beta}_3\right)e_{\mu\nu}\bigg] + \mathcal{O}(\{h,e\}^2)\,,\\
\bar{V}^f_{\mu\nu} &= \frac{1}{c^2}\bigg[-\left(\tilde{\beta}_1 + 3\tilde{\beta}_2 + 3\tilde{\beta}_3 + \tilde{\beta}_4\right)\eta_{\mu\nu} - \left(\frac{1}{4}\tilde{\beta}_1 + \frac{1}{2}\tilde{\beta}_2 + \frac{1}{4}\tilde{\beta}_3\right)\eta_{\mu\nu}\eta^{\rho\sigma}(h_{\rho\sigma} - e_{\rho\sigma})\nonumber\\
&\phantom{=}- \left(\frac{1}{2}\tilde{\beta}_1 + 2\tilde{\beta}_2 + \frac{5}{2}\tilde{\beta}_3 + \tilde{\beta}_4\right)e_{\mu\nu} - \left(\frac{1}{2}\tilde{\beta}_1 + \tilde{\beta}_2 + \frac{1}{2}\tilde{\beta}_3\right)h_{\mu\nu}\bigg] + \mathcal{O}(\{h,e\}^2)\,.
\end{align}
\end{subequations}
These expressions can now be used in the post-Newtonian field equations at the zeroth velocity order in section~\ref{ssec:background} and at the second velocity order in section~\ref{ssec:2ndorder}.

\section{Derivatives of the Yukawa potential}\label{app:yukawa}
During our calculation we have frequently encountered (mostly second order) derivatives of the Yukawa potential, for which we introduced the shorthand notation
\begin{equation}
\mathcal{Y}_k(r) = \frac{e^{-kr}}{r}\,.
\end{equation}
Taking into account the singularity at the origin, its second derivatives are given by
\begin{equation}
\partial_i\partial_j\mathcal{Y}_k = \left\{[kr(kr + 3) + 3]\frac{x_ix_j}{r^5} - (kr + 1)\frac{\delta_{ij}}{r^3}\right\}e^{-kr} - \frac{4\pi}{3}\delta_{ij}\delta(\vec{x})\,,
\end{equation}
which is a straightforward generalization of the well-known formula for the Coulomb potential~\cite{Frahm:1983}. Taking the trace yields the standard formula
\begin{equation}
\triangle\mathcal{Y}_k = k^2\frac{e^{-kr}}{r} - 4\pi\delta(\vec{x})\,.
\end{equation}
These formulas cover all expressions which appear in the final result for the Ricci tensor and the interaction potential. Note that during intermediate steps also fourth order derivatives of the Yukawa potential occur in derivatives of the metric perturbations. For completeness we also list the corresponding expressions. From the formula given above immediately follows
\begin{equation}
\partial_i\partial_j\triangle\mathcal{Y}_k = k^2\left\{[kr(kr + 3) + 3]\frac{x_ix_j}{r^5} - (kr + 1)\frac{\delta_{ij}}{r^3}\right\}e^{-kr} - \frac{4\pi k^2}{3}\delta_{ij}\delta(\vec{x}) - 4\pi\partial_i\partial_j\delta(\vec{x})
\end{equation}
and thus
\begin{equation}
\triangle\triangle\mathcal{Y}_k = k^4\frac{e^{-kr}}{r} - 4\pi k^2\delta(\vec{x}) - 4\pi\triangle\delta(\vec{x})\,.
\end{equation}
These are all terms which occur during our calculation.

\section{Checking the field equations in components}\label{app:feqcheck}
Since we have used a rather technical transformation of the field equations to gauge invariant potentials in section~\ref{ssec:feqdec} and the corresponding inverse transformation of their solution to metric components in section~\ref{ssec:metricsol}, it is appropriate to check the obtained result also using the field equations in their original component form as shown in section~\ref{ssec:2ndorder}. While this is rather cumbersome using the explicit expressions~\eqref{eqn:metsolr} and requires careful tracking of singular contributions from higher derivatives of Coulomb and Yukawa potentials, it becomes considerably simpler by using the abbreviations~\eqref{eqn:yukcoul}, starting from the expressions~\eqref{eqn:metsoly} and finally evaluating higher derivatives using the formulas shown in appendix~\ref{app:yukawa}.

From the expressions~\eqref{eqn:metsoly} one easily reads off the traces
\begin{subequations}
\begin{align}
h^{(2)}_{ii} &= 2[3\mathcal{I}_M\mathcal{Y}_0 - 3\mu^2(\mathcal{D} + 1)\mathcal{I}_-\mathcal{Y}_{\mu} + (\mathcal{I}_+ + \mathcal{I}_-)\triangle\mathcal{Y}_{\mu}]\,,\\
e^{(2)}_{ii} &= 2[3\mathcal{I}_M\mathcal{Y}_0 - 3\mu^2(\mathcal{D} - 1)\mathcal{I}_-\mathcal{Y}_{\mu} + (\mathcal{I}_+ - \mathcal{I}_-)\triangle\mathcal{Y}_{\mu}]
\end{align}
\end{subequations}
of the spatial components of the metric perturbations. Using the formulas~\eqref{eqn:pot2nd} for the potential at the second velocity order we then obtain
\begin{subequations}\label{eqn:potyuk}
\begin{align}
V^{g(2)}_{00} = -c^2V^{f(2)}_{00} &= -\tilde{\beta}\mathcal{I}_-(3\mu^2\mathcal{Y}_{\mu} + \triangle\mathcal{Y}_{\mu})\,,\\
V^{g(2)}_{ij} = -c^2V^{f(2)}_{ij} &= -\tilde{\beta}\mathcal{I}_-(3\mu^2\mathcal{Y}_{\mu}\delta_{ij} - \triangle\mathcal{Y}_{\mu}\delta_{ij} - 2\partial_i\partial_j\mathcal{Y}_{\mu})\,.
\end{align}
\end{subequations}
Further, we need to evaluate second order derivatives of the metric, which read
\begin{subequations}
\begin{align}
h^{(2)}_{00,ij} &= 2\mathcal{I}_M\partial_i\partial_j\mathcal{Y}_0 - 4\mu^2(\mathcal{D} + 1)\mathcal{I}_-\partial_i\partial_j\mathcal{Y}_{\mu}\,,\\
e^{(2)}_{00,ij} &= 2\mathcal{I}_M\partial_i\partial_j\mathcal{Y}_0 - 4\mu^2(\mathcal{D} - 1)\mathcal{I}_-\partial_i\partial_j\mathcal{Y}_{\mu}\,,\\
\triangle h^{(2)}_{00} &= 2\mathcal{I}_M\triangle\mathcal{Y}_0 - 4\mu^2(\mathcal{D} + 1)\mathcal{I}_-\triangle\mathcal{Y}_{\mu}\,,\\
\triangle e^{(2)}_{00} &= 2\mathcal{I}_M\triangle\mathcal{Y}_0 - 4\mu^2(\mathcal{D} - 1)\mathcal{I}_-\triangle\mathcal{Y}_{\mu}\,,\\
h^{(2)}_{kk,ij} &= 2[3\mathcal{I}_M\partial_i\partial_j\mathcal{Y}_0 - 3\mu^2(\mathcal{D} + 1)\mathcal{I}_-\partial_i\partial_j\mathcal{Y}_{\mu} + (\mathcal{I}_+ + \mathcal{I}_-)\partial_i\partial_j\triangle\mathcal{Y}_{\mu}]\,,\\
e^{(2)}_{kk,ij} &= 2[3\mathcal{I}_M\partial_i\partial_j\mathcal{Y}_0 - 3\mu^2(\mathcal{D} - 1)\mathcal{I}_-\partial_i\partial_j\mathcal{Y}_{\mu} + (\mathcal{I}_+ - \mathcal{I}_-)\partial_i\partial_j\triangle\mathcal{Y}_{\mu}]\,,\\
\triangle h^{(2)}_{ij} &= 2[\mathcal{I}_M\triangle\mathcal{Y}_0 - \mu^2(\mathcal{D} + 1)\mathcal{I}_-\triangle\mathcal{Y}_{\mu}]\delta_{ij} + 2(\mathcal{I}_+ + \mathcal{I}_-)\partial_i\partial_j\triangle\mathcal{Y}_{\mu}\,,\\
\triangle e^{(2)}_{ij} &= 2[\mathcal{I}_M\triangle\mathcal{Y}_0 - \mu^2(\mathcal{D} - 1)\mathcal{I}_-\triangle\mathcal{Y}_{\mu}]\delta_{ij} + 2(\mathcal{I}_+ - \mathcal{I}_-)\partial_i\partial_j\triangle\mathcal{Y}_{\mu}\,,\\
\triangle h^{(2)}_{ii} &= 2[3\mathcal{I}_M\triangle\mathcal{Y}_0 - 3\mu^2(\mathcal{D} + 1)\mathcal{I}_-\triangle\mathcal{Y}_{\mu} + (\mathcal{I}_+ + \mathcal{I}_-)\triangle\triangle\mathcal{Y}_{\mu}]\,,\\
\triangle e^{(2)}_{ii} &= 2[3\mathcal{I}_M\triangle\mathcal{Y}_0 - 3\mu^2(\mathcal{D} - 1)\mathcal{I}_-\triangle\mathcal{Y}_{\mu} + (\mathcal{I}_+ - \mathcal{I}_-)\triangle\triangle\mathcal{Y}_{\mu}]\,,\\
h^{(2)}_{ik,jk} &= 2[\mathcal{I}_M\partial_i\partial_j\mathcal{Y}_0 - \mu^2(\mathcal{D} + 1)\mathcal{I}_-\partial_i\partial_j\mathcal{Y}_{\mu} + (\mathcal{I}_+ + \mathcal{I}_-)\partial_i\partial_j\triangle\mathcal{Y}_{\mu}]\,,\\
e^{(2)}_{ik,jk} &= 2[\mathcal{I}_M\partial_i\partial_j\mathcal{Y}_0 - \mu^2(\mathcal{D} - 1)\mathcal{I}_-\partial_i\partial_j\mathcal{Y}_{\mu} + (\mathcal{I}_+ - \mathcal{I}_-)\partial_i\partial_j\triangle\mathcal{Y}_{\mu}]\,.
\end{align}
\end{subequations}
Inserting these expressions into the formulas~\eqref{eqn:ricci2nd} for the Ricci tensor then yields the components
\begin{subequations}\label{eqn:ricciyuk}
\begin{align}
R^{g,f(2)}_{00} &= -\mathcal{I}_M\triangle\mathcal{Y}_0 + 2\mu^2(\mathcal{D} \pm 1)\mathcal{I}_-\triangle\mathcal{Y}_{\mu}\,,\\
R^{g,f(2)}_{ij} &= -\mathcal{I}_M\triangle\mathcal{Y}_0\delta_{ij} + \mu^2(\mathcal{D} \pm 1)\mathcal{I}_-(\triangle\mathcal{Y}_{\mu}\delta_{ij} - \partial_i\partial_j\mathcal{Y}_{\mu})\,.
\end{align}
\end{subequations}
Inserting the expressions~\eqref{eqn:ricciyuk} and~\eqref{eqn:potyuk} into the second order field equations~\eqref{eqn:2ndorder}, applying the definitions~\eqref{eqn:potconst} and using the relations for the Coulomb and Yukawa potentials listed in appendix~\ref{app:yukawa} finally yields
\begin{subequations}
\begin{align}
\tilde{m}_g^2R^{g(2)}_{00} + m^4\bar{V}^{g(2)}_{00} &= -\frac{\tilde{m}_g^2(\tilde{M}^g + \tilde{M}^f)\triangle\mathcal{Y}_0 - (\tilde{m}_g^2\tilde{M}^f - \tilde{m}_f^2\tilde{M}^g)(\triangle\mathcal{Y}_{\mu} - \mu^2\mathcal{Y}_{\mu})}{8\pi(\tilde{m}_g^2 + \tilde{m}_f^2)}\nonumber\\
&= \frac{\tilde{M}^g}{2}\delta(\vec{x}) = \bar{T}^{g(2)}_{00}\,,\\
\frac{\tilde{m}_f^2}{c^2}R^{f(2)}_{00} + m^4\bar{V}^{f(2)}_{00} &= -\frac{\tilde{m}_f^2(\tilde{M}^g + \tilde{M}^f)\triangle\mathcal{Y}_0 + (\tilde{m}_g^2\tilde{M}^f - \tilde{m}_f^2\tilde{M}^g)(\triangle\mathcal{Y}_{\mu} - \mu^2\mathcal{Y}_{\mu})}{8\pi c^2(\tilde{m}_g^2 + \tilde{m}_f^2)}\nonumber\\
&= \frac{\tilde{M}^f}{2}\delta(\vec{x}) = \bar{T}^{f(2)}_{00}\,,\\
\tilde{m}_g^2R^{g(2)}_{ij} + m^4\bar{V}^{g(2)}_{ij} &= -\frac{\tilde{m}_g^2(\tilde{M}^g + \tilde{M}^f)\triangle\mathcal{Y}_0 - (\tilde{m}_g^2\tilde{M}^f - \tilde{m}_f^2\tilde{M}^g)(\triangle\mathcal{Y}_{\mu} - \mu^2\mathcal{Y}_{\mu})}{8\pi(\tilde{m}_g^2 + \tilde{m}_f^2)}\delta_{ij}\nonumber\\
&= \frac{\tilde{M}^g}{2}\delta(\vec{x})\delta_{ij} = \bar{T}^{g(2)}_{ij}\,,\\
\frac{\tilde{m}_f^2}{c^2}R^{f(2)}_{ij} + m^4\bar{V}^{f(2)}_{ij} &= -\frac{\tilde{m}_f^2(\tilde{M}^g + \tilde{M}^f)\triangle\mathcal{Y}_0 + (\tilde{m}_g^2\tilde{M}^f - \tilde{m}_f^2\tilde{M}^g)(\triangle\mathcal{Y}_{\mu} - \mu^2\mathcal{Y}_{\mu})}{8\pi c^2(\tilde{m}_g^2 + \tilde{m}_f^2)}\delta_{ij}\nonumber\\
&= \frac{\tilde{M}^f}{2}\delta(\vec{x})\delta_{ij} = \bar{T}^{f(2)}_{ij}\,.
\end{align}
\end{subequations}
This shows that the field equations are indeed satisfied.

\bibliographystyle{apsrev4-1}
\bibliography{bippn}

\begin{thebibliography}{102}%
\makeatletter
\providecommand \@ifxundefined [1]{%
 \@ifx{#1\undefined}
}%
\providecommand \@ifnum [1]{%
 \ifnum #1\expandafter \@firstoftwo
 \else \expandafter \@secondoftwo
 \fi
}%
\providecommand \@ifx [1]{%
 \ifx #1\expandafter \@firstoftwo
 \else \expandafter \@secondoftwo
 \fi
}%
\providecommand \natexlab [1]{#1}%
\providecommand \enquote  [1]{``#1''}%
\providecommand \bibnamefont  [1]{#1}%
\providecommand \bibfnamefont [1]{#1}%
\providecommand \citenamefont [1]{#1}%
\providecommand \href@noop [0]{\@secondoftwo}%
\providecommand \href [0]{\begingroup \@sanitize@url \@href}%
\providecommand \@href[1]{\@@startlink{#1}\@@href}%
\providecommand \@@href[1]{\endgroup#1\@@endlink}%
\providecommand \@sanitize@url [0]{\catcode `\\12\catcode `\$12\catcode
  `\&12\catcode `\#12\catcode `\^12\catcode `\_12\catcode `\%12\relax}%
\providecommand \@@startlink[1]{}%
\providecommand \@@endlink[0]{}%
\providecommand \url  [0]{\begingroup\@sanitize@url \@url }%
\providecommand \@url [1]{\endgroup\@href {#1}{\urlprefix }}%
\providecommand \urlprefix  [0]{URL }%
\providecommand \Eprint [0]{\href }%
\providecommand \doibase [0]{http://dx.doi.org/}%
\providecommand \selectlanguage [0]{\@gobble}%
\providecommand \bibinfo  [0]{\@secondoftwo}%
\providecommand \bibfield  [0]{\@secondoftwo}%
\providecommand \translation [1]{[#1]}%
\providecommand \BibitemOpen [0]{}%
\providecommand \bibitemStop [0]{}%
\providecommand \bibitemNoStop [0]{.\EOS\space}%
\providecommand \EOS [0]{\spacefactor3000\relax}%
\providecommand \BibitemShut  [1]{\csname bibitem#1\endcsname}%
\let\auto@bib@innerbib\@empty
%</preamble>
\bibitem [{\citenamefont {Ade}\ \emph {et~al.}(2016)\citenamefont {Ade} \emph
  {et~al.}}]{Ade:2015xua}%
  \BibitemOpen
  \bibfield  {author} {\bibinfo {author} {\bibfnamefont {P.~A.~R.}\
  \bibnamefont {Ade}} \emph {et~al.} (\bibinfo {collaboration} {Planck}),\
  }\href {\doibase 10.1051/0004-6361/201525830} {\bibfield  {journal} {\bibinfo
   {journal} {Astron. Astrophys.}\ }\textbf {\bibinfo {volume} {594}},\
  \bibinfo {pages} {A13} (\bibinfo {year} {2016})},\ \Eprint
  {http://arxiv.org/abs/1502.01589} {arXiv:1502.01589 [astro-ph.CO]}
  \BibitemShut {NoStop}%
%%CITATION = ARXIV:1502.01589;%%
\bibitem [{\citenamefont {Peebles}\ and\ \citenamefont
  {Ratra}(2003)}]{Peebles:2002gy}%
  \BibitemOpen
  \bibfield  {author} {\bibinfo {author} {\bibfnamefont {P.~J.~E.}\
  \bibnamefont {Peebles}}\ and\ \bibinfo {author} {\bibfnamefont
  {B.}~\bibnamefont {Ratra}},\ }\href {\doibase 10.1103/RevModPhys.75.559}
  {\bibfield  {journal} {\bibinfo  {journal} {Rev. Mod. Phys.}\ }\textbf
  {\bibinfo {volume} {75}},\ \bibinfo {pages} {559} (\bibinfo {year} {2003})},\
  \Eprint {http://arxiv.org/abs/astro-ph/0207347} {arXiv:astro-ph/0207347
  [astro-ph]} \BibitemShut {NoStop}%
%%CITATION = ASTRO-PH/0207347;%%
\bibitem [{\citenamefont {Riess}\ \emph {et~al.}(1998)\citenamefont {Riess}
  \emph {et~al.}}]{Riess:1998cb}%
  \BibitemOpen
  \bibfield  {author} {\bibinfo {author} {\bibfnamefont {A.~G.}\ \bibnamefont
  {Riess}} \emph {et~al.} (\bibinfo {collaboration} {Supernova Search Team}),\
  }\href {\doibase 10.1086/300499} {\bibfield  {journal} {\bibinfo  {journal}
  {Astron. J.}\ }\textbf {\bibinfo {volume} {116}},\ \bibinfo {pages} {1009}
  (\bibinfo {year} {1998})},\ \Eprint {http://arxiv.org/abs/astro-ph/9805201}
  {arXiv:astro-ph/9805201 [astro-ph]} \BibitemShut {NoStop}%
%%CITATION = ASTRO-PH/9805201;%%
\bibitem [{\citenamefont {Perlmutter}\ \emph {et~al.}(1999)\citenamefont
  {Perlmutter} \emph {et~al.}}]{Perlmutter:1998np}%
  \BibitemOpen
  \bibfield  {author} {\bibinfo {author} {\bibfnamefont {S.}~\bibnamefont
  {Perlmutter}} \emph {et~al.} (\bibinfo {collaboration} {Supernova Cosmology
  Project}),\ }\href {\doibase 10.1086/307221} {\bibfield  {journal} {\bibinfo
  {journal} {Astrophys. J.}\ }\textbf {\bibinfo {volume} {517}},\ \bibinfo
  {pages} {565} (\bibinfo {year} {1999})},\ \Eprint
  {http://arxiv.org/abs/astro-ph/9812133} {arXiv:astro-ph/9812133 [astro-ph]}
  \BibitemShut {NoStop}%
%%CITATION = ASTRO-PH/9812133;%%
\bibitem [{\citenamefont {Kowalski}\ \emph {et~al.}(2008)\citenamefont
  {Kowalski} \emph {et~al.}}]{Kowalski:2008ez}%
  \BibitemOpen
  \bibfield  {author} {\bibinfo {author} {\bibfnamefont {M.}~\bibnamefont
  {Kowalski}} \emph {et~al.} (\bibinfo {collaboration} {Supernova Cosmology
  Project}),\ }\href {\doibase 10.1086/589937} {\bibfield  {journal} {\bibinfo
  {journal} {Astrophys. J.}\ }\textbf {\bibinfo {volume} {686}},\ \bibinfo
  {pages} {749} (\bibinfo {year} {2008})},\ \Eprint
  {http://arxiv.org/abs/0804.4142} {arXiv:0804.4142 [astro-ph]} \BibitemShut
  {NoStop}%
%%CITATION = ARXIV:0804.4142;%%
\bibitem [{\citenamefont {Amanullah}\ \emph {et~al.}(2010)\citenamefont
  {Amanullah} \emph {et~al.}}]{Amanullah:2010vv}%
  \BibitemOpen
  \bibfield  {author} {\bibinfo {author} {\bibfnamefont {R.}~\bibnamefont
  {Amanullah}} \emph {et~al.},\ }\href {\doibase 10.1088/0004-637X/716/1/712}
  {\bibfield  {journal} {\bibinfo  {journal} {Astrophys. J.}\ }\textbf
  {\bibinfo {volume} {716}},\ \bibinfo {pages} {712} (\bibinfo {year}
  {2010})},\ \Eprint {http://arxiv.org/abs/1004.1711} {arXiv:1004.1711
  [astro-ph.CO]} \BibitemShut {NoStop}%
%%CITATION = ARXIV:1004.1711;%%
\bibitem [{\citenamefont {Suzuki}\ \emph {et~al.}(2012)\citenamefont {Suzuki}
  \emph {et~al.}}]{Suzuki:2011hu}%
  \BibitemOpen
  \bibfield  {author} {\bibinfo {author} {\bibfnamefont {N.}~\bibnamefont
  {Suzuki}} \emph {et~al.},\ }\href {\doibase 10.1088/0004-637X/746/1/85}
  {\bibfield  {journal} {\bibinfo  {journal} {Astrophys. J.}\ }\textbf
  {\bibinfo {volume} {746}},\ \bibinfo {pages} {85} (\bibinfo {year} {2012})},\
  \Eprint {http://arxiv.org/abs/1105.3470} {arXiv:1105.3470 [astro-ph.CO]}
  \BibitemShut {NoStop}%
%%CITATION = ARXIV:1105.3470;%%
\bibitem [{\citenamefont {Rubin}\ and\ \citenamefont
  {Ford}(1970)}]{Rubin:1970zza}%
  \BibitemOpen
  \bibfield  {author} {\bibinfo {author} {\bibfnamefont {V.~C.}\ \bibnamefont
  {Rubin}}\ and\ \bibinfo {author} {\bibfnamefont {W.~K.}\ \bibnamefont {Ford},
  \bibfnamefont {Jr.}},\ }\href {\doibase 10.1086/150317} {\bibfield  {journal}
  {\bibinfo  {journal} {Astrophys. J.}\ }\textbf {\bibinfo {volume} {159}},\
  \bibinfo {pages} {379} (\bibinfo {year} {1970})}\BibitemShut {NoStop}%
%%CITATION = ASJOA,159,379;%%
\bibitem [{\citenamefont {Rubin}\ \emph {et~al.}(1980)\citenamefont {Rubin},
  \citenamefont {Thonnard},\ and\ \citenamefont {Ford}}]{Rubin:1980zd}%
  \BibitemOpen
  \bibfield  {author} {\bibinfo {author} {\bibfnamefont {V.~C.}\ \bibnamefont
  {Rubin}}, \bibinfo {author} {\bibfnamefont {N.}~\bibnamefont {Thonnard}}, \
  and\ \bibinfo {author} {\bibfnamefont {W.~K.}\ \bibnamefont {Ford},
  \bibfnamefont {Jr.}},\ }\href {\doibase 10.1086/158003} {\bibfield  {journal}
  {\bibinfo  {journal} {Astrophys. J.}\ }\textbf {\bibinfo {volume} {238}},\
  \bibinfo {pages} {471} (\bibinfo {year} {1980})}\BibitemShut {NoStop}%
%%CITATION = ASJOA,238,471;%%
\bibitem [{\citenamefont {Persic}\ \emph {et~al.}(1996)\citenamefont {Persic},
  \citenamefont {Salucci},\ and\ \citenamefont {Stel}}]{Persic:1995ru}%
  \BibitemOpen
  \bibfield  {author} {\bibinfo {author} {\bibfnamefont {M.}~\bibnamefont
  {Persic}}, \bibinfo {author} {\bibfnamefont {P.}~\bibnamefont {Salucci}}, \
  and\ \bibinfo {author} {\bibfnamefont {F.}~\bibnamefont {Stel}},\ }\href
  {\doibase 10.1093/mnras/281.1.27, 10.1093/mnras/278.1.27} {\bibfield
  {journal} {\bibinfo  {journal} {Mon. Not. Roy. Astron. Soc.}\ }\textbf
  {\bibinfo {volume} {281}},\ \bibinfo {pages} {27} (\bibinfo {year} {1996})},\
  \Eprint {http://arxiv.org/abs/astro-ph/9506004} {arXiv:astro-ph/9506004
  [astro-ph]} \BibitemShut {NoStop}%
%%CITATION = ASTRO-PH/9506004;%%
\bibitem [{\citenamefont {Peebles}(1980)}]{Peebles:1980}%
  \BibitemOpen
  \bibfield  {author} {\bibinfo {author} {\bibfnamefont {P.~J.~E.}\
  \bibnamefont {Peebles}},\ }\href@noop {} {\emph {\bibinfo {title}
  {{Large-scale Structure of the Universe}}}}\ (\bibinfo  {publisher}
  {Princeton University Press},\ \bibinfo {year} {1980})\BibitemShut {NoStop}%
\bibitem [{\citenamefont {Padmanabhan}(1993)}]{Padmanabhan:1993}%
  \BibitemOpen
  \bibfield  {author} {\bibinfo {author} {\bibfnamefont {T.}~\bibnamefont
  {Padmanabhan}},\ }\href@noop {} {\emph {\bibinfo {title} {{Structure
  Formation in the Universe}}}}\ (\bibinfo  {publisher} {Cambridge, UK:
  Cambridge University Press},\ \bibinfo {year} {1993})\BibitemShut {NoStop}%
\bibitem [{\citenamefont {Springel}\ \emph {et~al.}(2005)\citenamefont
  {Springel} \emph {et~al.}}]{Springel:2005nw}%
  \BibitemOpen
  \bibfield  {author} {\bibinfo {author} {\bibfnamefont {V.}~\bibnamefont
  {Springel}} \emph {et~al.},\ }\href {\doibase 10.1038/nature03597} {\bibfield
   {journal} {\bibinfo  {journal} {Nature}\ }\textbf {\bibinfo {volume}
  {435}},\ \bibinfo {pages} {629} (\bibinfo {year} {2005})},\ \Eprint
  {http://arxiv.org/abs/astro-ph/0504097} {arXiv:astro-ph/0504097 [astro-ph]}
  \BibitemShut {NoStop}%
%%CITATION = ASTRO-PH/0504097;%%
\bibitem [{\citenamefont {Wambsganss}(1998)}]{Wambsganss:1998gg}%
  \BibitemOpen
  \bibfield  {author} {\bibinfo {author} {\bibfnamefont {J.}~\bibnamefont
  {Wambsganss}},\ }\href {\doibase 10.12942/lrr-1998-12} {\bibfield  {journal}
  {\bibinfo  {journal} {Living Rev. Rel.}\ }\textbf {\bibinfo {volume} {1}},\
  \bibinfo {pages} {12} (\bibinfo {year} {1998})},\ \Eprint
  {http://arxiv.org/abs/astro-ph/9812021} {arXiv:astro-ph/9812021 [astro-ph]}
  \BibitemShut {NoStop}%
%%CITATION = ASTRO-PH/9812021;%%
\bibitem [{\citenamefont {Zwicky}(1933)}]{Zwicky:1933gu}%
  \BibitemOpen
  \bibfield  {author} {\bibinfo {author} {\bibfnamefont {F.}~\bibnamefont
  {Zwicky}},\ }\href {\doibase 10.1007/s10714-008-0707-4} {\bibfield  {journal}
  {\bibinfo  {journal} {Helv. Phys. Acta}\ }\textbf {\bibinfo {volume} {6}},\
  \bibinfo {pages} {110} (\bibinfo {year} {1933})},\ \bibinfo {note} {[Gen.
  Rel. Grav.41,207(2009)]}\BibitemShut {NoStop}%
%%CITATION = HPACA,6,110;%%
\bibitem [{\citenamefont {Zwicky}(1937)}]{Zwicky:1937zza}%
  \BibitemOpen
  \bibfield  {author} {\bibinfo {author} {\bibfnamefont {F.}~\bibnamefont
  {Zwicky}},\ }\href {\doibase 10.1086/143864} {\bibfield  {journal} {\bibinfo
  {journal} {Astrophys. J.}\ }\textbf {\bibinfo {volume} {86}},\ \bibinfo
  {pages} {217} (\bibinfo {year} {1937})}\BibitemShut {NoStop}%
%%CITATION = ASJOA,86,217;%%
\bibitem [{\citenamefont {Boulanger}\ \emph {et~al.}(2001)\citenamefont
  {Boulanger}, \citenamefont {Damour}, \citenamefont {Gualtieri},\ and\
  \citenamefont {Henneaux}}]{Boulanger:2000rq}%
  \BibitemOpen
  \bibfield  {author} {\bibinfo {author} {\bibfnamefont {N.}~\bibnamefont
  {Boulanger}}, \bibinfo {author} {\bibfnamefont {T.}~\bibnamefont {Damour}},
  \bibinfo {author} {\bibfnamefont {L.}~\bibnamefont {Gualtieri}}, \ and\
  \bibinfo {author} {\bibfnamefont {M.}~\bibnamefont {Henneaux}},\ }\href
  {\doibase 10.1016/S0550-3213(00)00718-5} {\bibfield  {journal} {\bibinfo
  {journal} {Nucl. Phys.}\ }\textbf {\bibinfo {volume} {B597}},\ \bibinfo
  {pages} {127} (\bibinfo {year} {2001})},\ \Eprint
  {http://arxiv.org/abs/hep-th/0007220} {arXiv:hep-th/0007220 [hep-th]}
  \BibitemShut {NoStop}%
%%CITATION = HEP-TH/0007220;%%
\bibitem [{\citenamefont {Boulware}\ and\ \citenamefont
  {Deser}(1972)}]{Boulware:1973my}%
  \BibitemOpen
  \bibfield  {author} {\bibinfo {author} {\bibfnamefont {D.~G.}\ \bibnamefont
  {Boulware}}\ and\ \bibinfo {author} {\bibfnamefont {S.}~\bibnamefont
  {Deser}},\ }\href {\doibase 10.1103/PhysRevD.6.3368} {\bibfield  {journal}
  {\bibinfo  {journal} {Phys. Rev.}\ }\textbf {\bibinfo {volume} {D6}},\
  \bibinfo {pages} {3368} (\bibinfo {year} {1972})}\BibitemShut {NoStop}%
%%CITATION = PHRVA,D6,3368;%%
\bibitem [{\citenamefont {de~Rham}\ \emph {et~al.}(2011)\citenamefont
  {de~Rham}, \citenamefont {Gabadadze},\ and\ \citenamefont
  {Tolley}}]{deRham:2010kj}%
  \BibitemOpen
  \bibfield  {author} {\bibinfo {author} {\bibfnamefont {C.}~\bibnamefont
  {de~Rham}}, \bibinfo {author} {\bibfnamefont {G.}~\bibnamefont {Gabadadze}},
  \ and\ \bibinfo {author} {\bibfnamefont {A.~J.}\ \bibnamefont {Tolley}},\
  }\href {\doibase 10.1103/PhysRevLett.106.231101} {\bibfield  {journal}
  {\bibinfo  {journal} {Phys. Rev. Lett.}\ }\textbf {\bibinfo {volume} {106}},\
  \bibinfo {pages} {231101} (\bibinfo {year} {2011})},\ \Eprint
  {http://arxiv.org/abs/1011.1232} {arXiv:1011.1232 [hep-th]} \BibitemShut
  {NoStop}%
%%CITATION = ARXIV:1011.1232;%%
\bibitem [{\citenamefont {Hassan}\ and\ \citenamefont
  {Rosen}(2011)}]{Hassan:2011vm}%
  \BibitemOpen
  \bibfield  {author} {\bibinfo {author} {\bibfnamefont {S.~F.}\ \bibnamefont
  {Hassan}}\ and\ \bibinfo {author} {\bibfnamefont {R.~A.}\ \bibnamefont
  {Rosen}},\ }\href {\doibase 10.1007/JHEP07(2011)009} {\bibfield  {journal}
  {\bibinfo  {journal} {JHEP}\ }\textbf {\bibinfo {volume} {07}},\ \bibinfo
  {pages} {009} (\bibinfo {year} {2011})},\ \Eprint
  {http://arxiv.org/abs/1103.6055} {arXiv:1103.6055 [hep-th]} \BibitemShut
  {NoStop}%
%%CITATION = ARXIV:1103.6055;%%
\bibitem [{\citenamefont {Hassan}\ and\ \citenamefont
  {Rosen}(2012{\natexlab{a}})}]{Hassan:2011hr}%
  \BibitemOpen
  \bibfield  {author} {\bibinfo {author} {\bibfnamefont {S.~F.}\ \bibnamefont
  {Hassan}}\ and\ \bibinfo {author} {\bibfnamefont {R.~A.}\ \bibnamefont
  {Rosen}},\ }\href {\doibase 10.1103/PhysRevLett.108.041101} {\bibfield
  {journal} {\bibinfo  {journal} {Phys. Rev. Lett.}\ }\textbf {\bibinfo
  {volume} {108}},\ \bibinfo {pages} {041101} (\bibinfo {year}
  {2012}{\natexlab{a}})},\ \Eprint {http://arxiv.org/abs/1106.3344}
  {arXiv:1106.3344 [hep-th]} \BibitemShut {NoStop}%
%%CITATION = ARXIV:1106.3344;%%
\bibitem [{\citenamefont {de~Rham}\ \emph {et~al.}(2012)\citenamefont
  {de~Rham}, \citenamefont {Gabadadze},\ and\ \citenamefont
  {Tolley}}]{deRham:2011rn}%
  \BibitemOpen
  \bibfield  {author} {\bibinfo {author} {\bibfnamefont {C.}~\bibnamefont
  {de~Rham}}, \bibinfo {author} {\bibfnamefont {G.}~\bibnamefont {Gabadadze}},
  \ and\ \bibinfo {author} {\bibfnamefont {A.~J.}\ \bibnamefont {Tolley}},\
  }\href {\doibase 10.1016/j.physletb.2012.03.081} {\bibfield  {journal}
  {\bibinfo  {journal} {Phys. Lett.}\ }\textbf {\bibinfo {volume} {B711}},\
  \bibinfo {pages} {190} (\bibinfo {year} {2012})},\ \Eprint
  {http://arxiv.org/abs/1107.3820} {arXiv:1107.3820 [hep-th]} \BibitemShut
  {NoStop}%
%%CITATION = ARXIV:1107.3820;%%
\bibitem [{\citenamefont {Hassan}\ \emph
  {et~al.}(2012{\natexlab{a}})\citenamefont {Hassan}, \citenamefont {Rosen},\
  and\ \citenamefont {Schmidt-May}}]{Hassan:2011tf}%
  \BibitemOpen
  \bibfield  {author} {\bibinfo {author} {\bibfnamefont {S.~F.}\ \bibnamefont
  {Hassan}}, \bibinfo {author} {\bibfnamefont {R.~A.}\ \bibnamefont {Rosen}}, \
  and\ \bibinfo {author} {\bibfnamefont {A.}~\bibnamefont {Schmidt-May}},\
  }\href {\doibase 10.1007/JHEP02(2012)026} {\bibfield  {journal} {\bibinfo
  {journal} {JHEP}\ }\textbf {\bibinfo {volume} {02}},\ \bibinfo {pages} {026}
  (\bibinfo {year} {2012}{\natexlab{a}})},\ \Eprint
  {http://arxiv.org/abs/1109.3230} {arXiv:1109.3230 [hep-th]} \BibitemShut
  {NoStop}%
%%CITATION = ARXIV:1109.3230;%%
\bibitem [{\citenamefont {Hassan}\ and\ \citenamefont
  {Rosen}(2012{\natexlab{b}})}]{Hassan:2011zd}%
  \BibitemOpen
  \bibfield  {author} {\bibinfo {author} {\bibfnamefont {S.~F.}\ \bibnamefont
  {Hassan}}\ and\ \bibinfo {author} {\bibfnamefont {R.~A.}\ \bibnamefont
  {Rosen}},\ }\href {\doibase 10.1007/JHEP02(2012)126} {\bibfield  {journal}
  {\bibinfo  {journal} {JHEP}\ }\textbf {\bibinfo {volume} {02}},\ \bibinfo
  {pages} {126} (\bibinfo {year} {2012}{\natexlab{b}})},\ \Eprint
  {http://arxiv.org/abs/1109.3515} {arXiv:1109.3515 [hep-th]} \BibitemShut
  {NoStop}%
%%CITATION = ARXIV:1109.3515;%%
\bibitem [{\citenamefont {de~Rham}\ \emph
  {et~al.}(2014{\natexlab{a}})\citenamefont {de~Rham}, \citenamefont {Matas},\
  and\ \citenamefont {Tolley}}]{deRham:2013tfa}%
  \BibitemOpen
  \bibfield  {author} {\bibinfo {author} {\bibfnamefont {C.}~\bibnamefont
  {de~Rham}}, \bibinfo {author} {\bibfnamefont {A.}~\bibnamefont {Matas}}, \
  and\ \bibinfo {author} {\bibfnamefont {A.~J.}\ \bibnamefont {Tolley}},\
  }\href {\doibase 10.1088/0264-9381/31/16/165004} {\bibfield  {journal}
  {\bibinfo  {journal} {Class. Quant. Grav.}\ }\textbf {\bibinfo {volume}
  {31}},\ \bibinfo {pages} {165004} (\bibinfo {year} {2014}{\natexlab{a}})},\
  \Eprint {http://arxiv.org/abs/1311.6485} {arXiv:1311.6485 [hep-th]}
  \BibitemShut {NoStop}%
%%CITATION = ARXIV:1311.6485;%%
\bibitem [{\citenamefont {Hinterbichler}(2012)}]{Hinterbichler:2011tt}%
  \BibitemOpen
  \bibfield  {author} {\bibinfo {author} {\bibfnamefont {K.}~\bibnamefont
  {Hinterbichler}},\ }\href {\doibase 10.1103/RevModPhys.84.671} {\bibfield
  {journal} {\bibinfo  {journal} {Rev. Mod. Phys.}\ }\textbf {\bibinfo {volume}
  {84}},\ \bibinfo {pages} {671} (\bibinfo {year} {2012})},\ \Eprint
  {http://arxiv.org/abs/1105.3735} {arXiv:1105.3735 [hep-th]} \BibitemShut
  {NoStop}%
%%CITATION = ARXIV:1105.3735;%%
\bibitem [{\citenamefont {de~Rham}(2014)}]{deRham:2014zqa}%
  \BibitemOpen
  \bibfield  {author} {\bibinfo {author} {\bibfnamefont {C.}~\bibnamefont
  {de~Rham}},\ }\href {\doibase 10.12942/lrr-2014-7} {\bibfield  {journal}
  {\bibinfo  {journal} {Living Rev. Rel.}\ }\textbf {\bibinfo {volume} {17}},\
  \bibinfo {pages} {7} (\bibinfo {year} {2014})},\ \Eprint
  {http://arxiv.org/abs/1401.4173} {arXiv:1401.4173 [hep-th]} \BibitemShut
  {NoStop}%
%%CITATION = ARXIV:1401.4173;%%
\bibitem [{\citenamefont {Schmidt-May}\ and\ \citenamefont {von
  Strauss}(2016)}]{Schmidt-May:2015vnx}%
  \BibitemOpen
  \bibfield  {author} {\bibinfo {author} {\bibfnamefont {A.}~\bibnamefont
  {Schmidt-May}}\ and\ \bibinfo {author} {\bibfnamefont {M.}~\bibnamefont {von
  Strauss}},\ }\href {\doibase 10.1088/1751-8113/49/18/183001} {\bibfield
  {journal} {\bibinfo  {journal} {J. Phys.}\ }\textbf {\bibinfo {volume}
  {A49}},\ \bibinfo {pages} {183001} (\bibinfo {year} {2016})},\ \Eprint
  {http://arxiv.org/abs/1512.00021} {arXiv:1512.00021 [hep-th]} \BibitemShut
  {NoStop}%
%%CITATION = ARXIV:1512.00021;%%
\bibitem [{\citenamefont {Baccetti}\ \emph {et~al.}(2013)\citenamefont
  {Baccetti}, \citenamefont {Martin-Moruno},\ and\ \citenamefont
  {Visser}}]{Baccetti:2012bk}%
  \BibitemOpen
  \bibfield  {author} {\bibinfo {author} {\bibfnamefont {V.}~\bibnamefont
  {Baccetti}}, \bibinfo {author} {\bibfnamefont {P.}~\bibnamefont
  {Martin-Moruno}}, \ and\ \bibinfo {author} {\bibfnamefont {M.}~\bibnamefont
  {Visser}},\ }\href {\doibase 10.1088/0264-9381/30/1/015004} {\bibfield
  {journal} {\bibinfo  {journal} {Class. Quant. Grav.}\ }\textbf {\bibinfo
  {volume} {30}},\ \bibinfo {pages} {015004} (\bibinfo {year} {2013})},\
  \Eprint {http://arxiv.org/abs/1205.2158} {arXiv:1205.2158 [gr-qc]}
  \BibitemShut {NoStop}%
%%CITATION = ARXIV:1205.2158;%%
\bibitem [{\citenamefont {Hassan}\ \emph {et~al.}(2014)\citenamefont {Hassan},
  \citenamefont {Schmidt-May},\ and\ \citenamefont {von
  Strauss}}]{Hassan:2014vja}%
  \BibitemOpen
  \bibfield  {author} {\bibinfo {author} {\bibfnamefont {S.~F.}\ \bibnamefont
  {Hassan}}, \bibinfo {author} {\bibfnamefont {A.}~\bibnamefont {Schmidt-May}},
  \ and\ \bibinfo {author} {\bibfnamefont {M.}~\bibnamefont {von Strauss}},\
  }\href {\doibase 10.1142/S0218271814430020} {\bibfield  {journal} {\bibinfo
  {journal} {Int. J. Mod. Phys.}\ }\textbf {\bibinfo {volume} {D23}},\ \bibinfo
  {pages} {1443002} (\bibinfo {year} {2014})},\ \Eprint
  {http://arxiv.org/abs/1407.2772} {arXiv:1407.2772 [hep-th]} \BibitemShut
  {NoStop}%
%%CITATION = ARXIV:1407.2772;%%
\bibitem [{\citenamefont {Yamashita}\ \emph {et~al.}(2014)\citenamefont
  {Yamashita}, \citenamefont {De~Felice},\ and\ \citenamefont
  {Tanaka}}]{Yamashita:2014fga}%
  \BibitemOpen
  \bibfield  {author} {\bibinfo {author} {\bibfnamefont {Y.}~\bibnamefont
  {Yamashita}}, \bibinfo {author} {\bibfnamefont {A.}~\bibnamefont
  {De~Felice}}, \ and\ \bibinfo {author} {\bibfnamefont {T.}~\bibnamefont
  {Tanaka}},\ }\href {\doibase 10.1142/S0218271814430032} {\bibfield  {journal}
  {\bibinfo  {journal} {Int. J. Mod. Phys.}\ }\textbf {\bibinfo {volume}
  {D23}},\ \bibinfo {pages} {1443003} (\bibinfo {year} {2014})},\ \Eprint
  {http://arxiv.org/abs/1408.0487} {arXiv:1408.0487 [hep-th]} \BibitemShut
  {NoStop}%
%%CITATION = ARXIV:1408.0487;%%
\bibitem [{\citenamefont {de~Rham}\ \emph {et~al.}(2015)\citenamefont
  {de~Rham}, \citenamefont {Heisenberg},\ and\ \citenamefont
  {Ribeiro}}]{deRham:2014naa}%
  \BibitemOpen
  \bibfield  {author} {\bibinfo {author} {\bibfnamefont {C.}~\bibnamefont
  {de~Rham}}, \bibinfo {author} {\bibfnamefont {L.}~\bibnamefont {Heisenberg}},
  \ and\ \bibinfo {author} {\bibfnamefont {R.~H.}\ \bibnamefont {Ribeiro}},\
  }\href {\doibase 10.1088/0264-9381/32/3/035022} {\bibfield  {journal}
  {\bibinfo  {journal} {Class. Quant. Grav.}\ }\textbf {\bibinfo {volume}
  {32}},\ \bibinfo {pages} {035022} (\bibinfo {year} {2015})},\ \Eprint
  {http://arxiv.org/abs/1408.1678} {arXiv:1408.1678 [hep-th]} \BibitemShut
  {NoStop}%
%%CITATION = ARXIV:1408.1678;%%
\bibitem [{\citenamefont {Aoki}\ and\ \citenamefont
  {Maeda}(2014{\natexlab{a}})}]{Aoki:2013joa}%
  \BibitemOpen
  \bibfield  {author} {\bibinfo {author} {\bibfnamefont {K.}~\bibnamefont
  {Aoki}}\ and\ \bibinfo {author} {\bibfnamefont {K.-i.}\ \bibnamefont
  {Maeda}},\ }\href {\doibase 10.1103/PhysRevD.89.064051} {\bibfield  {journal}
  {\bibinfo  {journal} {Phys. Rev.}\ }\textbf {\bibinfo {volume} {D89}},\
  \bibinfo {pages} {064051} (\bibinfo {year} {2014}{\natexlab{a}})},\ \Eprint
  {http://arxiv.org/abs/1312.7040} {arXiv:1312.7040 [gr-qc]} \BibitemShut
  {NoStop}%
%%CITATION = ARXIV:1312.7040;%%
\bibitem [{\citenamefont {Aoki}\ and\ \citenamefont
  {Maeda}(2014{\natexlab{b}})}]{Aoki:2014cla}%
  \BibitemOpen
  \bibfield  {author} {\bibinfo {author} {\bibfnamefont {K.}~\bibnamefont
  {Aoki}}\ and\ \bibinfo {author} {\bibfnamefont {K.-i.}\ \bibnamefont
  {Maeda}},\ }\href {\doibase 10.1103/PhysRevD.90.124089} {\bibfield  {journal}
  {\bibinfo  {journal} {Phys. Rev.}\ }\textbf {\bibinfo {volume} {D90}},\
  \bibinfo {pages} {124089} (\bibinfo {year} {2014}{\natexlab{b}})},\ \Eprint
  {http://arxiv.org/abs/1409.0202} {arXiv:1409.0202 [gr-qc]} \BibitemShut
  {NoStop}%
%%CITATION = ARXIV:1409.0202;%%
\bibitem [{\citenamefont {Bernard}\ and\ \citenamefont
  {Blanchet}(2015)}]{Bernard:2014psa}%
  \BibitemOpen
  \bibfield  {author} {\bibinfo {author} {\bibfnamefont {L.}~\bibnamefont
  {Bernard}}\ and\ \bibinfo {author} {\bibfnamefont {L.}~\bibnamefont
  {Blanchet}},\ }\href {\doibase 10.1103/PhysRevD.91.103536} {\bibfield
  {journal} {\bibinfo  {journal} {Phys. Rev.}\ }\textbf {\bibinfo {volume}
  {D91}},\ \bibinfo {pages} {103536} (\bibinfo {year} {2015})},\ \Eprint
  {http://arxiv.org/abs/1410.7708} {arXiv:1410.7708 [astro-ph.CO]} \BibitemShut
  {NoStop}%
%%CITATION = ARXIV:1410.7708;%%
\bibitem [{\citenamefont {Blanchet}\ and\ \citenamefont
  {Heisenberg}(2015{\natexlab{a}})}]{Blanchet:2015sra}%
  \BibitemOpen
  \bibfield  {author} {\bibinfo {author} {\bibfnamefont {L.}~\bibnamefont
  {Blanchet}}\ and\ \bibinfo {author} {\bibfnamefont {L.}~\bibnamefont
  {Heisenberg}},\ }\href {\doibase 10.1103/PhysRevD.91.103518} {\bibfield
  {journal} {\bibinfo  {journal} {Phys. Rev.}\ }\textbf {\bibinfo {volume}
  {D91}},\ \bibinfo {pages} {103518} (\bibinfo {year} {2015}{\natexlab{a}})},\
  \Eprint {http://arxiv.org/abs/1504.00870} {arXiv:1504.00870 [gr-qc]}
  \BibitemShut {NoStop}%
%%CITATION = ARXIV:1504.00870;%%
\bibitem [{\citenamefont {Blanchet}\ and\ \citenamefont
  {Heisenberg}(2015{\natexlab{b}})}]{Blanchet:2015bia}%
  \BibitemOpen
  \bibfield  {author} {\bibinfo {author} {\bibfnamefont {L.}~\bibnamefont
  {Blanchet}}\ and\ \bibinfo {author} {\bibfnamefont {L.}~\bibnamefont
  {Heisenberg}},\ }\href {\doibase 10.1088/1475-7516/2015/12/026} {\bibfield
  {journal} {\bibinfo  {journal} {JCAP}\ }\textbf {\bibinfo {volume} {1512}},\
  \bibinfo {pages} {026} (\bibinfo {year} {2015}{\natexlab{b}})},\ \Eprint
  {http://arxiv.org/abs/1505.05146} {arXiv:1505.05146 [hep-th]} \BibitemShut
  {NoStop}%
%%CITATION = ARXIV:1505.05146;%%
\bibitem [{\citenamefont {Blanchet}\ and\ \citenamefont
  {Heisenberg}(2017)}]{Blanchet:2017duj}%
  \BibitemOpen
  \bibfield  {author} {\bibinfo {author} {\bibfnamefont {L.}~\bibnamefont
  {Blanchet}}\ and\ \bibinfo {author} {\bibfnamefont {L.}~\bibnamefont
  {Heisenberg}},\ }\href@noop {} {\  (\bibinfo {year} {2017})},\ \Eprint
  {http://arxiv.org/abs/1701.07747} {arXiv:1701.07747 [gr-qc]} \BibitemShut
  {NoStop}%
%%CITATION = ARXIV:1701.07747;%%
\bibitem [{\citenamefont {Schmidt-May}(2016)}]{Schmidt-May:2016hsx}%
  \BibitemOpen
  \bibfield  {author} {\bibinfo {author} {\bibfnamefont {A.}~\bibnamefont
  {Schmidt-May}},\ }\bibfield  {booktitle} {\emph {\bibinfo {booktitle}
  {{Proceedings, 15th Hellenic School and Workshops on Elementary Particle
  Physics and Gravity (CORFU2015): Corfu, Greece, September 1-25, 2015}}},\
  }\href@noop {} {\bibfield  {journal} {\bibinfo  {journal} {PoS}\ }\textbf
  {\bibinfo {volume} {CORFU2015}},\ \bibinfo {pages} {157} (\bibinfo {year}
  {2016})},\ \Eprint {http://arxiv.org/abs/1602.07520} {arXiv:1602.07520
  [gr-qc]} \BibitemShut {NoStop}%
%%CITATION = ARXIV:1602.07520;%%
\bibitem [{\citenamefont {Babichev}\ \emph
  {et~al.}(2016{\natexlab{a}})\citenamefont {Babichev}, \citenamefont
  {Marzola}, \citenamefont {Raidal}, \citenamefont {Schmidt-May}, \citenamefont
  {Urban}, \citenamefont {Veermäe},\ and\ \citenamefont {von
  Strauss}}]{Babichev:2016hir}%
  \BibitemOpen
  \bibfield  {author} {\bibinfo {author} {\bibfnamefont {E.}~\bibnamefont
  {Babichev}}, \bibinfo {author} {\bibfnamefont {L.}~\bibnamefont {Marzola}},
  \bibinfo {author} {\bibfnamefont {M.}~\bibnamefont {Raidal}}, \bibinfo
  {author} {\bibfnamefont {A.}~\bibnamefont {Schmidt-May}}, \bibinfo {author}
  {\bibfnamefont {F.}~\bibnamefont {Urban}}, \bibinfo {author} {\bibfnamefont
  {H.}~\bibnamefont {Veermäe}}, \ and\ \bibinfo {author} {\bibfnamefont
  {M.}~\bibnamefont {von Strauss}},\ }\href {\doibase
  10.1103/PhysRevD.94.084055} {\bibfield  {journal} {\bibinfo  {journal} {Phys.
  Rev.}\ }\textbf {\bibinfo {volume} {D94}},\ \bibinfo {pages} {084055}
  (\bibinfo {year} {2016}{\natexlab{a}})},\ \Eprint
  {http://arxiv.org/abs/1604.08564} {arXiv:1604.08564 [hep-ph]} \BibitemShut
  {NoStop}%
%%CITATION = ARXIV:1604.08564;%%
\bibitem [{\citenamefont {Aoki}\ and\ \citenamefont
  {Mukohyama}(2016)}]{Aoki:2016zgp}%
  \BibitemOpen
  \bibfield  {author} {\bibinfo {author} {\bibfnamefont {K.}~\bibnamefont
  {Aoki}}\ and\ \bibinfo {author} {\bibfnamefont {S.}~\bibnamefont
  {Mukohyama}},\ }\href {\doibase 10.1103/PhysRevD.94.024001} {\bibfield
  {journal} {\bibinfo  {journal} {Phys. Rev.}\ }\textbf {\bibinfo {volume}
  {D94}},\ \bibinfo {pages} {024001} (\bibinfo {year} {2016})},\ \Eprint
  {http://arxiv.org/abs/1604.06704} {arXiv:1604.06704 [hep-th]} \BibitemShut
  {NoStop}%
%%CITATION = ARXIV:1604.06704;%%
\bibitem [{\citenamefont {Babichev}\ \emph
  {et~al.}(2016{\natexlab{b}})\citenamefont {Babichev}, \citenamefont
  {Marzola}, \citenamefont {Raidal}, \citenamefont {Schmidt-May}, \citenamefont
  {Urban}, \citenamefont {Veermäe},\ and\ \citenamefont {von
  Strauss}}]{Babichev:2016bxi}%
  \BibitemOpen
  \bibfield  {author} {\bibinfo {author} {\bibfnamefont {E.}~\bibnamefont
  {Babichev}}, \bibinfo {author} {\bibfnamefont {L.}~\bibnamefont {Marzola}},
  \bibinfo {author} {\bibfnamefont {M.}~\bibnamefont {Raidal}}, \bibinfo
  {author} {\bibfnamefont {A.}~\bibnamefont {Schmidt-May}}, \bibinfo {author}
  {\bibfnamefont {F.}~\bibnamefont {Urban}}, \bibinfo {author} {\bibfnamefont
  {H.}~\bibnamefont {Veermäe}}, \ and\ \bibinfo {author} {\bibfnamefont
  {M.}~\bibnamefont {von Strauss}},\ }\href {\doibase
  10.1088/1475-7516/2016/09/016} {\bibfield  {journal} {\bibinfo  {journal}
  {JCAP}\ }\textbf {\bibinfo {volume} {1609}},\ \bibinfo {pages} {016}
  (\bibinfo {year} {2016}{\natexlab{b}})},\ \Eprint
  {http://arxiv.org/abs/1607.03497} {arXiv:1607.03497 [hep-th]} \BibitemShut
  {NoStop}%
%%CITATION = ARXIV:1607.03497;%%
\bibitem [{\citenamefont {Nordtvedt}(1968)}]{Nordtvedt:1968qs}%
  \BibitemOpen
  \bibfield  {author} {\bibinfo {author} {\bibfnamefont {K.}~\bibnamefont
  {Nordtvedt}},\ }\href {\doibase 10.1103/PhysRev.169.1017} {\bibfield
  {journal} {\bibinfo  {journal} {Phys. Rev.}\ }\textbf {\bibinfo {volume}
  {169}},\ \bibinfo {pages} {1017} (\bibinfo {year} {1968})}\BibitemShut
  {NoStop}%
%%CITATION = PHRVA,169,1017;%%
\bibitem [{\citenamefont {Thorne}\ and\ \citenamefont
  {Will}(1971)}]{Thorne:1970wv}%
  \BibitemOpen
  \bibfield  {author} {\bibinfo {author} {\bibfnamefont {K.~S.}\ \bibnamefont
  {Thorne}}\ and\ \bibinfo {author} {\bibfnamefont {C.~M.}\ \bibnamefont
  {Will}},\ }\href {\doibase 10.1086/150803} {\bibfield  {journal} {\bibinfo
  {journal} {Astrophys. J.}\ }\textbf {\bibinfo {volume} {163}},\ \bibinfo
  {pages} {595} (\bibinfo {year} {1971})}\BibitemShut {NoStop}%
%%CITATION = ASJOA,163,595;%%
\bibitem [{\citenamefont {Will}(1971{\natexlab{a}})}]{Will:1971zzb}%
  \BibitemOpen
  \bibfield  {author} {\bibinfo {author} {\bibfnamefont {C.~M.}\ \bibnamefont
  {Will}},\ }\href {\doibase 10.1086/150804} {\bibfield  {journal} {\bibinfo
  {journal} {Astrophys. J.}\ }\textbf {\bibinfo {volume} {163}},\ \bibinfo
  {pages} {611} (\bibinfo {year} {1971}{\natexlab{a}})}\BibitemShut {NoStop}%
%%CITATION = ASJOA,163,611;%%
\bibitem [{\citenamefont {Will}(1971{\natexlab{b}})}]{Will:1971wt}%
  \BibitemOpen
  \bibfield  {author} {\bibinfo {author} {\bibfnamefont {C.~M.}\ \bibnamefont
  {Will}},\ }\href {\doibase 10.1086/151124} {\bibfield  {journal} {\bibinfo
  {journal} {Astrophys. J.}\ }\textbf {\bibinfo {volume} {169}},\ \bibinfo
  {pages} {125} (\bibinfo {year} {1971}{\natexlab{b}})}\BibitemShut {NoStop}%
%%CITATION = ASJOA,169,125;%%
\bibitem [{\citenamefont {Will}(1993)}]{Will:1993ns}%
  \BibitemOpen
  \bibfield  {author} {\bibinfo {author} {\bibfnamefont {C.~M.}\ \bibnamefont
  {Will}},\ }\href@noop {} {\emph {\bibinfo {title} {{Theory and experiment in
  gravitational physics}}}}\ (\bibinfo  {publisher} {Cambridge, UK: Cambridge
  University Press},\ \bibinfo {year} {1993})\BibitemShut {NoStop}%
%%CITATION = ISBN-9780521439732;%%
\bibitem [{\citenamefont {Will}(2014)}]{Will:2014kxa}%
  \BibitemOpen
  \bibfield  {author} {\bibinfo {author} {\bibfnamefont {C.~M.}\ \bibnamefont
  {Will}},\ }\href {\doibase 10.12942/lrr-2014-4} {\bibfield  {journal}
  {\bibinfo  {journal} {Living Rev. Rel.}\ }\textbf {\bibinfo {volume} {17}},\
  \bibinfo {pages} {4} (\bibinfo {year} {2014})},\ \Eprint
  {http://arxiv.org/abs/1403.7377} {arXiv:1403.7377 [gr-qc]} \BibitemShut
  {NoStop}%
%%CITATION = ARXIV:1403.7377;%%
\bibitem [{\citenamefont {Shapiro}\ \emph {et~al.}(2004)\citenamefont
  {Shapiro}, \citenamefont {Davis}, \citenamefont {Lebach},\ and\ \citenamefont
  {Gregory}}]{Shapiro:2004zz}%
  \BibitemOpen
  \bibfield  {author} {\bibinfo {author} {\bibfnamefont {S.~S.}\ \bibnamefont
  {Shapiro}}, \bibinfo {author} {\bibfnamefont {J.~L.}\ \bibnamefont {Davis}},
  \bibinfo {author} {\bibfnamefont {D.~E.}\ \bibnamefont {Lebach}}, \ and\
  \bibinfo {author} {\bibfnamefont {J.~S.}\ \bibnamefont {Gregory}},\ }\href
  {\doibase 10.1103/PhysRevLett.92.121101} {\bibfield  {journal} {\bibinfo
  {journal} {Phys. Rev. Lett.}\ }\textbf {\bibinfo {volume} {92}},\ \bibinfo
  {pages} {121101} (\bibinfo {year} {2004})}\BibitemShut {NoStop}%
%%CITATION = PRLTA,92,121101;%%
\bibitem [{\citenamefont {Fomalont}\ \emph {et~al.}(2009)\citenamefont
  {Fomalont}, \citenamefont {Kopeikin}, \citenamefont {Lanyi},\ and\
  \citenamefont {Benson}}]{Fomalont:2009zg}%
  \BibitemOpen
  \bibfield  {author} {\bibinfo {author} {\bibfnamefont {E.}~\bibnamefont
  {Fomalont}}, \bibinfo {author} {\bibfnamefont {S.}~\bibnamefont {Kopeikin}},
  \bibinfo {author} {\bibfnamefont {G.}~\bibnamefont {Lanyi}}, \ and\ \bibinfo
  {author} {\bibfnamefont {J.}~\bibnamefont {Benson}},\ }\href {\doibase
  10.1088/0004-637X/699/2/1395} {\bibfield  {journal} {\bibinfo  {journal}
  {Astrophys. J.}\ }\textbf {\bibinfo {volume} {699}},\ \bibinfo {pages} {1395}
  (\bibinfo {year} {2009})},\ \Eprint {http://arxiv.org/abs/0904.3992}
  {arXiv:0904.3992 [astro-ph.CO]} \BibitemShut {NoStop}%
%%CITATION = ARXIV:0904.3992;%%
\bibitem [{\citenamefont {Lambert}\ and\ \citenamefont
  {Le~Poncin-Lafitte}(2009)}]{Lambert:2009xy}%
  \BibitemOpen
  \bibfield  {author} {\bibinfo {author} {\bibfnamefont {S.~B.}\ \bibnamefont
  {Lambert}}\ and\ \bibinfo {author} {\bibfnamefont {C.}~\bibnamefont
  {Le~Poncin-Lafitte}},\ }\href {\doibase 10.1051/0004-6361/200911714}
  {\bibfield  {journal} {\bibinfo  {journal} {Astron. Astrophys.}\ }\textbf
  {\bibinfo {volume} {499}},\ \bibinfo {pages} {331} (\bibinfo {year}
  {2009})},\ \Eprint {http://arxiv.org/abs/0903.1615} {arXiv:0903.1615 [gr-qc]}
  \BibitemShut {NoStop}%
%%CITATION = ARXIV:0903.1615;%%
\bibitem [{\citenamefont {Lambert}\ and\ \citenamefont
  {Le~Poncin-Lafitte}(2011)}]{Lambert:2011}%
  \BibitemOpen
  \bibfield  {author} {\bibinfo {author} {\bibfnamefont {S.~B.}\ \bibnamefont
  {Lambert}}\ and\ \bibinfo {author} {\bibfnamefont {C.}~\bibnamefont
  {Le~Poncin-Lafitte}},\ }\href {\doibase 10.1051/0004-6361/201016370}
  {\bibfield  {journal} {\bibinfo  {journal} {Astron. Astrophys.}\ }\textbf
  {\bibinfo {volume} {529}},\ \bibinfo {eid} {A70} (\bibinfo {year}
  {2011})}\BibitemShut {NoStop}%
\bibitem [{\citenamefont {Bertotti}\ \emph {et~al.}(2003)\citenamefont
  {Bertotti}, \citenamefont {Iess},\ and\ \citenamefont
  {Tortora}}]{Bertotti:2003rm}%
  \BibitemOpen
  \bibfield  {author} {\bibinfo {author} {\bibfnamefont {B.}~\bibnamefont
  {Bertotti}}, \bibinfo {author} {\bibfnamefont {L.}~\bibnamefont {Iess}}, \
  and\ \bibinfo {author} {\bibfnamefont {P.}~\bibnamefont {Tortora}},\ }\href
  {\doibase 10.1038/nature01997} {\bibfield  {journal} {\bibinfo  {journal}
  {Nature}\ }\textbf {\bibinfo {volume} {425}},\ \bibinfo {pages} {374}
  (\bibinfo {year} {2003})}\BibitemShut {NoStop}%
%%CITATION = NATUA,425,374;%%
\bibitem [{\citenamefont {Fienga}\ \emph {et~al.}(2011)\citenamefont {Fienga},
  \citenamefont {Laskar}, \citenamefont {Kuchynka}, \citenamefont {Manche},
  \citenamefont {Desvignes}, \citenamefont {Gastineau}, \citenamefont
  {Cognard},\ and\ \citenamefont {Theureau}}]{Fienga:2011qh}%
  \BibitemOpen
  \bibfield  {author} {\bibinfo {author} {\bibfnamefont {A.}~\bibnamefont
  {Fienga}}, \bibinfo {author} {\bibfnamefont {J.}~\bibnamefont {Laskar}},
  \bibinfo {author} {\bibfnamefont {P.}~\bibnamefont {Kuchynka}}, \bibinfo
  {author} {\bibfnamefont {H.}~\bibnamefont {Manche}}, \bibinfo {author}
  {\bibfnamefont {G.}~\bibnamefont {Desvignes}}, \bibinfo {author}
  {\bibfnamefont {M.}~\bibnamefont {Gastineau}}, \bibinfo {author}
  {\bibfnamefont {I.}~\bibnamefont {Cognard}}, \ and\ \bibinfo {author}
  {\bibfnamefont {G.}~\bibnamefont {Theureau}},\ }\href {\doibase
  10.1007/s10569-011-9377-8} {\bibfield  {journal} {\bibinfo  {journal}
  {Celest. Mech. Dyn. Astron.}\ }\textbf {\bibinfo {volume} {111}},\ \bibinfo
  {pages} {363} (\bibinfo {year} {2011})},\ \Eprint
  {http://arxiv.org/abs/1108.5546} {arXiv:1108.5546 [astro-ph.EP]} \BibitemShut
  {NoStop}%
%%CITATION = ARXIV:1108.5546;%%
\bibitem [{\citenamefont {Pitjeva}\ and\ \citenamefont
  {Pitjev}(2013)}]{Pitjeva:2013chs}%
  \BibitemOpen
  \bibfield  {author} {\bibinfo {author} {\bibfnamefont {E.~V.}\ \bibnamefont
  {Pitjeva}}\ and\ \bibinfo {author} {\bibfnamefont {N.~P.}\ \bibnamefont
  {Pitjev}},\ }\href {\doibase 10.1093/mnras/stt695} {\bibfield  {journal}
  {\bibinfo  {journal} {Mon. Not. Roy. Astron. Soc.}\ }\textbf {\bibinfo
  {volume} {432}},\ \bibinfo {pages} {3431} (\bibinfo {year} {2013})},\ \Eprint
  {http://arxiv.org/abs/1306.3043} {arXiv:1306.3043 [astro-ph.EP]} \BibitemShut
  {NoStop}%
%%CITATION = ARXIV:1306.3043;%%
\bibitem [{\citenamefont {Verma}\ \emph {et~al.}(2014)\citenamefont {Verma},
  \citenamefont {Fienga}, \citenamefont {Laskar}, \citenamefont {Manche},\ and\
  \citenamefont {Gastineau}}]{Verma:2013ata}%
  \BibitemOpen
  \bibfield  {author} {\bibinfo {author} {\bibfnamefont {A.}~\bibnamefont
  {Verma}}, \bibinfo {author} {\bibfnamefont {A.}~\bibnamefont {Fienga}},
  \bibinfo {author} {\bibfnamefont {J.}~\bibnamefont {Laskar}}, \bibinfo
  {author} {\bibfnamefont {H.}~\bibnamefont {Manche}}, \ and\ \bibinfo {author}
  {\bibfnamefont {M.}~\bibnamefont {Gastineau}},\ }\href {\doibase
  10.1051/0004-6361/201322124} {\bibfield  {journal} {\bibinfo  {journal}
  {Astron. Astrophys.}\ }\textbf {\bibinfo {volume} {561}},\ \bibinfo {pages}
  {A115} (\bibinfo {year} {2014})},\ \Eprint {http://arxiv.org/abs/1306.5569}
  {arXiv:1306.5569 [astro-ph.EP]} \BibitemShut {NoStop}%
%%CITATION = ARXIV:1306.5569;%%
\bibitem [{\citenamefont {{Fienga}}\ \emph {et~al.}(2014)\citenamefont
  {{Fienga}}, \citenamefont {{Manche}}, \citenamefont {{Laskar}}, \citenamefont
  {{Gastineau}},\ and\ \citenamefont {{Verma}}}]{Fienga:2014}%
  \BibitemOpen
  \bibfield  {author} {\bibinfo {author} {\bibfnamefont {A.}~\bibnamefont
  {{Fienga}}}, \bibinfo {author} {\bibfnamefont {H.}~\bibnamefont {{Manche}}},
  \bibinfo {author} {\bibfnamefont {J.}~\bibnamefont {{Laskar}}}, \bibinfo
  {author} {\bibfnamefont {M.}~\bibnamefont {{Gastineau}}}, \ and\ \bibinfo
  {author} {\bibfnamefont {A.}~\bibnamefont {{Verma}}},\ }\href@noop {} {\
  (\bibinfo {year} {2014})},\ \Eprint {http://arxiv.org/abs/1405.0484}
  {arXiv:1405.0484 [astro-ph.EP]} \BibitemShut {NoStop}%
\bibitem [{\citenamefont {Bolton}\ \emph {et~al.}(2006)\citenamefont {Bolton},
  \citenamefont {Rappaport},\ and\ \citenamefont {Burles}}]{Bolton:2006yz}%
  \BibitemOpen
  \bibfield  {author} {\bibinfo {author} {\bibfnamefont {A.~S.}\ \bibnamefont
  {Bolton}}, \bibinfo {author} {\bibfnamefont {S.}~\bibnamefont {Rappaport}}, \
  and\ \bibinfo {author} {\bibfnamefont {S.}~\bibnamefont {Burles}},\ }\href
  {\doibase 10.1103/PhysRevD.74.061501} {\bibfield  {journal} {\bibinfo
  {journal} {Phys. Rev.}\ }\textbf {\bibinfo {volume} {D74}},\ \bibinfo {pages}
  {061501} (\bibinfo {year} {2006})},\ \Eprint
  {http://arxiv.org/abs/astro-ph/0607657} {arXiv:astro-ph/0607657 [astro-ph]}
  \BibitemShut {NoStop}%
%%CITATION = ASTRO-PH/0607657;%%
\bibitem [{\citenamefont {Schwab}\ \emph {et~al.}(2010)\citenamefont {Schwab},
  \citenamefont {Bolton},\ and\ \citenamefont {Rappaport}}]{Schwab:2009nz}%
  \BibitemOpen
  \bibfield  {author} {\bibinfo {author} {\bibfnamefont {J.}~\bibnamefont
  {Schwab}}, \bibinfo {author} {\bibfnamefont {A.~S.}\ \bibnamefont {Bolton}},
  \ and\ \bibinfo {author} {\bibfnamefont {S.~A.}\ \bibnamefont {Rappaport}},\
  }\href {\doibase 10.1088/0004-637X/708/1/750} {\bibfield  {journal} {\bibinfo
   {journal} {Astrophys. J.}\ }\textbf {\bibinfo {volume} {708}},\ \bibinfo
  {pages} {750} (\bibinfo {year} {2010})},\ \Eprint
  {http://arxiv.org/abs/0907.4992} {arXiv:0907.4992 [astro-ph.CO]} \BibitemShut
  {NoStop}%
%%CITATION = ARXIV:0907.4992;%%
\bibitem [{\citenamefont {Cao}\ \emph {et~al.}(2017)\citenamefont {Cao},
  \citenamefont {Li}, \citenamefont {Biesiada}, \citenamefont {Xu},
  \citenamefont {Cai},\ and\ \citenamefont {Zhu}}]{Cao:2017nnq}%
  \BibitemOpen
  \bibfield  {author} {\bibinfo {author} {\bibfnamefont {S.}~\bibnamefont
  {Cao}}, \bibinfo {author} {\bibfnamefont {X.}~\bibnamefont {Li}}, \bibinfo
  {author} {\bibfnamefont {M.}~\bibnamefont {Biesiada}}, \bibinfo {author}
  {\bibfnamefont {T.}~\bibnamefont {Xu}}, \bibinfo {author} {\bibfnamefont
  {Y.}~\bibnamefont {Cai}}, \ and\ \bibinfo {author} {\bibfnamefont {Z.-H.}\
  \bibnamefont {Zhu}},\ }\href@noop {} {\  (\bibinfo {year} {2017})},\ \Eprint
  {http://arxiv.org/abs/1701.00357} {arXiv:1701.00357 [astro-ph.CO]}
  \BibitemShut {NoStop}%
%%CITATION = ARXIV:1701.00357;%%
\bibitem [{\citenamefont {Tucker}\ \emph {et~al.}(1998)\citenamefont {Tucker},
  \citenamefont {Blanco}, \citenamefont {Rappoport}, \citenamefont {David},
  \citenamefont {Fabricant}, \citenamefont {Falco}, \citenamefont {Forman},
  \citenamefont {Dressler},\ and\ \citenamefont {Ramella}}]{Tucker:1998tp}%
  \BibitemOpen
  \bibfield  {author} {\bibinfo {author} {\bibfnamefont {W.}~\bibnamefont
  {Tucker}}, \bibinfo {author} {\bibfnamefont {P.}~\bibnamefont {Blanco}},
  \bibinfo {author} {\bibfnamefont {S.}~\bibnamefont {Rappoport}}, \bibinfo
  {author} {\bibfnamefont {L.}~\bibnamefont {David}}, \bibinfo {author}
  {\bibfnamefont {D.}~\bibnamefont {Fabricant}}, \bibinfo {author}
  {\bibfnamefont {E.~E.}\ \bibnamefont {Falco}}, \bibinfo {author}
  {\bibfnamefont {W.}~\bibnamefont {Forman}}, \bibinfo {author} {\bibfnamefont
  {A.}~\bibnamefont {Dressler}}, \ and\ \bibinfo {author} {\bibfnamefont
  {M.}~\bibnamefont {Ramella}},\ }\href {\doibase 10.1086/311234} {\bibfield
  {journal} {\bibinfo  {journal} {Astrophys. J.}\ }\textbf {\bibinfo {volume}
  {496}},\ \bibinfo {pages} {L5} (\bibinfo {year} {1998})},\ \Eprint
  {http://arxiv.org/abs/astro-ph/9801120} {arXiv:astro-ph/9801120 [astro-ph]}
  \BibitemShut {NoStop}%
%%CITATION = ASTRO-PH/9801120;%%
\bibitem [{\citenamefont {Markevitch}\ \emph {et~al.}(2004)\citenamefont
  {Markevitch}, \citenamefont {Gonzalez}, \citenamefont {Clowe}, \citenamefont
  {Vikhlinin}, \citenamefont {David}, \citenamefont {Forman}, \citenamefont
  {Jones}, \citenamefont {Murray},\ and\ \citenamefont
  {Tucker}}]{Markevitch:2003at}%
  \BibitemOpen
  \bibfield  {author} {\bibinfo {author} {\bibfnamefont {M.}~\bibnamefont
  {Markevitch}}, \bibinfo {author} {\bibfnamefont {A.~H.}\ \bibnamefont
  {Gonzalez}}, \bibinfo {author} {\bibfnamefont {D.}~\bibnamefont {Clowe}},
  \bibinfo {author} {\bibfnamefont {A.}~\bibnamefont {Vikhlinin}}, \bibinfo
  {author} {\bibfnamefont {L.}~\bibnamefont {David}}, \bibinfo {author}
  {\bibfnamefont {W.}~\bibnamefont {Forman}}, \bibinfo {author} {\bibfnamefont
  {C.}~\bibnamefont {Jones}}, \bibinfo {author} {\bibfnamefont
  {S.}~\bibnamefont {Murray}}, \ and\ \bibinfo {author} {\bibfnamefont
  {W.}~\bibnamefont {Tucker}},\ }\href {\doibase 10.1086/383178} {\bibfield
  {journal} {\bibinfo  {journal} {Astrophys. J.}\ }\textbf {\bibinfo {volume}
  {606}},\ \bibinfo {pages} {819} (\bibinfo {year} {2004})},\ \Eprint
  {http://arxiv.org/abs/astro-ph/0309303} {arXiv:astro-ph/0309303 [astro-ph]}
  \BibitemShut {NoStop}%
%%CITATION = ASTRO-PH/0309303;%%
\bibitem [{\citenamefont {Clowe}\ \emph {et~al.}(2004)\citenamefont {Clowe},
  \citenamefont {Gonzalez},\ and\ \citenamefont {Markevitch}}]{Clowe:2003tk}%
  \BibitemOpen
  \bibfield  {author} {\bibinfo {author} {\bibfnamefont {D.}~\bibnamefont
  {Clowe}}, \bibinfo {author} {\bibfnamefont {A.}~\bibnamefont {Gonzalez}}, \
  and\ \bibinfo {author} {\bibfnamefont {M.}~\bibnamefont {Markevitch}},\
  }\href {\doibase 10.1086/381970} {\bibfield  {journal} {\bibinfo  {journal}
  {Astrophys. J.}\ }\textbf {\bibinfo {volume} {604}},\ \bibinfo {pages} {596}
  (\bibinfo {year} {2004})},\ \Eprint {http://arxiv.org/abs/astro-ph/0312273}
  {arXiv:astro-ph/0312273 [astro-ph]} \BibitemShut {NoStop}%
%%CITATION = ASTRO-PH/0312273;%%
\bibitem [{\citenamefont {Clowe}\ \emph {et~al.}(2006)\citenamefont {Clowe},
  \citenamefont {Bradac}, \citenamefont {Gonzalez}, \citenamefont {Markevitch},
  \citenamefont {Randall}, \citenamefont {Jones},\ and\ \citenamefont
  {Zaritsky}}]{Clowe:2006eq}%
  \BibitemOpen
  \bibfield  {author} {\bibinfo {author} {\bibfnamefont {D.}~\bibnamefont
  {Clowe}}, \bibinfo {author} {\bibfnamefont {M.}~\bibnamefont {Bradac}},
  \bibinfo {author} {\bibfnamefont {A.~H.}\ \bibnamefont {Gonzalez}}, \bibinfo
  {author} {\bibfnamefont {M.}~\bibnamefont {Markevitch}}, \bibinfo {author}
  {\bibfnamefont {S.~W.}\ \bibnamefont {Randall}}, \bibinfo {author}
  {\bibfnamefont {C.}~\bibnamefont {Jones}}, \ and\ \bibinfo {author}
  {\bibfnamefont {D.}~\bibnamefont {Zaritsky}},\ }\href {\doibase
  10.1086/508162} {\bibfield  {journal} {\bibinfo  {journal} {Astrophys. J.}\
  }\textbf {\bibinfo {volume} {648}},\ \bibinfo {pages} {L109} (\bibinfo {year}
  {2006})},\ \Eprint {http://arxiv.org/abs/astro-ph/0608407}
  {arXiv:astro-ph/0608407 [astro-ph]} \BibitemShut {NoStop}%
%%CITATION = ASTRO-PH/0608407;%%
\bibitem [{\citenamefont {Randall}\ \emph {et~al.}(2008)\citenamefont
  {Randall}, \citenamefont {Markevitch}, \citenamefont {Clowe}, \citenamefont
  {Gonzalez},\ and\ \citenamefont {Bradac}}]{Randall:2007ph}%
  \BibitemOpen
  \bibfield  {author} {\bibinfo {author} {\bibfnamefont {S.~W.}\ \bibnamefont
  {Randall}}, \bibinfo {author} {\bibfnamefont {M.}~\bibnamefont {Markevitch}},
  \bibinfo {author} {\bibfnamefont {D.}~\bibnamefont {Clowe}}, \bibinfo
  {author} {\bibfnamefont {A.~H.}\ \bibnamefont {Gonzalez}}, \ and\ \bibinfo
  {author} {\bibfnamefont {M.}~\bibnamefont {Bradac}},\ }\href {\doibase
  10.1086/587859} {\bibfield  {journal} {\bibinfo  {journal} {Astrophys. J.}\
  }\textbf {\bibinfo {volume} {679}},\ \bibinfo {pages} {1173} (\bibinfo {year}
  {2008})},\ \Eprint {http://arxiv.org/abs/0704.0261} {arXiv:0704.0261
  [astro-ph]} \BibitemShut {NoStop}%
%%CITATION = ARXIV:0704.0261;%%
\bibitem [{\citenamefont {Mahdavi}\ \emph {et~al.}(2007)\citenamefont
  {Mahdavi}, \citenamefont {Hoekstra}, \citenamefont {Babul}, \citenamefont
  {Balam},\ and\ \citenamefont {Capak}}]{Mahdavi:2007yp}%
  \BibitemOpen
  \bibfield  {author} {\bibinfo {author} {\bibfnamefont {A.}~\bibnamefont
  {Mahdavi}}, \bibinfo {author} {\bibfnamefont {H.~y.}\ \bibnamefont
  {Hoekstra}}, \bibinfo {author} {\bibfnamefont {A.~y.}\ \bibnamefont {Babul}},
  \bibinfo {author} {\bibfnamefont {D.~y.}\ \bibnamefont {Balam}}, \ and\
  \bibinfo {author} {\bibfnamefont {P.}~\bibnamefont {Capak}},\ }\href
  {\doibase 10.1086/521383} {\bibfield  {journal} {\bibinfo  {journal}
  {Astrophys. J.}\ }\textbf {\bibinfo {volume} {668}},\ \bibinfo {pages} {806}
  (\bibinfo {year} {2007})},\ \Eprint {http://arxiv.org/abs/0706.3048}
  {arXiv:0706.3048 [astro-ph]} \BibitemShut {NoStop}%
%%CITATION = ARXIV:0706.3048;%%
\bibitem [{\citenamefont {Jee}\ \emph {et~al.}(2014)\citenamefont {Jee},
  \citenamefont {Hoekstra}, \citenamefont {Mahdavi},\ and\ \citenamefont
  {Babul}}]{Jee:2014hja}%
  \BibitemOpen
  \bibfield  {author} {\bibinfo {author} {\bibfnamefont {M.~J.}\ \bibnamefont
  {Jee}}, \bibinfo {author} {\bibfnamefont {H.}~\bibnamefont {Hoekstra}},
  \bibinfo {author} {\bibfnamefont {A.}~\bibnamefont {Mahdavi}}, \ and\
  \bibinfo {author} {\bibfnamefont {A.}~\bibnamefont {Babul}},\ }\href
  {\doibase 10.1088/0004-637X/783/2/78} {\bibfield  {journal} {\bibinfo
  {journal} {Astrophys. J.}\ }\textbf {\bibinfo {volume} {783}},\ \bibinfo
  {pages} {78} (\bibinfo {year} {2014})},\ \Eprint
  {http://arxiv.org/abs/1401.3356} {arXiv:1401.3356 [astro-ph.CO]} \BibitemShut
  {NoStop}%
%%CITATION = ARXIV:1401.3356;%%
\bibitem [{\citenamefont {Bradač}\ \emph {et~al.}(2008)\citenamefont
  {Bradač}, \citenamefont {Allen}, \citenamefont {Treu}, \citenamefont
  {Ebeling}, \citenamefont {Massey}, \citenamefont {Morris}, \citenamefont
  {von~der Linden},\ and\ \citenamefont {Applegate}}]{Bradac:2008eu}%
  \BibitemOpen
  \bibfield  {author} {\bibinfo {author} {\bibfnamefont {M.}~\bibnamefont
  {Bradač}}, \bibinfo {author} {\bibfnamefont {S.~W.}\ \bibnamefont {Allen}},
  \bibinfo {author} {\bibfnamefont {T.}~\bibnamefont {Treu}}, \bibinfo {author}
  {\bibfnamefont {H.}~\bibnamefont {Ebeling}}, \bibinfo {author} {\bibfnamefont
  {R.}~\bibnamefont {Massey}}, \bibinfo {author} {\bibfnamefont {R.~G.}\
  \bibnamefont {Morris}}, \bibinfo {author} {\bibfnamefont {A.}~\bibnamefont
  {von~der Linden}}, \ and\ \bibinfo {author} {\bibfnamefont {D.}~\bibnamefont
  {Applegate}},\ }\href {\doibase 10.1086/591246} {\bibfield  {journal}
  {\bibinfo  {journal} {Astrophys. J.}\ }\textbf {\bibinfo {volume} {687}},\
  \bibinfo {pages} {959} (\bibinfo {year} {2008})},\ \Eprint
  {http://arxiv.org/abs/0806.2320} {arXiv:0806.2320 [astro-ph]} \BibitemShut
  {NoStop}%
%%CITATION = ARXIV:0806.2320;%%
\bibitem [{\citenamefont {Massey}\ \emph {et~al.}(2015)\citenamefont {Massey}
  \emph {et~al.}}]{Massey:2015dkw}%
  \BibitemOpen
  \bibfield  {author} {\bibinfo {author} {\bibfnamefont {R.}~\bibnamefont
  {Massey}} \emph {et~al.},\ }\href {\doibase 10.1093/mnras/stv467} {\bibfield
  {journal} {\bibinfo  {journal} {Mon. Not. Roy. Astron. Soc.}\ }\textbf
  {\bibinfo {volume} {449}},\ \bibinfo {pages} {3393} (\bibinfo {year}
  {2015})},\ \Eprint {http://arxiv.org/abs/1504.03388} {arXiv:1504.03388
  [astro-ph.CO]} \BibitemShut {NoStop}%
%%CITATION = ARXIV:1504.03388;%%
\bibitem [{\citenamefont {Clifton}\ \emph {et~al.}(2010)\citenamefont
  {Clifton}, \citenamefont {Bañados},\ and\ \citenamefont
  {Skordis}}]{Clifton:2010hz}%
  \BibitemOpen
  \bibfield  {author} {\bibinfo {author} {\bibfnamefont {T.}~\bibnamefont
  {Clifton}}, \bibinfo {author} {\bibfnamefont {M.}~\bibnamefont {Bañados}}, \
  and\ \bibinfo {author} {\bibfnamefont {C.}~\bibnamefont {Skordis}},\ }\href
  {\doibase 10.1088/0264-9381/27/23/235020} {\bibfield  {journal} {\bibinfo
  {journal} {Class. Quant. Grav.}\ }\textbf {\bibinfo {volume} {27}},\ \bibinfo
  {pages} {235020} (\bibinfo {year} {2010})},\ \Eprint
  {http://arxiv.org/abs/1006.5619} {arXiv:1006.5619 [gr-qc]} \BibitemShut
  {NoStop}%
%%CITATION = ARXIV:1006.5619;%%
\bibitem [{\citenamefont {Hohmann}\ and\ \citenamefont
  {Wohlfarth}(2010)}]{Hohmann:2010ni}%
  \BibitemOpen
  \bibfield  {author} {\bibinfo {author} {\bibfnamefont {M.}~\bibnamefont
  {Hohmann}}\ and\ \bibinfo {author} {\bibfnamefont {M.~N.~R.}\ \bibnamefont
  {Wohlfarth}},\ }\href {\doibase 10.1103/PhysRevD.82.084028} {\bibfield
  {journal} {\bibinfo  {journal} {Phys. Rev.}\ }\textbf {\bibinfo {volume}
  {D82}},\ \bibinfo {pages} {084028} (\bibinfo {year} {2010})},\ \Eprint
  {http://arxiv.org/abs/1007.4945} {arXiv:1007.4945 [gr-qc]} \BibitemShut
  {NoStop}%
%%CITATION = ARXIV:1007.4945;%%
\bibitem [{\citenamefont {Hohmann}(2014)}]{Hohmann:2013oca}%
  \BibitemOpen
  \bibfield  {author} {\bibinfo {author} {\bibfnamefont {M.}~\bibnamefont
  {Hohmann}},\ }\href {\doibase 10.1088/0264-9381/31/13/135003} {\bibfield
  {journal} {\bibinfo  {journal} {Class. Quant. Grav.}\ }\textbf {\bibinfo
  {volume} {31}},\ \bibinfo {pages} {135003} (\bibinfo {year} {2014})},\
  \Eprint {http://arxiv.org/abs/1309.7787} {arXiv:1309.7787 [gr-qc]}
  \BibitemShut {NoStop}%
%%CITATION = ARXIV:1309.7787;%%
\bibitem [{\citenamefont {Vainshtein}(1972)}]{Vainshtein:1972sx}%
  \BibitemOpen
  \bibfield  {author} {\bibinfo {author} {\bibfnamefont {A.~I.}\ \bibnamefont
  {Vainshtein}},\ }\href {\doibase 10.1016/0370-2693(72)90147-5} {\bibfield
  {journal} {\bibinfo  {journal} {Phys. Lett.}\ }\textbf {\bibinfo {volume}
  {B39}},\ \bibinfo {pages} {393} (\bibinfo {year} {1972})}\BibitemShut
  {NoStop}%
%%CITATION = PHLTA,B39,393;%%
\bibitem [{\citenamefont {Babichev}\ and\ \citenamefont
  {Crisostomi}(2013)}]{Babichev:2013pfa}%
  \BibitemOpen
  \bibfield  {author} {\bibinfo {author} {\bibfnamefont {E.}~\bibnamefont
  {Babichev}}\ and\ \bibinfo {author} {\bibfnamefont {M.}~\bibnamefont
  {Crisostomi}},\ }\href {\doibase 10.1103/PhysRevD.88.084002} {\bibfield
  {journal} {\bibinfo  {journal} {Phys. Rev.}\ }\textbf {\bibinfo {volume}
  {D88}},\ \bibinfo {pages} {084002} (\bibinfo {year} {2013})},\ \Eprint
  {http://arxiv.org/abs/1307.3640} {arXiv:1307.3640 [gr-qc]} \BibitemShut
  {NoStop}%
%%CITATION = ARXIV:1307.3640;%%
\bibitem [{\citenamefont {Babichev}\ and\ \citenamefont
  {Deffayet}(2013)}]{Babichev:2013usa}%
  \BibitemOpen
  \bibfield  {author} {\bibinfo {author} {\bibfnamefont {E.}~\bibnamefont
  {Babichev}}\ and\ \bibinfo {author} {\bibfnamefont {C.}~\bibnamefont
  {Deffayet}},\ }\href {\doibase 10.1088/0264-9381/30/18/184001} {\bibfield
  {journal} {\bibinfo  {journal} {Class. Quant. Grav.}\ }\textbf {\bibinfo
  {volume} {30}},\ \bibinfo {pages} {184001} (\bibinfo {year} {2013})},\
  \Eprint {http://arxiv.org/abs/1304.7240} {arXiv:1304.7240 [gr-qc]}
  \BibitemShut {NoStop}%
%%CITATION = ARXIV:1304.7240;%%
\bibitem [{\citenamefont {Bardeen}(1980)}]{Bardeen:1980kt}%
  \BibitemOpen
  \bibfield  {author} {\bibinfo {author} {\bibfnamefont {J.~M.}\ \bibnamefont
  {Bardeen}},\ }\href {\doibase 10.1103/PhysRevD.22.1882} {\bibfield  {journal}
  {\bibinfo  {journal} {Phys. Rev.}\ }\textbf {\bibinfo {volume} {D22}},\
  \bibinfo {pages} {1882} (\bibinfo {year} {1980})}\BibitemShut {NoStop}%
%%CITATION = PHRVA,D22,1882;%%
\bibitem [{\citenamefont {Malik}\ and\ \citenamefont
  {Wands}(2009)}]{Malik:2008im}%
  \BibitemOpen
  \bibfield  {author} {\bibinfo {author} {\bibfnamefont {K.~A.}\ \bibnamefont
  {Malik}}\ and\ \bibinfo {author} {\bibfnamefont {D.}~\bibnamefont {Wands}},\
  }\href {\doibase 10.1016/j.physrep.2009.03.001} {\bibfield  {journal}
  {\bibinfo  {journal} {Phys. Rept.}\ }\textbf {\bibinfo {volume} {475}},\
  \bibinfo {pages} {1} (\bibinfo {year} {2009})},\ \Eprint
  {http://arxiv.org/abs/0809.4944} {arXiv:0809.4944 [astro-ph]} \BibitemShut
  {NoStop}%
%%CITATION = ARXIV:0809.4944;%%
\bibitem [{\citenamefont {Stewart}(1990)}]{Stewart:1990fm}%
  \BibitemOpen
  \bibfield  {author} {\bibinfo {author} {\bibfnamefont {J.~M.}\ \bibnamefont
  {Stewart}},\ }\href {\doibase 10.1088/0264-9381/7/7/013} {\bibfield
  {journal} {\bibinfo  {journal} {Class. Quant. Grav.}\ }\textbf {\bibinfo
  {volume} {7}},\ \bibinfo {pages} {1169} (\bibinfo {year} {1990})}\BibitemShut
  {NoStop}%
%%CITATION = CQGRD,7,1169;%%
\bibitem [{\citenamefont {Hohmann}\ \emph {et~al.}(2013)\citenamefont
  {Hohmann}, \citenamefont {Jarv}, \citenamefont {Kuusk},\ and\ \citenamefont
  {Randla}}]{Hohmann:2013rba}%
  \BibitemOpen
  \bibfield  {author} {\bibinfo {author} {\bibfnamefont {M.}~\bibnamefont
  {Hohmann}}, \bibinfo {author} {\bibfnamefont {L.}~\bibnamefont {Jarv}},
  \bibinfo {author} {\bibfnamefont {P.}~\bibnamefont {Kuusk}}, \ and\ \bibinfo
  {author} {\bibfnamefont {E.}~\bibnamefont {Randla}},\ }\href {\doibase
  10.1103/PhysRevD.89.069901, 10.1103/PhysRevD.88.084054} {\bibfield  {journal}
  {\bibinfo  {journal} {Phys. Rev.}\ }\textbf {\bibinfo {volume} {D88}},\
  \bibinfo {pages} {084054} (\bibinfo {year} {2013})},\ \bibinfo {note}
  {[Erratum: Phys. Rev.D89,no.6,069901(2014)]},\ \Eprint
  {http://arxiv.org/abs/1309.0031} {arXiv:1309.0031 [gr-qc]} \BibitemShut
  {NoStop}%
%%CITATION = ARXIV:1309.0031;%%
\bibitem [{\citenamefont {Schärer}\ \emph {et~al.}(2014)\citenamefont
  {Schärer}, \citenamefont {Angélil}, \citenamefont {Bondarescu},
  \citenamefont {Jetzer},\ and\ \citenamefont {Lundgren}}]{Scharer:2014kya}%
  \BibitemOpen
  \bibfield  {author} {\bibinfo {author} {\bibfnamefont {A.}~\bibnamefont
  {Schärer}}, \bibinfo {author} {\bibfnamefont {R.}~\bibnamefont {Angélil}},
  \bibinfo {author} {\bibfnamefont {R.}~\bibnamefont {Bondarescu}}, \bibinfo
  {author} {\bibfnamefont {P.}~\bibnamefont {Jetzer}}, \ and\ \bibinfo {author}
  {\bibfnamefont {A.}~\bibnamefont {Lundgren}},\ }\href {\doibase
  10.1103/PhysRevD.90.123005} {\bibfield  {journal} {\bibinfo  {journal} {Phys.
  Rev.}\ }\textbf {\bibinfo {volume} {D90}},\ \bibinfo {pages} {123005}
  (\bibinfo {year} {2014})},\ \Eprint {http://arxiv.org/abs/1410.7914}
  {arXiv:1410.7914 [gr-qc]} \BibitemShut {NoStop}%
%%CITATION = ARXIV:1410.7914;%%
\bibitem [{\citenamefont {Hohmann}(2015)}]{Hohmann:2015kra}%
  \BibitemOpen
  \bibfield  {author} {\bibinfo {author} {\bibfnamefont {M.}~\bibnamefont
  {Hohmann}},\ }\href {\doibase 10.1103/PhysRevD.92.064019} {\bibfield
  {journal} {\bibinfo  {journal} {Phys. Rev.}\ }\textbf {\bibinfo {volume}
  {D92}},\ \bibinfo {pages} {064019} (\bibinfo {year} {2015})},\ \Eprint
  {http://arxiv.org/abs/1506.04253} {arXiv:1506.04253 [gr-qc]} \BibitemShut
  {NoStop}%
%%CITATION = ARXIV:1506.04253;%%
\bibitem [{\citenamefont {Bueno}\ \emph {et~al.}(2016)\citenamefont {Bueno},
  \citenamefont {Cano}, \citenamefont {Min},\ and\ \citenamefont
  {Visser}}]{Bueno:2016ypa}%
  \BibitemOpen
  \bibfield  {author} {\bibinfo {author} {\bibfnamefont {P.}~\bibnamefont
  {Bueno}}, \bibinfo {author} {\bibfnamefont {P.~A.}\ \bibnamefont {Cano}},
  \bibinfo {author} {\bibfnamefont {V.~S.}\ \bibnamefont {Min}}, \ and\
  \bibinfo {author} {\bibfnamefont {M.~R.}\ \bibnamefont {Visser}},\
  }\href@noop {} {\  (\bibinfo {year} {2016})},\ \Eprint
  {http://arxiv.org/abs/1610.08519} {arXiv:1610.08519 [hep-th]} \BibitemShut
  {NoStop}%
%%CITATION = ARXIV:1610.08519;%%
\bibitem [{\citenamefont {Deng}\ and\ \citenamefont
  {Xie}(2016)}]{Deng:2016moh}%
  \BibitemOpen
  \bibfield  {author} {\bibinfo {author} {\bibfnamefont {X.-M.}\ \bibnamefont
  {Deng}}\ and\ \bibinfo {author} {\bibfnamefont {Y.}~\bibnamefont {Xie}},\
  }\href {\doibase 10.1103/PhysRevD.93.044013} {\bibfield  {journal} {\bibinfo
  {journal} {Phys. Rev.}\ }\textbf {\bibinfo {volume} {D93}},\ \bibinfo {pages}
  {044013} (\bibinfo {year} {2016})}\BibitemShut {NoStop}%
%%CITATION = PHRVA,D93,044013;%%
\bibitem [{\citenamefont {Harvey}\ \emph {et~al.}(2015)\citenamefont {Harvey},
  \citenamefont {Massey}, \citenamefont {Kitching}, \citenamefont {Taylor},\
  and\ \citenamefont {Tittley}}]{Harvey:2015hha}%
  \BibitemOpen
  \bibfield  {author} {\bibinfo {author} {\bibfnamefont {D.}~\bibnamefont
  {Harvey}}, \bibinfo {author} {\bibfnamefont {R.}~\bibnamefont {Massey}},
  \bibinfo {author} {\bibfnamefont {T.}~\bibnamefont {Kitching}}, \bibinfo
  {author} {\bibfnamefont {A.}~\bibnamefont {Taylor}}, \ and\ \bibinfo {author}
  {\bibfnamefont {E.}~\bibnamefont {Tittley}},\ }\href {\doibase
  10.1126/science.1261381} {\bibfield  {journal} {\bibinfo  {journal}
  {Science}\ }\textbf {\bibinfo {volume} {347}},\ \bibinfo {pages} {1462}
  (\bibinfo {year} {2015})},\ \Eprint {http://arxiv.org/abs/1503.07675}
  {arXiv:1503.07675 [astro-ph.CO]} \BibitemShut {NoStop}%
%%CITATION = ARXIV:1503.07675;%%
\bibitem [{\citenamefont {Heikinheimo}\ \emph {et~al.}(2015)\citenamefont
  {Heikinheimo}, \citenamefont {Raidal}, \citenamefont {Spethmann},\ and\
  \citenamefont {Veermäe}}]{Heikinheimo:2015kra}%
  \BibitemOpen
  \bibfield  {author} {\bibinfo {author} {\bibfnamefont {M.}~\bibnamefont
  {Heikinheimo}}, \bibinfo {author} {\bibfnamefont {M.}~\bibnamefont {Raidal}},
  \bibinfo {author} {\bibfnamefont {C.}~\bibnamefont {Spethmann}}, \ and\
  \bibinfo {author} {\bibfnamefont {H.}~\bibnamefont {Veermäe}},\ }\href
  {\doibase 10.1016/j.physletb.2015.08.012} {\bibfield  {journal} {\bibinfo
  {journal} {Phys. Lett.}\ }\textbf {\bibinfo {volume} {B749}},\ \bibinfo
  {pages} {236} (\bibinfo {year} {2015})},\ \Eprint
  {http://arxiv.org/abs/1504.04371} {arXiv:1504.04371 [hep-ph]} \BibitemShut
  {NoStop}%
%%CITATION = ARXIV:1504.04371;%%
\bibitem [{\citenamefont {Kahlhoefer}\ \emph {et~al.}(2015)\citenamefont
  {Kahlhoefer}, \citenamefont {Schmidt-Hoberg}, \citenamefont {Kummer},\ and\
  \citenamefont {Sarkar}}]{Kahlhoefer:2015vua}%
  \BibitemOpen
  \bibfield  {author} {\bibinfo {author} {\bibfnamefont {F.}~\bibnamefont
  {Kahlhoefer}}, \bibinfo {author} {\bibfnamefont {K.}~\bibnamefont
  {Schmidt-Hoberg}}, \bibinfo {author} {\bibfnamefont {J.}~\bibnamefont
  {Kummer}}, \ and\ \bibinfo {author} {\bibfnamefont {S.}~\bibnamefont
  {Sarkar}},\ }\href {\doibase 10.1093/mnrasl/slv088} {\bibfield  {journal}
  {\bibinfo  {journal} {Mon. Not. Roy. Astron. Soc.}\ }\textbf {\bibinfo
  {volume} {452}},\ \bibinfo {pages} {L54} (\bibinfo {year} {2015})},\ \Eprint
  {http://arxiv.org/abs/1504.06576} {arXiv:1504.06576 [astro-ph.CO]}
  \BibitemShut {NoStop}%
%%CITATION = ARXIV:1504.06576;%%
\bibitem [{\citenamefont {Sepp}\ \emph {et~al.}(2016)\citenamefont {Sepp},
  \citenamefont {Deshev}, \citenamefont {Heikinheimo}, \citenamefont {Hektor},
  \citenamefont {Raidal}, \citenamefont {Spethmann}, \citenamefont {Tempel},\
  and\ \citenamefont {Veermäe}}]{Sepp:2016tfs}%
  \BibitemOpen
  \bibfield  {author} {\bibinfo {author} {\bibfnamefont {T.}~\bibnamefont
  {Sepp}}, \bibinfo {author} {\bibfnamefont {B.}~\bibnamefont {Deshev}},
  \bibinfo {author} {\bibfnamefont {M.}~\bibnamefont {Heikinheimo}}, \bibinfo
  {author} {\bibfnamefont {A.}~\bibnamefont {Hektor}}, \bibinfo {author}
  {\bibfnamefont {M.}~\bibnamefont {Raidal}}, \bibinfo {author} {\bibfnamefont
  {C.}~\bibnamefont {Spethmann}}, \bibinfo {author} {\bibfnamefont
  {E.}~\bibnamefont {Tempel}}, \ and\ \bibinfo {author} {\bibfnamefont
  {H.}~\bibnamefont {Veermäe}},\ }\href@noop {} {\  (\bibinfo {year}
  {2016})},\ \Eprint {http://arxiv.org/abs/1603.07324} {arXiv:1603.07324
  [astro-ph.CO]} \BibitemShut {NoStop}%
%%CITATION = ARXIV:1603.07324;%%
\bibitem [{\citenamefont {Bolton}\ \emph {et~al.}(2008)\citenamefont {Bolton},
  \citenamefont {Burles}, \citenamefont {Koopmans}, \citenamefont {Treu},
  \citenamefont {Gavazzi}, \citenamefont {Moustakas}, \citenamefont {Wayth},\
  and\ \citenamefont {Schlegel}}]{Bolton:2008xf}%
  \BibitemOpen
  \bibfield  {author} {\bibinfo {author} {\bibfnamefont {A.~S.}\ \bibnamefont
  {Bolton}}, \bibinfo {author} {\bibfnamefont {S.}~\bibnamefont {Burles}},
  \bibinfo {author} {\bibfnamefont {L.~V.~E.}\ \bibnamefont {Koopmans}},
  \bibinfo {author} {\bibfnamefont {T.}~\bibnamefont {Treu}}, \bibinfo {author}
  {\bibfnamefont {R.}~\bibnamefont {Gavazzi}}, \bibinfo {author} {\bibfnamefont
  {L.~A.}\ \bibnamefont {Moustakas}}, \bibinfo {author} {\bibfnamefont
  {R.}~\bibnamefont {Wayth}}, \ and\ \bibinfo {author} {\bibfnamefont {D.~J.}\
  \bibnamefont {Schlegel}},\ }\href {\doibase 10.1086/589327} {\bibfield
  {journal} {\bibinfo  {journal} {Astrophys. J.}\ }\textbf {\bibinfo {volume}
  {682}},\ \bibinfo {pages} {964} (\bibinfo {year} {2008})},\ \Eprint
  {http://arxiv.org/abs/0805.1931} {arXiv:0805.1931 [astro-ph]} \BibitemShut
  {NoStop}%
%%CITATION = ARXIV:0805.1931;%%
\bibitem [{\citenamefont {Cao}\ \emph {et~al.}(2015)\citenamefont {Cao},
  \citenamefont {Biesiada}, \citenamefont {Gavazzi}, \citenamefont
  {Piórkowska},\ and\ \citenamefont {Zhu}}]{Cao:2015qja}%
  \BibitemOpen
  \bibfield  {author} {\bibinfo {author} {\bibfnamefont {S.}~\bibnamefont
  {Cao}}, \bibinfo {author} {\bibfnamefont {M.}~\bibnamefont {Biesiada}},
  \bibinfo {author} {\bibfnamefont {R.}~\bibnamefont {Gavazzi}}, \bibinfo
  {author} {\bibfnamefont {A.}~\bibnamefont {Piórkowska}}, \ and\ \bibinfo
  {author} {\bibfnamefont {Z.-H.}\ \bibnamefont {Zhu}},\ }\href {\doibase
  10.1088/0004-637X/806/2/185} {\bibfield  {journal} {\bibinfo  {journal}
  {Astrophys. J.}\ }\textbf {\bibinfo {volume} {806}},\ \bibinfo {pages} {185}
  (\bibinfo {year} {2015})},\ \Eprint {http://arxiv.org/abs/1509.07649}
  {arXiv:1509.07649 [astro-ph.CO]} \BibitemShut {NoStop}%
%%CITATION = ARXIV:1509.07649;%%
\bibitem [{\citenamefont {Sanghai}\ and\ \citenamefont
  {Clifton}(2016)}]{Sanghai:2016tbi}%
  \BibitemOpen
  \bibfield  {author} {\bibinfo {author} {\bibfnamefont {V.~A.~A.}\
  \bibnamefont {Sanghai}}\ and\ \bibinfo {author} {\bibfnamefont
  {T.}~\bibnamefont {Clifton}},\ }\href@noop {} {\  (\bibinfo {year} {2016})},\
  \Eprint {http://arxiv.org/abs/1610.08039} {arXiv:1610.08039 [gr-qc]}
  \BibitemShut {NoStop}%
%%CITATION = ARXIV:1610.08039;%%
\bibitem [{\citenamefont {Hinterbichler}\ and\ \citenamefont
  {Rosen}(2012)}]{Hinterbichler:2012cn}%
  \BibitemOpen
  \bibfield  {author} {\bibinfo {author} {\bibfnamefont {K.}~\bibnamefont
  {Hinterbichler}}\ and\ \bibinfo {author} {\bibfnamefont {R.~A.}\ \bibnamefont
  {Rosen}},\ }\href {\doibase 10.1007/JHEP07(2012)047} {\bibfield  {journal}
  {\bibinfo  {journal} {JHEP}\ }\textbf {\bibinfo {volume} {07}},\ \bibinfo
  {pages} {047} (\bibinfo {year} {2012})},\ \Eprint
  {http://arxiv.org/abs/1203.5783} {arXiv:1203.5783 [hep-th]} \BibitemShut
  {NoStop}%
%%CITATION = ARXIV:1203.5783;%%
\bibitem [{\citenamefont {Hassan}\ \emph
  {et~al.}(2012{\natexlab{b}})\citenamefont {Hassan}, \citenamefont
  {Schmidt-May},\ and\ \citenamefont {von Strauss}}]{Hassan:2012wt}%
  \BibitemOpen
  \bibfield  {author} {\bibinfo {author} {\bibfnamefont {S.~F.}\ \bibnamefont
  {Hassan}}, \bibinfo {author} {\bibfnamefont {A.}~\bibnamefont {Schmidt-May}},
  \ and\ \bibinfo {author} {\bibfnamefont {M.}~\bibnamefont {von Strauss}},\
  }\href@noop {} {\  (\bibinfo {year} {2012}{\natexlab{b}})},\ \Eprint
  {http://arxiv.org/abs/1204.5202} {arXiv:1204.5202 [hep-th]} \BibitemShut
  {NoStop}%
%%CITATION = ARXIV:1204.5202;%%
\bibitem [{\citenamefont {Nomura}\ and\ \citenamefont
  {Soda}(2012)}]{Nomura:2012xr}%
  \BibitemOpen
  \bibfield  {author} {\bibinfo {author} {\bibfnamefont {K.}~\bibnamefont
  {Nomura}}\ and\ \bibinfo {author} {\bibfnamefont {J.}~\bibnamefont {Soda}},\
  }\href {\doibase 10.1103/PhysRevD.86.084052} {\bibfield  {journal} {\bibinfo
  {journal} {Phys. Rev.}\ }\textbf {\bibinfo {volume} {D86}},\ \bibinfo {pages}
  {084052} (\bibinfo {year} {2012})},\ \Eprint {http://arxiv.org/abs/1207.3637}
  {arXiv:1207.3637 [hep-th]} \BibitemShut {NoStop}%
%%CITATION = ARXIV:1207.3637;%%
\bibitem [{\citenamefont {Noller}\ and\ \citenamefont
  {Melville}(2015)}]{Noller:2014sta}%
  \BibitemOpen
  \bibfield  {author} {\bibinfo {author} {\bibfnamefont {J.}~\bibnamefont
  {Noller}}\ and\ \bibinfo {author} {\bibfnamefont {S.}~\bibnamefont
  {Melville}},\ }\href {\doibase 10.1088/1475-7516/2015/01/003} {\bibfield
  {journal} {\bibinfo  {journal} {JCAP}\ }\textbf {\bibinfo {volume} {1501}},\
  \bibinfo {pages} {003} (\bibinfo {year} {2015})},\ \Eprint
  {http://arxiv.org/abs/1408.5131} {arXiv:1408.5131 [hep-th]} \BibitemShut
  {NoStop}%
%%CITATION = ARXIV:1408.5131;%%
\bibitem [{\citenamefont {Scargill}\ \emph {et~al.}(2014)\citenamefont
  {Scargill}, \citenamefont {Noller},\ and\ \citenamefont
  {Ferreira}}]{Scargill:2014wya}%
  \BibitemOpen
  \bibfield  {author} {\bibinfo {author} {\bibfnamefont {J.~H.~C.}\
  \bibnamefont {Scargill}}, \bibinfo {author} {\bibfnamefont {J.}~\bibnamefont
  {Noller}}, \ and\ \bibinfo {author} {\bibfnamefont {P.~G.}\ \bibnamefont
  {Ferreira}},\ }\href {\doibase 10.1007/JHEP12(2014)160} {\bibfield  {journal}
  {\bibinfo  {journal} {JHEP}\ }\textbf {\bibinfo {volume} {12}},\ \bibinfo
  {pages} {160} (\bibinfo {year} {2014})},\ \Eprint
  {http://arxiv.org/abs/1410.7774} {arXiv:1410.7774 [hep-th]} \BibitemShut
  {NoStop}%
%%CITATION = ARXIV:1410.7774;%%
\bibitem [{\citenamefont {Hinterbichler}\ and\ \citenamefont
  {Rosen}(2015)}]{Hinterbichler:2015yaa}%
  \BibitemOpen
  \bibfield  {author} {\bibinfo {author} {\bibfnamefont {K.}~\bibnamefont
  {Hinterbichler}}\ and\ \bibinfo {author} {\bibfnamefont {R.~A.}\ \bibnamefont
  {Rosen}},\ }\href {\doibase 10.1103/PhysRevD.92.024030} {\bibfield  {journal}
  {\bibinfo  {journal} {Phys. Rev.}\ }\textbf {\bibinfo {volume} {D92}},\
  \bibinfo {pages} {024030} (\bibinfo {year} {2015})},\ \Eprint
  {http://arxiv.org/abs/1503.06796} {arXiv:1503.06796 [hep-th]} \BibitemShut
  {NoStop}%
%%CITATION = ARXIV:1503.06796;%%
\bibitem [{\citenamefont {Noller}\ and\ \citenamefont
  {Scargill}(2015)}]{Noller:2015eda}%
  \BibitemOpen
  \bibfield  {author} {\bibinfo {author} {\bibfnamefont {J.}~\bibnamefont
  {Noller}}\ and\ \bibinfo {author} {\bibfnamefont {J.~H.~C.}\ \bibnamefont
  {Scargill}},\ }\href {\doibase 10.1007/JHEP05(2015)034} {\bibfield  {journal}
  {\bibinfo  {journal} {JHEP}\ }\textbf {\bibinfo {volume} {05}},\ \bibinfo
  {pages} {034} (\bibinfo {year} {2015})},\ \Eprint
  {http://arxiv.org/abs/1503.02700} {arXiv:1503.02700 [hep-th]} \BibitemShut
  {NoStop}%
%%CITATION = ARXIV:1503.02700;%%
\bibitem [{\citenamefont {Akrami}\ \emph {et~al.}(2013)\citenamefont {Akrami},
  \citenamefont {Koivisto}, \citenamefont {Mota},\ and\ \citenamefont
  {Sandstad}}]{Akrami:2013ffa}%
  \BibitemOpen
  \bibfield  {author} {\bibinfo {author} {\bibfnamefont {Y.}~\bibnamefont
  {Akrami}}, \bibinfo {author} {\bibfnamefont {T.~S.}\ \bibnamefont
  {Koivisto}}, \bibinfo {author} {\bibfnamefont {D.~F.}\ \bibnamefont {Mota}},
  \ and\ \bibinfo {author} {\bibfnamefont {M.}~\bibnamefont {Sandstad}},\
  }\href {\doibase 10.1088/1475-7516/2013/10/046} {\bibfield  {journal}
  {\bibinfo  {journal} {JCAP}\ }\textbf {\bibinfo {volume} {1310}},\ \bibinfo
  {pages} {046} (\bibinfo {year} {2013})},\ \Eprint
  {http://arxiv.org/abs/1306.0004} {arXiv:1306.0004 [hep-th]} \BibitemShut
  {NoStop}%
%%CITATION = ARXIV:1306.0004;%%
\bibitem [{\citenamefont {de~Rham}\ \emph
  {et~al.}(2014{\natexlab{b}})\citenamefont {de~Rham}, \citenamefont
  {Heisenberg},\ and\ \citenamefont {Ribeiro}}]{deRham:2014fha}%
  \BibitemOpen
  \bibfield  {author} {\bibinfo {author} {\bibfnamefont {C.}~\bibnamefont
  {de~Rham}}, \bibinfo {author} {\bibfnamefont {L.}~\bibnamefont {Heisenberg}},
  \ and\ \bibinfo {author} {\bibfnamefont {R.~H.}\ \bibnamefont {Ribeiro}},\
  }\href {\doibase 10.1103/PhysRevD.90.124042} {\bibfield  {journal} {\bibinfo
  {journal} {Phys. Rev.}\ }\textbf {\bibinfo {volume} {D90}},\ \bibinfo {pages}
  {124042} (\bibinfo {year} {2014}{\natexlab{b}})},\ \Eprint
  {http://arxiv.org/abs/1409.3834} {arXiv:1409.3834 [hep-th]} \BibitemShut
  {NoStop}%
%%CITATION = ARXIV:1409.3834;%%
\bibitem [{\citenamefont {Huang}\ \emph {et~al.}(2015)\citenamefont {Huang},
  \citenamefont {Ribeiro}, \citenamefont {Xing}, \citenamefont {Zhang},\ and\
  \citenamefont {Zhou}}]{Huang:2015yga}%
  \BibitemOpen
  \bibfield  {author} {\bibinfo {author} {\bibfnamefont {Q.-G.}\ \bibnamefont
  {Huang}}, \bibinfo {author} {\bibfnamefont {R.~H.}\ \bibnamefont {Ribeiro}},
  \bibinfo {author} {\bibfnamefont {Y.-H.}\ \bibnamefont {Xing}}, \bibinfo
  {author} {\bibfnamefont {K.-C.}\ \bibnamefont {Zhang}}, \ and\ \bibinfo
  {author} {\bibfnamefont {S.-Y.}\ \bibnamefont {Zhou}},\ }\href {\doibase
  10.1016/j.physletb.2015.07.003} {\bibfield  {journal} {\bibinfo  {journal}
  {Phys. Lett.}\ }\textbf {\bibinfo {volume} {B748}},\ \bibinfo {pages} {356}
  (\bibinfo {year} {2015})},\ \Eprint {http://arxiv.org/abs/1505.02616}
  {arXiv:1505.02616 [hep-th]} \BibitemShut {NoStop}%
%%CITATION = ARXIV:1505.02616;%%
\bibitem [{\citenamefont {Melville}\ and\ \citenamefont
  {Noller}(2016)}]{Melville:2015dba}%
  \BibitemOpen
  \bibfield  {author} {\bibinfo {author} {\bibfnamefont {S.}~\bibnamefont
  {Melville}}\ and\ \bibinfo {author} {\bibfnamefont {J.}~\bibnamefont
  {Noller}},\ }\href {\doibase 10.1007/JHEP01(2016)094} {\bibfield  {journal}
  {\bibinfo  {journal} {JHEP}\ }\textbf {\bibinfo {volume} {01}},\ \bibinfo
  {pages} {094} (\bibinfo {year} {2016})},\ \Eprint
  {http://arxiv.org/abs/1511.01485} {arXiv:1511.01485 [hep-th]} \BibitemShut
  {NoStop}%
%%CITATION = ARXIV:1511.01485;%%
\bibitem [{\citenamefont {Frahm}(1983)}]{Frahm:1983}%
  \BibitemOpen
  \bibfield  {author} {\bibinfo {author} {\bibfnamefont {C.~P.}\ \bibnamefont
  {Frahm}},\ }\href {\doibase 10.1119/1.13127} {\bibfield  {journal} {\bibinfo
  {journal} {American Journal of Physics}\ }\textbf {\bibinfo {volume} {51}},\
  \bibinfo {pages} {826} (\bibinfo {year} {1983})}\BibitemShut {NoStop}%
\end{thebibliography}%
\end{document}